\documentclass[12pt]{article}
\pdfoutput=1

\usepackage{cite} 
\usepackage{amsmath}
\usepackage{amssymb}
\usepackage{amsfonts}
\usepackage{amsthm}
\usepackage{pstricks}
\usepackage{graphicx}
\usepackage{dcolumn}
\usepackage{bm}
\usepackage{slashed}
\usepackage{bigints}
\usepackage{fullpage}
\usepackage{authblk}
\usepackage{empheq}
\usepackage{xcolor}
\usepackage{mathrsfs}
\usepackage{sectsty}
\allsectionsfont{\centering\normalfont}

\usepackage[hidelinks]{hyperref}

\hypersetup{
    colorlinks,
    linkcolor={blue!60!black},
    citecolor={blue!60!black},
    urlcolor={blue!60!black}
}

\newcommand*\wc{{\mkern 1mu\cdot\mkern 1mu}}

\renewcommand{\Re}{\operatorname{Re}}
\renewcommand{\Im}{\operatorname{Im}}
\renewcommand{\v}[0]{\bm}

\DeclareMathOperator{\Tr}{Tr}

\def\be#1\ee{\begin{equation}#1\end{equation}}
\def\ba#1\ea{\begin{align}#1\end{align}}
\def\[#1\]{\begin{align}#1\end{align}}

\title{QED as a many-body theory of worldlines:\\ I. General formalism and infrared structure}

\author[1]{Xabier Feal}
\author[2]{Andrey Tarasov}
\author[1]{Raju Venugopalan}
\affil[1]{\small \emph{Physics Department, Bldg. 510A , 
Brookhaven National Laboratory, 
 Upton, NY 11973, USA}
}%
\affil[2]{\small \emph{
 Department of Physics, The Ohio State University, Columbus, OH 43210, USA\protect\\
Joint BNL-SBU Center for Frontiers in Nuclear Science (CFNS) at Stony Brook University\protect\\
Stony Brook, NY 11794, USA}}

\date{\today}

\begin{document}

\maketitle

\begin{abstract}
We discuss a reformulation of QED in which matter and gauge fields are integrated out explicitly, resulting in a many-body Lorentz covariant theory of 0+1 dimensional worldlines describing super-pairs of spinning charges interacting through  Lorentz forces. This provides a powerful, string inspired definition of amplitudes to all loop orders. In particular, one obtains a more general formulation of Wilson loops and lines, with exponentiated dynamical fields and spin precession contributions, 
and worldline contour averages exactly defined through first quantized path integrals. We discuss in detail the attractive features of this formalism for high order perturbative computations. 
We show that worldline S-matrix elements, to all loop orders in perturbation theory, can be constructed to be manifestly free of soft singularities, with infrared (IR) divergences captured and removed by endpoint photon exchanges at infinity that are equivalent to the soft coherent dressings of the Dyson S-matrix proposed by Faddeev and Kulish.  We discuss these IR structures and make connections with soft theorems, the Abelian exponentiation of IR divergences and cusp anomalous dimensions. Follow-up papers will discuss the efficient computation of cusp anomalous dimensions and universal features of soft theorems.

\end{abstract}

\newpage
\begingroup
\hypersetup{linkcolor=black}
\tableofcontents
\endgroup

\newpage 

\section{Introduction}

The infrared (IR) and ultraviolet (UV) divergences that appear in perturbative computations of amplitudes in gauge theories and gravity 
are essential to understanding the fundamental structure of quantum field theories. 
While the UV structure of amplitudes in the standard model and gravity,   are the primary focus of attention due to their sensitivity to novel physics, IR divergences are also fascinating because of their importance in understanding the overall consistency of the respective quantum field theories and the possible existence of universal features and emergent dynamics that they may share.

The IR structure of amplitudes in QCD and gravity have a number of features that are not fully understood and are a subject of active research. In the Bjorken asymptotics of QCD, understanding the absorption of IR divergences into universal non-perturbative structures are key to active work on factorization theorems that are essential to the predictive power of amplitudes~\cite{Collins:1989gx}. In the Regge limit of the theory, a ``reggeization" of amplitudes is seen to occur, with the emergent  Reggeon and Pomeron degrees of freedom capturing the leading IR dynamics of amplitudes~\cite{Lipatov:1996ts}. A further striking feature is the phenomenon of gluon saturation, where an emergent semi-hard saturation scale characterizes the maximal close packing of soft gluons~\cite{Gelis:2010nm,Kovchegov:2012mbw}.  In gravity, as in QED, soft theorems govern the structure of perturbative scattering amplitudes~\cite{PhysRev.140.B516}. However, unlike QED, the self-interactions of soft gravitons produced in trans-Planckian scattering can also lead to reggeization~\cite{Lipatov:1982vv,Melville:2013qca} and novel emergent structures, an example of which are a description of Black Holes as overoccupied self-bound gravitons~\cite{Dvali:2011aa}, in close correspondence with gluon saturation~\cite{Dvali:2021ooc}. Further, it has been argued that the existence of Black Holes is sufficient for the UV completion of gravity~\cite{Dvali:2010jz}, directly relating the UV to the IR, and signalling the breakdown of Effective Field Theory~\cite{tHooft:1993dmi,Cohen:1998zx}.

In contrast, the IR structure of QED is much simpler and does not pose any practical problem for the computation of scattering amplitudes. Nevertheless, there are nagging conceptual issues, a clearer  understanding of which may provide deeper insight into their more challenging counterparts in QCD and gravity. 
One formulation of the IR problem~\cite{Bethe:1934za} states that an infinite number of real gauge
bosons, with arbitrarily low energies, accompany the
\textit{in} and \textit{out} states in any process. Hence it is not
possible to precisely define, in the Fock sense, the interacting state of a system of charged particles. Moreover, virtual IR divergences
in any scattering transition set the corresponding S-matrix elements to zero~\cite{PhysRev.52.54}.

In Abelian theories, 
the spectrum of real and virtual IR quanta agrees with the classical expectation: it is independent of the spin dimension of matter fields and dominated by long wavelength bosons of small
momenta $k^\mu$ that are sensitive only to large spacetime regions $x^\mu$ of the interaction, $k_\mu x^\mu \sim 1$. The amplitude for the  emission or absorption of  IR quanta can therefore be computed  without precise knowledge of the physics inside the region $x^\mu$~\cite{Low:1958sn}. Its momenta are peaked along the momenta of the charged incoming and outgoing particles at the boundaries of that region, and its contribution to the amplitude can be isolated as an overall soft factor, order-by-order in perturbation theory, from the amplitude without the IR bosons attached~\cite{PhysRev.140.B516}. In non-Abelian theories, the self-interactions of gauge bosons  makes isolating such soft factors, a much more involved problem in general.

The solution to the stated IR problem in QED is usually understood to come from the unitarity of scattering amplitudes. Bloch and 
Nordsieck \cite{PhysRev.52.54} showed that transitions to final states containing an arbitrary number of soft
photons are given by the exponentiation of the single photon emission cross-section, which diverges logarithmically as $\sim
d\omega_k/\omega_k$ in the IR. When virtual photons are also attached, and one
sums over all possible configurations with real and virtual quanta
above the IR cut-off, the IR divergences cancel\footnote{Most generally, as argued by Kinoshita, Lee and Nauenberg, for theories with massless charged particles, this requires a summation over both initial and final degenerate energy states ~\cite{Kinoshita:1962ur,Lee:1964is}. For recent discussions of this KLN theorem, see \cite{Akhoury:1997pb,Frye:2018xjj} .}. Hence the total cross-section evolves depending only on the residual UV cut-off $\Lambda$ to some power $a(\gamma_{ij})$. This function depends on the initial and final momenta of the charges through geometrical factors $\gamma_{ij}$ known as cusp anomalous dimensions.

The radiative cross-sections are thus
renormalizable: for any other $\Lambda'$, they scale by a factor $(\Lambda'/\Lambda)^a$, independently of the physics of the IR. 
For instance, if one wishes to represent the radiative cross-section in terms of the non-radiative one, one can choose as $\Lambda$ the mass of the charges in the interaction \cite{PhysRev.140.B516}. Hence through renormalization, radiative cross-sections are finite, with the relevant scales set by the cross-section without real or virtual bosons. This cancellation of IR
singularities was proved to all orders by Jauch and Rohrlich
\cite{Jauch1954,Jauch1955} and by Yennie, Frautschi and Suura
\cite{YENNIE1961379}. The modern treatment demonstrating 
the classical features of soft factors in Abelian gauge
theories is due to Weinberg~\cite{PhysRev.140.B516}. In particular, he showed that this behavior is universal to gravity even though gravitons can themselves emit softer gravitons; this is because the effective coupling of soft gravitons is proportional to their energy, and therefore vanishes in the IR limit\footnote{This is not true for 
gluons emitting softer gluons, for which case, as Weinberg states, (the) {\it elimination of such complicated interlocking infrared divergences would certainly be a Herculean task, and might even not be possible.} This is the problem of reggeization and gluon saturation in QCD alluded to previously; while much progress has been made, the treatment of IR divergences is still by no means fully understood.}.

The remarkable thing about this cancellation of IR singularities in the cross-section is that it does not naturally occur at the amplitude level. IR singularities perturbatively are still infinite order by order, or if summed to all orders,  set all S-matrix elements to zero. So despite powerful proofs of consistency, the ubiquitous cancellation of IR divergences raises the question of why observables are always found IR finite in nature but this finiteness is not manifest at the amplitude level. In perturbative
calculations, interactions are typically switched off at very late and early times relative to the timescales of the process under consideration. However because softer and softer virtual and real bosons can be attached to the process under consideration,  no practical definition of the \textit{in} and
\textit{out} states can be accomplished without taking into account these interactions at asymptotic times.

Building on the work of Dollard \cite{Dollard:1964},
Chung \cite{PhysRev.140.B1110} and Kibble
\cite{Kibble:1968sfb,PhysRev.173.1527,PhysRev.174.1882,PhysRev.175.1624}, Faddeev and Kulish (FK) \cite{Kulish:1970ut}  realized\footnote{Followed by Zwanzinger in~\cite{Zwanziger:1973if}. See also \cite{Bagan:1999jf,Hannesdottir:2019opa}
for modern treatments of the problem.} that when the \textit{in} and
\textit{out} states are suitably 
dressed to take into account asymptotic
interactions, IR divergences can be systematically absorbed in these dressed states, order to
order in perturbation theory. 
Specifically, in this FK picture, the wave
functions of the incoming and outgoing charged particles are dressed with coherent states corresponding to a cloud of infinite numbers of soft photons.  The corresponding S-matrices of the dressed states are infrared finite.

The handling of IR singularities in many practical circumstances
should benefit from the definition of IR finite S-matrix
elements. As we will discuss at length in this paper, and in follow-up work, both vacuum matrix elements of operators and scattering matrix elements can be expressed in an exact worldline formulation of QED as expectation values of so-called Wilson loops, and lines, respectively, to all orders in perturbation theory. The cusp anomalous dimensions we mentioned previously in the context of total cross-sections have a natural interpretation at the amplitude level in terms of the IR and UV properties of Wilson lines, which can be shown explicitly to be IR finite to all orders in perturbation 
theory. Thus scattering cross-sections can be understood in QED entirely in terms of these IR finite structures, with all  non-trivial information contained in their UV behavior, as sketched previously.

Besides the intrinsic interest in such a manifestly IR finite formulation of amplitudes in QED, the Wilson line is also a fundamental object in Yang-Mills theories~\cite{Dotsenko:1979wb,Polyakov:1980ca}, where their UV and IR properties can be formulated in terms of evolution equations for the cusp anomalous dimensions \cite{Korchemsky:1985xj,Korchemsky:1987wg}. The Wilson line defined as such can be employed to derive the DGLAP and BFKL evolution equations for parton distributions \cite{Korchemskaya:1992je,Korchemsky:1992xv,Korchemsky:1988si}, the low energy behavior of form factors \cite{Collins:1989gx}, Drell-Yan processes and Higgs production or the re-summation of large Sudakov logarithms \cite{Collins:1981uw,Catani:1989ne}; for recent work on computing cusp anomalous dimensions to high orders in QED, QCD and ${\cal N}=4$ supersymmetric Yang-Mills theory, see \cite{PhysRevLett.126.021601,Grozin:2014hna,Grozin:2015kna,Catani:2019rvy,Henn:2019swt}. 
There is likewise work in parallel on understanding the cusp anomalous dimensions of gravity amplitudes~\cite{White:2011yy,Miller:2012an,Melville:2013qca} and possible color-kinematics dualities between such  infrared structures in QCD and gravity~\cite{Bern:2019prr}, which show great potential in simplifying computations of gravitational radiation amplitudes~\cite{Goldberger:2017frp}.

In addition to their key role in precision computations, the resurgence of interest in  soft theorems and FK
dressings in gauge theories arose independently motivated by developments in the context of gravity. It was shown around the same time as Weinberg's work by Bondi, Metzner, Van der  Burg, and by Sachs ~\cite{Bondi:1962px,Sachs:1962wk} (BMS), that degenerate vacua in gravity separated by differing numbers of soft gravitons are connected by supertranslations corresponding to spontaneously broken symmetries of the infinite dimensional BMS group of asymptotic spacetime symmetries\footnote{For a nice review of the BMS group, and subtleties in interpretations in infrared effects in gravity, see \cite{Ashtekar:2018lor}. }. More recently, it was shown by Strominger and collaborators that Weinberg's soft graviton theorem could be understood as the Ward identity corresponding to BMS invariance of gravitational amplitudes~\cite{He:2014cra,Strominger:2013jfa}. They showed subsequently that the Weinberg formula is equivalent to that of gravitational displacement (or memory) experienced by two inertial detectors following the passage of a pulse of gravitational radiation~\cite{Strominger:2014pwa}. This ``triangle" of connections  between asymptotic symmetries, soft theorems and memory has been conjectured by Strominger to be a universal feature of the infrared behavior of gauge theories and gravity~\cite{Strominger:2017zoo}.

Indeed, it was shown in \cite{He:2014cra} that scattering amplitudes in QED can be understood to satisfy an infinite dimensional symmetry 
group of large $U(1)$ gauge transformations that approach angle dependent constants on the celestial sphere at null infinity. As with the BMS group in gravity, the spontaneous breaking of this symmetry by the QED vacuum results in an infinite number of degenerate vacua characterized by differing numbers of soft photons, which are Goldstone modes on the celestial sphere. The Ward identity~\cite{Campiglia:2015qka,Kapec:2015ena,Kapec:2017tkm} corresponding to the QED BMS-like symmetries is the soft photon theorem~\cite{Low:1958sn,PhysRev.140.B516}. Completing the triangle is the QED memory effect~\cite{Susskind:2015hpa,Bieri:2013hqa}.

This approach reimagines our conventional understanding of infrared divergences in the language of symmetries of the S-matrix of charged particles. Specifically, the usual S-matrix vanishes for the emission of ultrasoft photons precisely because it does not allow for vacuum transitions between the degenerate vacua which are allowed by the BMS-like symmetry. Likewise, the FK dressed states are coherent states of Fock vacua that are eigenstates of the BMS charges at null infinity. 
The FK cancellations of infrared divergences of the S-matrix by the soft photon dressings of the incoming and outgoing charges is simply  reinterpreted as the statement that matrix elements of the S-matrix between the dressed eigenstates is finite~\cite{Kapec:2017tkm,Choi:2017ylo}. This FK phase is the previously noted Goldstone mode of large gauge transformations  satisfying a two-dimensional current algebra on the celestial sphere at null infinity~\cite{Arkani-Hamed:2020gyp}. A further intriguing prospect is the possibility of a clear dictionary between this language of asymptotic symmetries and that of  Wilson lines and cusp anomalous dimensions~\cite{Nande:2017dba}.

A question that one can ask about this program of constructing and exploring infrared finite amplitudes is their validity to all orders in perturbation theory and, as a corollary, its relative efficacy compared to conventional approaches in the computation of cross-sections for physical processes.  As indicated earlier, we will show here that this question can be cleanly formulated and addressed within the worldline formulation of QED, which is exact. While this framework for computing cross-sections in terms of manifestly IR safe structures is robust, its advantage remains to be established, though, as we will discuss, there are good reasons to believe that this may be the case.

The formulation of QED as a theory of first quantized 0+1-dimensional worldlines can be traced back to the original work of Feynman and Schwinger~\cite{PhysRev.80.440,PhysRev.84.108,Schwinger:1951nm}. However a complete treatment of spin (and color) in terms of Grassmann worldlines was only developed in the 1970s~\cite{Berezin:1976eg,Casalbuoni:1975bj,Brink:1976uf,Barducci:1976qu,Balachandran:1976ya,Barducci:1976xq,Ohnuki:1978jv}. The modern treatment of worldlines in gauge theories, following Polyakov's pioneering work~\cite{Polyakov:1980ca},  is due to Strassler \cite{Strassler:1992zr}, who also showed its equivalence to the powerful string amplitude formalism of Bern, Dixon, Dunbar and Kosower~\cite{Bern:1990cu,Bern:1991aq,Bern:1994zx,Bern:1996je} for multi-leg loop amplitudes. A classic review of the subject is due to Schubert \cite{Schubert:2001he}; for a recent update, see \cite{Edwards:2019eby}.

In the worldline formulation of QED, matter and gauge fields can be integrated out explicitly, resulting in a many-body theory of point particle bosonic and Grassmann worldlines, with the multiplicity of the closed worldlines growing with each loop order and with the external particles described by a fixed number of open worldlines. The propertime variable controls the evolution of these first quantized worldlines; UV and IR scales are encoded respectively in the short and long time behavior of the  Abelian classical Coulomb interaction, that emerges naturally exponentiated.

The UV singularities found in these worldline interactions can be mapped on to the UV divergences found order by order in perturbation theory and treated with standard renormalization methods. The striking feature of S-matrix elements, when expressed as worldlines, is that they can be constructed to be  manifestly free of soft singularities, with IR divergences captured and removed by  endpoint photon exchanges at infinity. As we will see, these asymptotic contributions exactly match the soft coherent dressings of the Dyson S-matrix proposed by Faddeev and Kulish \cite{Kulish:1970ut} to define IR finite amplitudes.

It can be argued  that the Faddeev-Kulish dressings only reshuffle IR divergences from the amplitudes to the states--see \cite{Hannesdottir:2019opa} for a nice discussion. However the same conceptual issue can be raised for any renormalization procedure eliminating singularities, be they IR or UV.  Worldline amplitudes offer a covariant proper-time  method to introduce IR and UV cut-offs in perturbation theory, with a transparent physical interpretation: the possibility of further interpretation in the language of symmetries makes it all the more attractive. In the worldline approach, both self-energy corrections to individual worldlines, and virtual  exchanges between different worldlines, are treated on a common footing. 
As noted in  \cite{Weinberg:1995mt}, the improper treatment of these leaves the cancellation of IR divergences incomplete.

Worldline methods in QED have been employed previously to compute the Schwinger pair production rate in background fields~\cite{Gelis:2015kya}, utilizing the method of worldline ``instantons"  (in propertime) to extend the computation to arbitrary coupling strengths when the background fields are weak~\cite{Affleck:1981bma,Dunne:2005sx,Dunne:2006ff}. Besides such semi-classical approaches, there have also been significant developments in worldline numerical Monte Carlo techniques~\cite{Gies:2001tj,Gies:2005sb,Schmidt:2002mt}. Though these were previously applied primarily for studies of pair production in background fields, applying these techniques to compute high order contributions scattering amplitudes in the formalism we will develop is worthy of further exploration.

The formulation of QED (in particular its IR structure) as a many-body theory of worldlines may find an important application in quantum computing where such methods were previously discussed in the context of a systematic mapping the Wigner-Weyl-Moyal formalism~\cite{Kocia-Love,Mueller:2019gjj} for the phase space distributions we discussed earlier to the full set of Clifford gates.  Similar single particle digitization frameworks have since been developed for scattering in a $\phi^4$ scalar field theory~\cite{Barata:2020jtq} and a hybrid framework for deeply inelastic scattering in Regge asymptotics~\cite{Mueller:2019qqj}. Our many-body framework allows one to extend these ideas to QED; for earlier work in this direction, see \cite{Stetina:2020abi}. Non-perturbative worldline approaches~\cite{Chandrasekharan:2008gp} 
are also under active study in the context of quantum computing. Relating these approaches to our discussion offers a possible path towards quantum computation of scattering amplitudes.

Another attractive feature of this formalism is its immediate generalization to Yang-Mills
theories, allowing one to treat on common footing the exponentiation of spinor, helicity as well as 
color degrees of freedom through the use of Grassmann variables  \cite{Brink:1976uf,10.1143/PTP.60.548,DHoker:1995aat}. In particular, 
the fact that colored Wilson lines can also be expressed in terms of the  Grassmannian counterparts~\cite{Gervais:1979fv,Arefeva:1980zd,Barducci:1980xk,Korchemsky:1987wg,DHoker:1995uyv} of the spin degrees of degrees of freedom does not appear to be widely appreciated. The Grassmann framework also admits an elegant semi-classical interpretation\footnote{These are the Bargmann-Michel-Telegdi (BMT)~\cite{PhysRevLett.2.435} equations in QED (discussed at length in Appendix~\ref{appendix_d}) and Wong's equations in QCD~\cite{Wong:1970fu}. In gravity, their worldline counterparts~\cite{Barducci:1976wc} are the  Papapetrou-Mathisson-Dixon equations~\cite{Papapetrou:1951pa,Mathisson:1937zz,Dixon:1970zza}; for a recent discussion and development of the latter in the language of effective field theories, we refer the reader to \cite{Goldberger:2020fot}.} in terms of spinning, colored particles in background fields~\cite{Barducci:1976xq,Balachandran:1976ya,Bastianelli:2013pta,Edwards:2016acz}; a canonical coordinate transformation of these variables to Grassmann bilinears representing semi-classical spin vector and color charges allows one to construct, from first principles, extended phase space distributions including these so-called\footnote{Ref.~\cite{Mueller:2019gjj} established the  equivalence of the semi-classical limit of the worldline formalism to  the Darboux phase space approach of Alekseev, Faddeev and Shatashvili~\cite{Alekseev:1988vx}.} Darboux variables~\cite{Mueller:2019gjj}.

While a detailed discussion is beyond the scope of our discussion here, we note that the worldline representation of colored Wilson lines in terms of a path integral over a local one-dimensional Grassmann action  presents potential advantages in wide range of practical QCD computations ranging from a chiral kinetic theory of the Quark-Gluon Plasma~\cite{Mueller:2017arw,Mueller:2017lzw}, to deeply inelastic scattering in the Regge limit~\cite{Tarasov:2019rfp}, to the role of the chiral anomaly in the proton's spin~\cite{Tarasov:2020cwl,Tarasov:2021yll}. For computations of IR soft factors for the problems we mentioned earlier,  unlike the QED case, one does not have an exact 
exponentiation in QCD~\cite{Dotsenko:1979wb,FRENKEL1984231,GATHERAL198390}, with IR factorization requiring a systematic eikonal and next-to-eikonal exponentiation of  so-called \textit{web} connected diagrams~\cite{Erdogan:2011yc,Laenen:2010uz,Gardi:2010rn,Gardi:2013ita}.
The web structure of non-Abelian exponentiation is reproduced in a generating functional formalism which allows for matrix exponentiation of the products of Wilson lines~\cite{Vladimirov:2015fea}. The representation of Wilson lines in terms of a local Grassmann action suggests that our framework can aid in simplifying such computations.

We envisage this paper as the first of several that will explore in some depth the infrared structure of QED as a many-body theory of worldlines. We will develop the formalism in this paper and outline some of the key features that are relevant for the IR issues we have noted. Paper II will discuss at length the computation of real soft infrared divergences and present a detailed and efficient computation of the two loop cusp anomalous dimension. Future papers will discuss the extensions of this formalism to higher orders and to QCD.  An intriguing offshoot of this work that we noted, and will pursue separately, is in the  application of non-perturbative worldline methods to quantum computing.

The paper is structured as follows. In Sections \ref{section_2}
and \ref{section_3}, we will present a formal derivation of 
worldline amplitudes in QED to all orders in perturbation theory.
In Section  \ref{section_2}, we show that the QED path integral,  involving closed vacuum loops to all orders, can equivalently be expressed in terms of first-quantized generalized Wilson loops of 
``super-pairs" of pointlike boson and Grassmann variables. 
This framework is then extended in Section \ref{section_3} to consider general scattering processes, which can be formulated exactly, to all orders, as first-quantized point-like boson and Grassmann generalized Wilson lines and loops. 
In Section \ref{section_4}, we will take these results to familiar ground by reinterpreting them in the language of  conventional Feynman diagrams. We will first show in Section \ref{section_4_1} how the worldine structures obtained in the previous sections lead to universal ``master" expressions for generalized polarization tensors that allow one to  construct compactly QED amplitudes order-by-order in perturbation theory. This  allows, in principle, efficient computations of QED processes at higher loops, where the factorial growth in the number of diagrams with each loop order in conventional perturbation theory significantly hinders the quest for precision. In Section \ref{section_4_2}, our results are expressed equivalently in terms of a generating functional for arbitrary background fields. As we will discuss, this approach is useful in strong field settings and potentially, in the use of worldline Monte Carlo methods in high order computations. 
Section \ref{section_5} examines the IR structure of the QED Dyson S-matrix in the worldline formalism. We demonstrate in Section \ref{section_5_1} a formal proof of the IR finiteness of the Faddeev-Kulish S-matrix to all orders in perturbation theory. The proof follows fundamentally from the simple asymptotic classical structure of the first-quantized worldline currents.  
Our result is illustrated for the case of M\"{o}ller scattering in 
Section \ref{section_5_2}. In particular, we show how the asymptotic real and virtual contributions of the FK S-matrix arise in this approach, and can be expressed naturally in the first-quantized language of cusps in worldline currents. In Section \ref{section_5_3}, we connect our discussion of IR safety in QED to recent discussions and alternative approaches to this problem. In Section \ref{section_6}, we recover tbe well-known result for the one loop QED cusp anomalous dimension, and discuss its computation to higher orders. As noted earlier, a further detailed discussion will be the subject of Paper II. 
We briefly summarize our conclusions in Section \ref{section_7}, and provide an outlook on extensions and applications of our work. 

The paper contains five appendices. In Appendix \ref{appendix_a}, 
we state the conventions that are used in this work. Appendix \ref{appendix_b} provides a detailed derivation of the all-order dressed spin-$1/2$ propagator, which clarified subtle features of a previous computation by Fradkin and Gitman \cite{Fradkin:1991ci}. The compact worldline result for $n^{\rm th}$ rank polarization tensors is derived in Appendix \ref{appendix_c}. As a limiting case of these general results, as previously shown by Strassler~\cite{Strassler:1992zr}, one recovers the Bern-Kosower expression for the one-loop polarization tensor with arbitrary numbers of external photon legs~\cite{Bern:1991aq}. Appendix \ref{appendix_d} discusses the classical worldline equations of motion which, for homogeneous backgrounds, reduce to the well-known Bargmann-Michel-Telegdi equations~\cite{PhysRevLett.2.435} for spinning charges in background electromagnetic fields. The semi-classical worldline instanton method for arbitrary background fields is also outlined in this appendix. Finally, Appendix \ref{appendix_e} provides details of key aspects of the derivation of the IR dressings in Section \ref{section_5_2}. 


\section{\label{section_2}The QED vacuum as a many-body theory of closed  worldlines}

Following pioneering work on the worldline description of quantum field theory~\cite{Strassler:1992zr}, there have been a number of significant advances in this framework, which are nicely summarized in \cite{Schubert:2001he,Edwards:2019eby}. However the fact that QED can be expressed exactly as a many-body theory of open and closed first-quantized worldlines is not widely appreciated, though, as noted, this possibility was already considered in Feynman and Schwinger's classic papers~\cite{PhysRev.80.440,PhysRev.84.108,Schwinger:1951nm}. In this section, we will systematically develop the many-body worldline framework for the QED vacuum-vacuum amplitude. This will be extended to scattering processes in the next section. Our approach is pedagogical, requiring for the most part only basic familiarity with quantum field theory.

Consider  the vacuum-vacuum matrix element
$\text{Z}=\langle 0 | 0 \rangle$ in QED, encoding the dynamics
of an infinite sea of virtual particles,
\ba
   \text{Z}= \int \mathcal{D}A\mathcal{D}\bar{\Psi}\mathcal{D}\Psi\exp\bigg\{&-\frac{1}{4}\int d^4x F_{\mu\nu}^2-\frac{1}{2\zeta}\int d^4x \,(\partial_\mu A_\mu)^2-\int d^4x \bar{\Psi}\big(\slashed{D}+m\big)\Psi\bigg\}.
   \label{QED_partition_function}
\ea
Here $g=\pm e$, $\zeta$ is the gauge fixing parameter, $\slashed{D}=D_\mu\gamma^\mu$ with $D_\mu = \partial_\mu-igA_\mu(x)$, and 
$[D_\mu,D_\nu]=-igF_{\mu\nu}(x)$. We will work in Euclidean 
spacetime, wherein this non-perturbative amplitude is well defined. Perturbative amplitudes 
can be Wick rotated back to Minkowski spacetime  using the conventions given in Appendix~\ref{appendix_a}.

To express $\text{Z}$ entirely in terms of worldlines,  we must integrate out the $A_\mu(x)$ and $\Psi(x)$ fields. Integrating  out the matter fields first,  we get
\ba
   \text{Z}= \int \mathcal{D}A\exp\bigg\{&-\frac{1}{4}\int d^4x F_{\mu\nu}^2-\frac{1}{2\zeta}\int d^4x \,(\partial_\mu A_\mu)^2+\ln\det\big(\slashed{D}+m\big)\bigg\}.\label{QED_partition_function_matterintegrated}
\ea
Before integrating out the gauge fields  $A_\mu(x)$, we will first write the fermion determinant in worldline form. To do so, we use the identity\footnote{In the first step, we introduced $\gamma_5^2=1$ and then used
$\{\gamma_5,\gamma_\mu\}=0$.}, 
\ba
\det\big(\slashed{D}+m\big)=\bigg\{\det\big(D_\mu\gamma^\mu+m\big)\det\big(-D_\mu\gamma^\mu+m\big)\bigg\}^{\frac{1}{2}}=\bigg\{\det\Big(m^2-\slashed{D}^2\Big)\bigg\}^{\frac{1}{2}}\,.
\ea
with 
\ba
 \slashed{D}^2= D_\mu \gamma_\mu D_\nu \gamma_\nu = D^2-\frac{g}{2}\sigma_{\mu\nu}F_{\mu\nu}, \medspace\medspace\medspace\medspace\medspace\medspace  \sigma_{\mu\nu}=\frac{i}{2}[\gamma_\mu,\gamma_{\nu}]\, .
\ea
The fermion determinant term in the path integral then becomes
\ba
  \ln\det\left(\slashed{D}+m\right)=\frac{1}{2}\Tr\ln\left(m^2-\slashed{D}^2\right)=\frac{1}{2}\Tr \ln\left(m^2-D^2+\frac{g}{2}\sigma_{\mu\nu}F_{\mu\nu}\right) . \label{determinant_trace}
\ea
As is customary, to compute the trace of this operator, we introduce the  Schwinger propertime parameter $\varepsilon_0$ and noting that 
\ba
\frac{\partial}{\partial z}\int^\infty_0 \frac{d\varepsilon_0}{\varepsilon_0}e^{-\varepsilon_0z}=-
\frac{1}{z}\,,
\ea
(valid for $\Re{z}>0$), we integrate both sides between $z=1$ and $z=a$, to obtain 
\ba
\ln(a)= \int_0^{\infty} \frac{d\varepsilon_0}{\varepsilon_0}\left\{e^{-\varepsilon_0}-e^{-\varepsilon_0a}\right\}\,.
\ea
The square of the Hermitian operator $\slashed{D}+m$ is positive, and the previous heat-kernel regularization method can be applied. The functional trace in Eq.~\eqref{determinant_trace} becomes
\ba
\frac{1}{2}\Tr \ln\left(m^2-D^2+\frac{g}{2}\sigma_{\mu\nu}F_{\mu\nu}\right)=\frac{1}{2}\Tr
  \int^{\infty}_0
  \frac{d\varepsilon_0}{\varepsilon_0}e^{-\varepsilon_0}-\frac{1}{2}\Tr\int_0^{\infty}
  \frac{d\varepsilon_0}{\varepsilon_0}e^{-\hat{H}}\,,
  \label{trace_propertime}
\ea
allowing one to reduce the problem of computing the fermion  determinant to one of finding the matrix elements of a first-quantized fermion traversing a loop in the background $A_\mu(x)$, governed by the Hamiltonian
operator
\ba
  \hat{H}=\varepsilon_0\bigg\{m^2-\hat{D}^2+\frac{g}{2}\hat{\sigma}_{\mu\nu}\hat{F}_{\mu\nu}\bigg\}=\varepsilon_0\bigg\{\big(\hat{p}_\mu+gA_\mu(\hat{x})\big)^2+m^2+\frac{g}{2}\hat{\sigma}_{\mu\nu} F_{\mu\nu}(\hat{x})\bigg\}\,,
  \label{hamiltonian_operator}
\ea
where we identified $\hat{p}_\mu=i\partial_\mu$.
The first term in
the r.h.s of Eq.~\eqref{trace_propertime} acts as the required
UV cut-off of the effective action, renormalizing the genuine UV divergence of the zero-point
energy of the vacuum. The vacuum-vacuum
amplitude now reads,
\ba
  \text{Z}=\int \mathcal{D}A\exp\bigg\{-\frac{1}{4}\int d^4x F_{\mu\nu}^2-\frac{1}{2\zeta}\int d^4x\,(\partial_\mu A_\mu)^2
  +\frac{1}{2}\Tr\int^\infty_0 \frac{d\varepsilon_0}{\varepsilon_0}e^{-\varepsilon_0}-\frac{1}{2}\Tr\int^{\infty}_0 \frac{d\varepsilon_0}{\varepsilon_0} e^{-\hat{H}}\bigg\}\,.
  \label{QED_partition_function_matterintegrated2}
\ea
To perform the trace in the expression above, the
matrix elements of $e^{-\hat{H}}$ can be expressed as point particle boson and fermion path integrals
\cite{PhysRev.84.108,Fradkin:1966zz,Fradkin:1991ci,Strassler:1992zr,DHoker:1995aat,vanHolten:1995ds,Reuter:1996zm,Schubert:2001he,Ahmadiniaz:2020wlm}. The eigenfunctions of the fermion operator $\hat{\sigma}_{\mu\nu}=i[\gamma_\mu,\gamma_\nu]/2$ can be treated on equal footing with the particle
4-position $\hat{x}$ and 4-momentum $\hat{p}$ bosonic operators \cite{Brink:1976uf,10.1143/PTP.60.548}: 
\ba
\hat{x}_\mu \to x_\mu(\tau), \medspace\medspace\medspace\medspace\medspace\medspace\medspace\medspace \hat{p}_\mu = i\partial_\mu \to p_\mu(\tau),\medspace\medspace\medspace\medspace\medspace\medspace\medspace\medspace \hat{\sigma}_{\mu\nu} \to \sigma_{\mu\nu}(\tau)\,,
\ea
spanning the space of intermediate states of
worldlines propagating in propertime $\tau$ in the gauge field background $A_\mu(x)$. 

Our starting point is the result in
\cite{Strassler:1992zr}, 
which can be expressed in general as 
\ba
\label{trace_pathintegral}
\Tr e^{-\hat{H}}&=
\int_{P}\mathcal{D}^4x\int_{AP}\mathcal{D}^4\psi\exp\bigg\{-\epsilon_0m^2-\frac{1}{4\varepsilon_0}\int_{0}^{1}
  d\tau \dot{x}_\mu^2(\tau)-\frac{1}{4}\int^1_0 d\tau \psi_\mu(\tau)\dot{\psi}_\mu(\tau)
  \\ 
  &+ig\int^1_0 d\tau\, \dot{x}_\mu  (\tau) A_\mu\big(x(\tau)\big)-i\frac{g\,\varepsilon_0}{2}\int^1_0 d\tau \psi_{\mu}(\tau)\psi_\nu(\tau)F_{\mu\nu}\big(x(\tau)\big)\bigg\}\,,
  \nonumber
\ea
where we gauge-fixed the \textit{einbein} $\varepsilon_0$
so that $\tau\in [0,1]$. The 0+1-dimensional bosonic commuting worldline $x_\mu(\tau)$ encodes the particle
4-position and its fermion anticommuting worldline counterpart 
$\psi_\mu(\tau)$ encodes spin precession. The latter inherits the anti-commutation
properties of the Dirac matrices and their action on a single
particle state. The two currents coupling to $A_\mu(x)$ are respectively, i) of the standard form $g\,\dot{x}_\mu(\tau)$ in the Wilson loop/line formulation of gauge theories
(denoting the Lorentz force of a scalar charge
in a background gauge field), and ii) the coupling of the local spin tensor
$\sigma_{\mu\nu}(\tau)=i[\psi_\mu(\tau),\psi_\nu(\tau)]/2$ to the local magnetic and (boosted) electric components in
$F_{\mu\nu}(x)$.

If we impose 
periodic boundary conditions (P) $x_\mu(1)=x_\mu(0)$ 
and anti-periodic boundary conditions (AP) $\psi_\mu(1)=-\psi_\mu(0)$ 
respectively on the boson and fermion worldline trajectories in Eq.~\eqref{trace_pathintegral}, they describe the sum over all possible configurations in propertime of worldlines 
representing the dynamics of virtual spin-$1/2$ charges in 
an arbitrary background $A_\mu(x)$. Inserting Eq.~\eqref{trace_pathintegral} in
Eq.~\eqref{QED_partition_function_matterintegrated2}, and expanding in loops, one obtains 
\ba
  &\text{Z}=\exp\bigg\{\frac{1}{2}\Tr\int^\infty_0 \frac{d\varepsilon_0}{\varepsilon_0}e^{-\varepsilon_0}\bigg\}\sum_{\ell=0}^{\infty}\frac{(-1)^{\ell}}{\ell!}\prod_{i=1}^\ell\Bigg[ \int_0^{\infty} \frac{d\varepsilon_0^{i}}{2\varepsilon_0^{i}} \int_{\rm P}\mathcal{D}^4x_{i}\int_{\rm AP}\mathcal{D}^4\psi_{i}
 \label{QED_partition_function_loopexpansion2}\\&\times\exp\bigg\{-m^2\varepsilon_0^{i}-\frac{1}{4\varepsilon_0^{i}}\int_{0}^{1}
  d\tau\dot{x}_{i}^2(\tau)-\frac{1}{4}\int^1_0d\tau \psi_\mu^{i}(\tau)\dot{\psi}^{i}_\mu(\tau)\bigg\}\Bigg]\nonumber\\
  &\times  \int \mathcal{D}A\exp\bigg\{-\frac{1}{4}\int d^4x F_{\mu\nu}^2(x) -\frac{1}{2\zeta}\int d^4x \big(\partial_\mu A_\mu (x)\big)^2\nonumber\\
  &+ig\sum_{i=1}^\ell \int_0^{1} d\tau\, \dot{x}^{i}_\mu(\tau) A_\mu\big(x_{i}(\tau)\big)-i\frac{g}{2}\sum_{i=1}^\ell  \varepsilon_0^{i}\int_0^{1} d\tau\psi_{\mu}^{i}(\tau)\psi_\nu^{i}(\tau) F_{\mu\nu}\big(x_{i}(\tau)\big)\bigg\}\,.\nonumber
\ea

To perform the remaining integration
in $A_\mu(x)$, we express the free
Maxwell action as
\ba
-\frac{1}{4}&\int d^4x F_{\mu\nu}^2(x)-\frac{1}{2\zeta} \int d^4x \,(\partial_\mu A_\mu(x))^2=-\frac{1}{2}\int d^4x \int d^4y\,A_\mu(x) \Big(D_{\mu\nu}^{B}\Big)^{-1}(x-y)A_\nu(y)\,,
\label{maxwell_action_byparts}
\ea
where we integrated by parts to find the inverse of the gauge-unfixed Euclidean photon propagator, with the matrix elements
\ba
\Big( D^{B}_{\mu\nu}\Big)^{-1}(x)=\bigg\{-\eta_{\mu\nu}\partial^2+\left(1-\frac{1}{\zeta}\right)\partial_\mu\partial_\nu\bigg\}\delta^4(x)\,.
\ea
Taking its Fourier transform, and inverting the result, the propagator in  $d$ dimensions is given by 
\ba
D_{\mu\nu}^B(x)=\int \frac{d^dk}{(2\pi)^{d}}e^{-ik\wc x}\tilde{D}_{\mu\nu}^B(k),\medspace\medspace\medspace \tilde{D}_{\mu\nu}^B(k)=\frac{1}{k^2}\bigg\{\eta_{\mu\nu}-\left(1-\zeta\right)\frac{k_\mu k_\nu}{k^2} \bigg\}\,.
\label{photon_propagator_momentum}
\ea
Finally, performing the momentum integrals for arbitrary values of the gauge parameter, its coordinate space counterpart is 
\ba
D_{\mu\nu}^B(x)=\frac{1}{4\pi^{{d}/2}}\Gamma\bigg(\frac{{d}-2}{2}\bigg)\frac{1}{(x^2)^{\frac{{d}}{2}-1}}\bigg\{\frac{1+\zeta}{2}\eta_{\mu\nu}+({d}-2)\frac{1-\zeta}{2}\frac{x_\mu x_\nu}{x^2}\bigg\}\,.
\label{photon_propagator_coordinate}
\ea
We shall also express the worldline currents
as 4-densities to integrate out the gauge field. Eq.~\eqref{QED_partition_function_loopexpansion2} can
be cast as
\ba
 &g\sum_{i=1}^\ell\int_0^{1} d\tau \dot{x}^{i}_\mu(\tau) A_\mu(x_{i}(\tau))-\frac{g}{2}\sum_{i=1}^\ell\varepsilon_0^{i}\int_0^{1} d\tau \medspace \psi_\mu^{i}(\tau)\psi_\nu^{i}(\tau)F_{\mu\nu}(x_{i}(\tau))= \int  d^4xJ_\mu^{(\ell)}(x)A_\mu(x)\,,
\ea
where we express the sum over the 
currents of the $\ell$ virtual charges as 
\ba
  J^{(\ell)}_\mu(x)=\sum_{i=1}^\ell J_\mu^i(x)=\sum_{i=1}^\ell \left(j^{i}_\mu(x)+h^{i}_\mu(x)\right)\,,
  \label{total_current}
\ea
with the charged scalar and fermion current densities, respectively, 
\ba
    j^{i}_\mu(x)&=g\int_0^{1} d\tau \dot{x}^{i}_\mu(\tau)\delta^4(x-x_{i}(\tau))\,\,;\,\,
    h^{i}_{\mu}(x)=-g\varepsilon_0^{i}\int_0^{1} d\tau \psi_{\mu}^{i}(\tau)\psi_\nu^{i}(\tau)\frac{\partial}{\partial x_\nu}\delta^4(x-x_{i}(\tau))\,.
    \label{auxiliary_currents_jh}
\ea
Thus with the help of Eqs.~\eqref{maxwell_action_byparts} and
\eqref{total_current}, the vacuum-vacuum amplitude
in Eq.~\eqref{QED_partition_function_loopexpansion2} can be reexpressed as
\ba
&\text{Z}=\exp\bigg\{\frac{1}{2}\Tr\int^\infty_0 \frac{d\varepsilon_0}{\varepsilon_0}e^{-\varepsilon_0}\bigg\}\sum_{\ell=0}^{\infty}\frac{(-1)^{\ell}}{\ell!}\nonumber\\\times&\prod_{i=1}^\ell\bigg[ \int_0^{\infty} \frac{d\varepsilon_0^{i}}{2\varepsilon_0^{i}} \int_{\rm P}\mathcal{D}^4x_{i}\int_{\rm AP}\mathcal{D}^4\psi_{i}
 \exp\bigg\{-m^2\varepsilon_0^{i}-\frac{1}{4\varepsilon_0^{i}}\int_{0}^{1}
  d\tau\dot{x}_{i}^2(\tau)-\frac{1}{4}\int^1_0d\tau \psi_\mu^{i}(\tau)\dot{\psi}^{i}_\mu(\tau)\bigg\}\bigg]\nonumber\\\times&\int \mathcal{D}A\exp\bigg\{-\frac{1}{2}\int d^4x \int d^4 y A_\mu(x) \Big(D_{\mu\nu}^{B}\Big)^{-1}(x,y)A_\nu(y)+i\int d^4x A_\mu(x) J_\mu^{(\ell)}(y)\bigg\}\,.
  \label{QED_partition_function_loopexpansion3}
\ea
We can now integrate over $A_\mu(x)$ and write this expression (normalizing with respect to the pure gauge vacuum sea of disconnected photon loops),
\ba
\frac{\text{Z}}{\text{Z}_\text{MW}}=\sum_{\ell=0}^\infty
\text{Z}^{(\ell)}=\sum_{\ell=0}^\infty \frac{(-1)^\ell}{\ell!}\mathrm{W}^{(\ell)}, \medspace\medspace\medspace\medspace\medspace\medspace\medspace\medspace\medspace\medspace
\label{QED_partition_function_worldline}\text{Z}_\text{MW}^{-1}=\sqrt{{\rm det}D_{B}^{-1}}\,.
\ea
The $\ell$-th contribution ($\text{W}^{(\ell)}$) in
this loop expansion is the amplitude for finding the QED vacuum in a configuration with $\ell$ virtual charged particles. It can be represented as the  sum over all possible worldline paths $x_\mu^{i}(\tau)$ and $\psi_\mu^{i}(\tau)$ with periodic and anti-periodic boundary conditions,  respectively, of $\ell$ virtual worldlines describing  loops in propertime $\varepsilon_0^{i}$, summed over all possible propertimes, and takes the form 
\ba
&\text{W}^{(\ell)}=\exp\bigg\{\frac{1}{2}\Tr\int^\infty_0 \frac{d\varepsilon_0}{\varepsilon_0}e^{-\varepsilon_0}\bigg\}\prod_{i=1}^\ell\bigg\{\int_0^{\infty} \frac{d\varepsilon_0^{i}}{2\varepsilon_0^{i}} \int_{\rm P}\mathcal{D}^4x_{i}\int_{\rm AP}\mathcal{D}^4\psi_{i}\exp\bigg\{-\frac{1}{4\varepsilon_0^{i}}\int_{0}^{1}
 d\tau \dot{x}_{i}^2(\tau)\nonumber
\\\label{w_l_definition}
& - m^2\varepsilon_0^{i}-\frac{1}{4}\int^1_0 d\tau \psi_\mu^{i}(\tau)\dot{\psi}^{i}_\mu(\tau)\bigg\}\bigg\}\exp\bigg\{-\frac{1}{2} \int d^4x \int d^4y J^{(\ell)}_\mu(x)D_{\mu\nu}^B(x-y)J^{(\ell)}_\nu(y)\bigg\}\,.
\ea
The dynamics of these $\ell$ virtual fermions are
encoded in the exponentiation of the photon exchanges between the $\ell$-loop currents in Eq.~\eqref{total_current}, functionals of the set of $\ell$ super-pairs of virtual worldlines $\{x_\mu(\tau),\psi_\mu(\tau)\}$. Using the r.h.s of Eq.~\eqref{total_current} one may express the exponential as that of the sum of contributions due to the exchanges between pairs of worldline currents $i$ and $j$:

\ba
 &\exp\bigg\{-\frac{1}{2}\int d^4x \int d^4y J^{(\ell)}_\mu(x)D_{\mu\nu}^B(x-y)J^{(\ell)}_\nu(y)\bigg\}\\
 &=\exp\bigg\{-\frac{1}{2}
\sum_{i,j=1}^{\ell}\int d^4x \int d^4y \big(j_\mu^{i}(x)+h_\mu^{i}(x)\big)  D_{\mu\nu}^B(x-y)
  \big(j_\nu^{j}(y)+h_\nu^j(y)\big)\bigg\}\nonumber\,.
\ea
Represented thus, one observes that the worldline dynamics includes the exponentiation of both the self-energy subgraphs $i=j$ and the non-diagonal exchanges $i\neq j$ on equal
footing, consisting of the Lorentz forces between two spinning
charges $i$ and $j$. This will be important to note for future reference when we discuss in Section \ref{section_5} the FK dressings of worldlines.

Even though we ``got rid" of the gauge fields, we can reintroduce them in a purely first quantized interpretation of the $\ell^\text{th}$-loop contribution to the QED vacuum-vacuum transition amplitude as a many-body theory of worldlines. Specifically, with Eq.~\eqref{auxiliary_currents_jh}, the boson-boson  coupling between the coordinate space currents 
can be rewritten as

\ba
\label{j_i_j_j_coupling}
\int d^4x \int d^4y \,j^{i}_\mu(x) D_{\mu\nu}^B(x-y) j^{j}_\nu(y)= g\int_0^{1} d\tau \frac{dx^{i}_\mu}{d\tau}A_\mu^{x_{j}}(x_{i}(\tau))\,,
\ea
where (for $\zeta=1$ in $d=4$)  using Eqs.~\eqref{photon_propagator_momentum} and \eqref{auxiliary_currents_jh},
\ba
  A_\mu^{x_{j}}(x)=\int d^4y &\,D_{\mu\nu}^B(x-y)j_\nu^j(y)= \frac{g}{4\pi^2}\int_0^{1} d\tau \frac{1}{(x-x_{j}(\tau))^2}\frac{dx^j_\mu}{d\tau}\label{a_mu_j_j}\,.
\ea
Here $A_\mu^{x_{j}}(x)$ the field at point $x$ created by the
coordinate space current $g\dot{x}^{j}_\mu(\tau)$ of the $j^\text{th}$ charge moving along $x^{j}_\mu(\tau)$.

Likewise, the spin precession of the $i^\text{th}$ worldline couples independently to the
magnetic and (boosted) electric components of the field created by the scalar current $g\dot{x}^{j}_\mu(\tau)$:
\ba
  \int  d^4x \,h^{i}_\mu(x) &\int d^4y \medspace D^B_{\mu\nu}(x-y) \,j^{j}_\nu(y)=i \frac{g\varepsilon_0^{i}}{2}
  \int_0^{1} d\tau \sigma_{\mu\nu}^{i}(\tau)F_{\mu\nu}^{x_j}(x_{i}(\tau))\,,
  \label{h_i_j_j_coupling}
\ea
where we used Eq.~\eqref{auxiliary_currents_jh}, integrated by parts, and antisymmetrized to define the local
spin tensor of the $i^\text{th}$ virtual charge $\sigma_{\mu\nu}^{i}(\tau)=i[\psi_\mu^{i}(\tau),\psi_\nu^{i}(\tau)]/2$. The field strength tensor created by
the vector current $x_{j}(\tau)$ refers to
\ba
F_{\mu\nu}^{x_j}(x)=\partial_\mu
A_\nu^{x_j}(x)-\partial_\nu A_\mu^{x_j}(x) \,,
\label{f_munu_j_j}
\ea
with $A_\mu^{x_j}(x)$ given in Eq.~\eqref{a_mu_j_j}. 

The couplings of the spin of worldline $j$ with the field emerging from the spatial
current of $i$ can be rewritten using 
Eq.~\eqref{auxiliary_currents_jh} as
\ba
  \int d^4x \,j_\mu^{i}(x)\int d^4y D_{\mu\nu}^B(x-y)h_\nu^{j}(y)=g\int_0^{1} d\tau \frac{dx^{i}_\mu}{d\tau} A_\mu^{\psi_j}(x_{i}(\tau))\,,
  \label{j_i_h_j_coupling}
\ea
where the field created by the spin precession of $j$ is 
\ba
  A_\mu^{\psi_j}(x)=\int d^4y &D_{\mu\nu}^F(x-y)h_\nu^j(y) \label{a_mu_j_h}= -\frac{ig\varepsilon_0^{j}}{4\pi^2}\int_0^{1} d\tau \sigma_{\mu\nu}^{j}(\tau) \frac{\partial}{\partial x_\nu}\frac{1}{(x-x_{j}(\tau))^2}\,.
\ea
We used here Eq.~\eqref{photon_propagator_momentum} with 
$\zeta=1$ and $d=4$. 

Lastly, the pure spin-spin interaction of the worldlines, using
Eq.~\eqref{auxiliary_currents_jh} again, can be written as 
\ba
  \int d^4x\, h_\mu^{i}(x)&\int d^4y\, D_{\mu\nu}^B(x-y)\, h_\nu^{j}(y)=i\frac{g\varepsilon_0^{i}}{2}\int_0^{1}d\tau \sigma_{\mu\nu}^{i}(\tau)F_{\mu\nu}^{\psi_j}(x_{i}(\tau))\,,
  \label{h_i_h_j_coupling}
\ea
where the field strength tensor created by the spin precession of worldline $j$ at the point $x$ is 
\ba
F_{\mu\nu}^{\psi_j}(x)=\partial_\mu A_\nu^{\psi_j}(x)-\partial_\nu A_\mu^{\psi_j}(x)\,,
\ea
with $A_\mu^{\psi_j}(x)$ given in Eq.~\eqref{a_mu_j_h}. 

Putting together the contributions from 
Eqs.~\eqref{j_i_j_j_coupling}, \eqref{h_i_j_j_coupling},
\eqref{j_i_h_j_coupling} and \eqref{h_i_h_j_coupling}, 
the interaction functional in Eq.~\eqref{w_l_definition} can now be rearranged as a more general Wilson loop describing the dynamics of particle $i$: 
\ba
  &\text{W}^{(\ell)}= \exp\bigg\{\frac{1}{2}\Tr\int^\infty_0 \frac{d\varepsilon_0}{\varepsilon_0}e^{-\varepsilon_0}\bigg\}\prod_{i=1}^\ell\bigg\{\int_0^{\infty} \frac{d\varepsilon_0^{i}}{2\varepsilon_0^{i}} \int_{\rm P}\mathcal{D}^4x_{i}\int_{\rm AP}\mathcal{D}^4\psi_{i}\exp\bigg\{-m^2\varepsilon_0^{i}
\nonumber\\
&
 -\frac{1}{4\varepsilon_0^{i}}\int_{0}^{1}
 d\tau \dot{x}_{i}^2(\tau) -\frac{1}{4}\int^1_0 d\tau \psi_\mu^{i}(\tau)\dot{\psi}^{i}_\mu(\tau)-\frac{g}{2} \int_0^{1} d\tau \frac{dx^{i}_\mu}{d\tau} \sum_{j=1}^\ell\Big(A_\mu^{x_j}(x_{i}(\tau))+A_\mu^{\psi_j}(x_{i}(\tau))\Big)\nonumber\\
  &-i\frac{g\varepsilon_0^{i}}{4}\int_0^{1} d\tau\sigma_{\mu\nu}^{i}(\tau)\sum_{j=1}^\ell\Big(F_{\mu\nu}^{x_j}(x_{i}(\tau))+F_{\mu\nu}^{\psi_j}(x_{i}(\tau))\Big)\bigg\}\bigg\}\label{w_l_definition_2} \,.
\ea
It is important to note that in this generalized Wilson loop\footnote{Note that the generalized Wilson loop and lines (which we will soon discuss) defined here are, despite some similarities,  qualitatively different from those discussed in \cite{Laenen:2008gt,Bahjat-Abbas:2019fqa,Bonocore:2020xuj}. This is because we have integrated out the fundamental gauge fields explicitly.} the 
background gauge fields in Eq.~\eqref{a_mu_j_j} and \eqref{a_mu_j_h} are replaced by dynamical fields.  They include the self-interaction terms  $i=j$, the spin precession is exponentiated on the same footing as the dynamical fields, and worldline contour averages exactly defined through path integrals. These features are not accessible in the conventional Wilson loop formulation of gauge theories.  

Using Eq.~\eqref{a_mu_j_j} and \eqref{a_mu_j_h}, with  Feynman gauge $\zeta=1$ and $d=4$, the generalized Wilson loop takes the elegant form 
\ba
  &\text{W}^{(\ell)}=\bigg\langle \exp\bigg\{
  -\frac{g^2}{8\pi^2}\sum_{i,j=1}^\ell \int_0^{1} d\tau_i \bigg(\frac{dx^{i}_\mu}{d\tau_i}-i\varepsilon_0^{i}\sigma^{i}_{\mu\rho}(\tau_i)\frac{\partial}{\partial x_{\rho}^i}\bigg)\nonumber\\
&  \times\int_0^{1} d\tau_j\bigg(\frac{dx^j_\mu}{d\tau_j}-i\varepsilon_0^{j}\sigma^{j}_{\mu\eta}(\tau_j)\frac{\partial}{\partial x_{\eta}^{j}}\bigg)\frac{1}{(x_{i}(\tau_i)-x_{j}(\tau_j))^2}\bigg\}\bigg\rangle\label{w_l_definition_3}\,,
\ea
where the structure of the Wilson loop is that of a normalized expectation value $\langle {\cal O}^{(\ell)}\rangle$ of a many-body functional ${\cal O}^{(\ell)}$ of the $\ell$ super-pairs of worldlines $\{x_\mu(\tau),\psi_\mu(\tau)\}$, with the expectation value defined as
\ba
  & \Big\langle {\cal O}^{(\ell)}\big[\{x_\mu(\tau),\psi_\mu(\tau)\}\big]\Big\rangle =\exp\bigg\{\frac{1}{2}\Tr\int^\infty_0 \frac{d\varepsilon_0}{\varepsilon_0}e^{-\varepsilon_0}\bigg\}\prod_{i=1}^\ell\bigg\{\int_0^{\infty} \frac{d\varepsilon_0^{i}}{2\varepsilon_0^{i}} \int_{\rm P}\mathcal{D}^4x_{i}\int_{\rm AP}\mathcal{D}^4\psi_{i}
 \nonumber\\
 &\times\exp\bigg\{- \frac{1}{4\varepsilon_0^{i}}\int_{0}^{1}
 d\tau \dot{x}_{i}^2(\tau)-m^2\varepsilon_0^{i}-\frac{1}{4}\int^1_0 d\tau \psi_\mu^{i}(\tau)\dot{\psi}^{i}_\mu(\tau)\bigg\}
 {\cal O}^{(\ell)}\big[\{x_\mu(\tau),\psi_\mu(\tau)\}\big]
 \label{normalized_expectation_value_closedworldlines}\,.
\ea
This allows us to write
the QED partition function (with the shorthands  $x_\mu^i=x_\mu^i(\tau_i)$ and $\sigma_{\mu\nu}^i=\sigma_{\mu\nu}^i(\tau_i)$) in the following compact and suggestive form:
\ba
&\frac{\text{Z}}{\text{Z}_\text{MW}}=\sum_{\ell=0}^\infty\frac{(-1)^\ell}{\ell!}\bigg\langle \exp\bigg\{
  -\frac{g^2}{8\pi^2}\sum_{i,j=1}^\ell \int_0^{1} d\tau_i \bigg(\dot{x}^{i}_\mu-i\varepsilon_0^{i}\sigma^{i}_{\mu\rho}\frac{\partial}{\partial x_{\rho}^i}\bigg)\nonumber\\
&  \times\int_0^{1} d\tau_j\bigg(\dot{x}^j_\mu-i\varepsilon_0^{j}\sigma^{j}_{\mu\eta}\frac{\partial}{\partial x_{\eta}^{j}}\bigg)\frac{1}{(x_\mu^i-x_\mu^j)^2}\bigg\}\bigg\rangle
\,.
\label{notation_wilsonlines}
\ea

\begin{figure}[ht]
    \centering
    \includegraphics[scale=0.6]{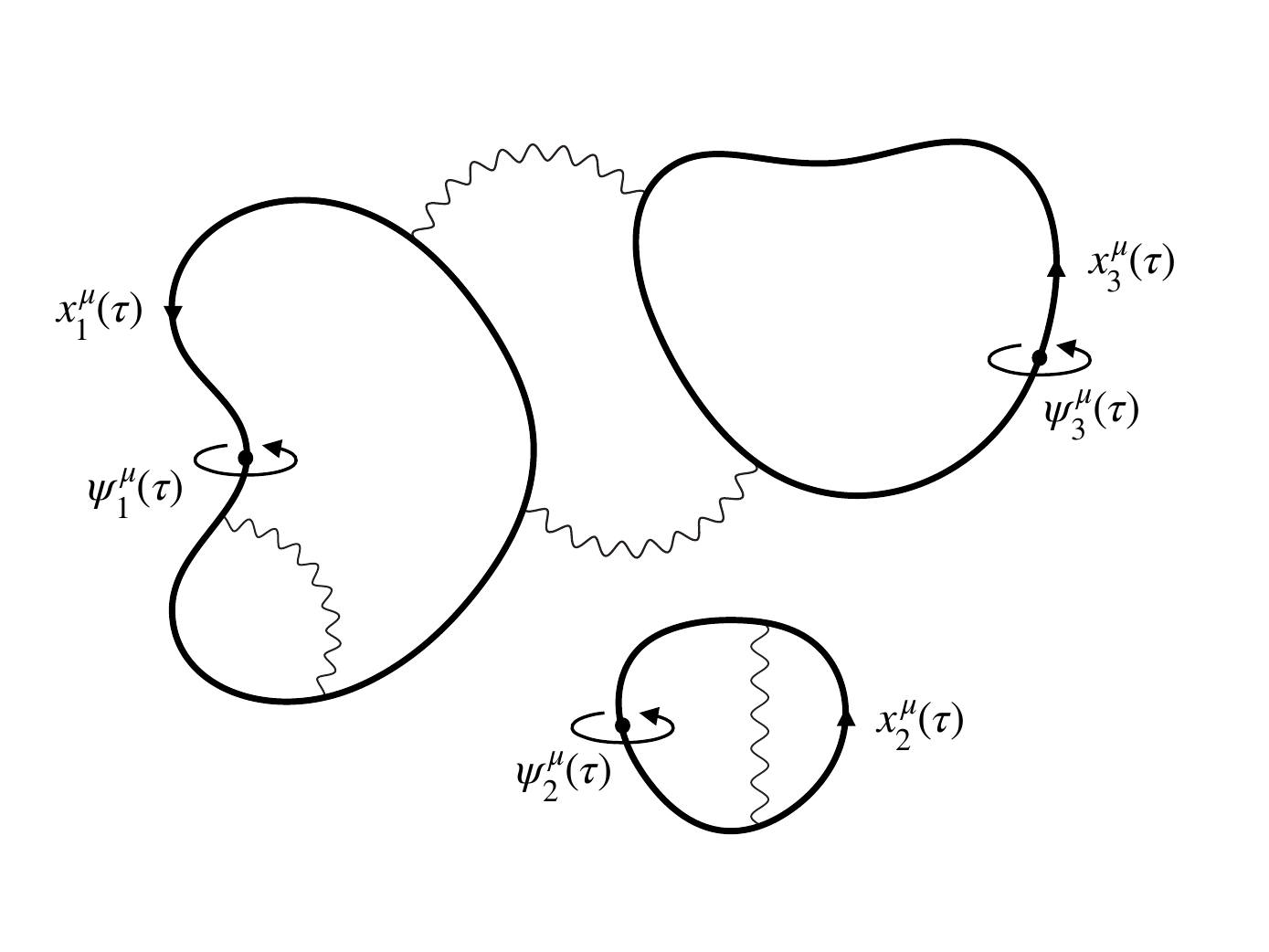}
    \caption{A three fermion worldline loop ($\ell=3$) contribution to $\text{Z}_{(3)}$  in Eq.~\eqref{notation_wilsonlines} for the vacuum-vacuum amplitude. Each of these 0+1-dimensional point particles are fully described by a super-pair $\{x^\mu_i(\tau),\psi^\mu_i(\tau)\}$ of closed worldlines in propertime, created at $\tau=0$ and destroyed at $\tau=1$; these emit, reabsorb,  and exchange an arbitrary number of photons that transmit the Lorentz forces between spin-$1/2$ charges.}
    \label{figure_1}
\end{figure}

Eq.~\eqref{notation_wilsonlines} is the central result\footnote{We will not discuss here UV renormalization in the sense of mass and charge renormalization. Worldline perturbation theory, as we will discuss in Section~\ref{section_4}, consists of explicit computations of generalized polarization tensors.  Employing dimensional regularization in this framework,  one can identify the corresponding  UV divergences as the usual ones extracted from Feynman diagram computations~\cite{Schubert:2001he}. Renormalization of generalized Wilson lines~\cite{Polyakov:1980ca} will be discussed in Section~\ref{section_6}.} of this
section. It expresses the QED vacuum 
as a many-body theory of pointlike 0+1 dimensional particles  represented by ``super-pairs" of closed 
boson and fermion  worldlines\footnote{Note that the QED worldline Lagrangian satisfies an ${\cal N}=1$ SUSY algebra~\cite{Mueller:2017arw,BastianelliLectures}. } $(x_\mu^{i}(\tau),\psi_\mu^{i}(\tau))$ with typical virtualities $1/\varepsilon_0^{i}$. Eq.~\eqref{notation_wilsonlines} is exact and 
generalizes the notion of a Wilson loop to spinning, precessing particles in fully dynamical backgrounds, with the field of  the $i$th worldline 
sourced by self-interactions as well as those generated by the other  particles at any given order in the loop expansion. An illustrative
representation of one of these amplitudes is depicted in Fig. \ref{figure_1}.

While it is perilous to claim anything novel in such a highly developed field, we have not, to the best of our knowledge,  seen the explicit and compact many-body worldline representation given in  Eq.~\eqref{notation_wilsonlines} elsewhere in the literature. In the next section, we will extend this framework to  open worldlines, thereby  enabling one to compute many-body scattering processes.

\section{\label{section_3} S-matrix in the many-body worldline formalism}

We will consider here the scattering amplitudes for transitions 
between \textit{in} and \textit{out} charged states via the emission of virtual photons\footnote{Transitions involving the emission of real photons and  corresponding soft theorems will be addressed later in the paper and in greater detail in Paper II.}. 
{We denote $n_i$ and $n_o$ ($\bar{n}_i$ and $\bar{n}_o$) to be the number of incoming and outgoing charged particles (antiparticles) and $N_i=n_i+\bar{n}_i$ and $N_o=n_o+\bar{n}_o$ the total number of \textit{in} and \textit{out} charges. 
We also identify a pair of external particles with indices
$n$ and $m$, with $n,m=1,\ldots,r$, with $r=(N_o+N_i)/2$ being the number of real worldlines in the scattering process. 
A pair of virtual particles is identified with indices $i$ and $j$ with $i,j=1,\ldots,\ell$, 
where $\ell$ represents the number of virtual worldlines. 

The Dyson  S-matrix element for the transition amplitude 
$i\to f$ is defined as
\ba
\mathcal{S}_{fi} = \big\langle \medspace p_f^{n_o},s_f^{n_o},\ldots,p_f^1,s_f^1,\bar{p}_f^{\bar{n}_o},\bar{s}_f^{\bar{n}_o},\ldots,\bar{p}_f^1,\bar{s}_f^1\medspace;\medspace t_f \medspace \big|\medspace p_i^n,s_i^n,\ldots,p_i^1,s_i^1,\bar{p}_i^{\bar{n}_i},\bar{s}_i^{\bar{n}_o},\ldots,\bar{p}_i^1,\bar{s}_i^1\medspace;\medspace t_i \medspace\big\rangle\,,
\label{amplitude_external_particles}
\ea
and the limits $t_f\to+\infty$ and $t_i\to-\infty$ are assumed. 
Here $|p,s\rangle$ and $|\bar{p},\bar{s}\rangle$ are single particle and antiparticle states of momentum $p$ and $s$.
The vacuum-vacuum amplitude calculated in Section \ref{section_2}
corresponds to the particular case $r=0$ in Eq.~\eqref{amplitude_external_particles} with $|\mathrm{in}\rangle=|0 \rangle$ and $\langle \mathrm{out}|=\langle 0 |$. 

Our strategy here is to continue working in Euclidean spacetime to construct the non-perturbative building blocks, which can then be Wick rotated to Minkowski spacetime, as needed, to compute scattering amplitudes to a given order in perturbation theory. Our starting point is the Euclidean QED path integral
\ba
\label{QED_partition_function_externalfermions}
&\text{Z}[\bar{\eta},\eta] = \int \mathcal{D}A \mathcal{D}\bar{\Psi}\mathcal{D}\Psi 
\exp\bigg\{-\frac{1}{4}\int d^4x F_{\mu\nu}^2-\frac{1}{2\zeta}\int d^4x
  \,(\partial_\mu A_\mu)^2
  \\
&-\int
  d^4x\bar{\Psi}(\slashed{D}+m)\Psi+ \int d^4x \bar{\eta}\Psi \int d^4x \bar{\Psi}\eta\bigg\}\,,\nonumber
\ea
where $\bar{\eta}(x)$ and $\eta(x)$ are anti-commuting  external sources.  Proceeding along the same lines as in the previous section, and first 
integrating out the Dirac fields, we get
\ba
&\text{Z}[\bar{\eta},\eta]= \int \mathcal{D}A 
\exp\bigg\{-\frac{1}{4}\int  d^4x F_{\mu\nu}^2-\frac{1}{2\zeta}\int d^4x
  \,(\partial_\mu A_\mu)^2\\
  &+\ln\det\big(\slashed{D}+m\big)+\int d^4x \int d^4y \medspace \bar{\eta}(x)D_F^A(x,y)\eta(y)\bigg\}\,.\nonumber
\ea
The dressed Euclidean fermion Green function $D_F^A(x,y)$ satisfies
\ba
(\slashed{D}+m)D_F^A(x,y)=\delta^{(4)}(x-y)\,, \medspace\medspace\medspace {\rm with}\medspace\medspace\medspace D_\mu = \partial_\mu-igA_\mu(x)\,.
\ea
We shall now construct dressed $N$-point Green functions and normalize the amplitudes for these with respect to the infinite sea of virtual loops in $\text{Z}=\langle 0 | 0\rangle\equiv\text{Z}[0,0]$.
The 2-point function is given by 
\ba
 \frac{1}{\text{Z}}\frac{\delta
    \text{Z}[\bar{\eta},\eta]}{\delta \eta(x_i^{1})\delta
    \bar{\eta}(x_f^{1})}\Bigg|_{\substack{\bar{\eta}=0 \\ \eta=0}}  \label{2_point_function}= \medspace\Big\langle D_F^A(x_f^{1},x_i^{1}) \Big\rangle_{A}\,,
\ea
where the r.h.s,  defined as
\ba
  &\Big\langle D_F^A(x_f^{1},x_i^{1}) \Big\rangle_{A}=  \frac{1}{\text{Z}}\int
 \mathcal{D}A \exp\bigg\{-\frac{1}{4}\int d^4x 
   F_{\mu\nu}^2-\frac{1}{2\zeta}\int d^4x \,(\partial_\mu
   A_\mu)^2\nonumber\\
   &+\ln\det\big(\slashed{D}+m\big)\bigg\} \medspace D_F^A(x_f^{1},x_i^{1}) \,,
   \label{functional_average}
\ea
corresponds to the propagator for a single real fermion to go from position $x_i^{1}$ to $x_f^{1}$ while coupled to a fully dynamical gauge field. It contains arbitrary photon insertions and  fermion loops, the latter arising from the fermion determinant. 

Higher order $N$-point functions 
describe the interacting problem of $r=(N_i+N_o)/2$
real spinning charges. (Odd $N$-point functions vanish.) The $4$-point function gives
\ba
 & \frac{1}{\text{Z}}\frac{\delta
    \text{Z}[\bar{\eta},\eta]}{\delta \eta(x^i_1)\delta
    {\eta}(x^i_2)\delta
    \bar{\eta}(x^f_2)\delta\bar{\eta}(x^f_1)}\Bigg|_{\substack{\bar{\eta}=0 \\ \eta=0}}
  =  \left\langle
  D_F^A(x^f_2,x^i_2)D_F^A(x^f_1,x^i_1)\right\rangle_{A}-
  \left\langle
  D_F^A(x^f_1,x^i_2)D_F^A(x^f_2,x^i_1)\right\rangle_{A}\,.
\ea
The relative sign between the two terms follows from  functional differentiation, yielding 
the odd symmetry factor of the final state under the interchange of single final states.
Likewise, higher $N$-point functions and their matrix elements can be built as functional averages of products of 2-point functions given by 
$D_F^A(x_f,x_i)$. Thus to compute $N$-point amplitudes in the worldline formalism, it is sufficient to express $D_F^A(x_f,x_i)$ in terms of worldlines, and subsequently perform the functional average over the gauge fields.

Our starting point here is a result first obtained by Fradkin and Gitman \cite{Fradkin:1991ci}, which we will rederive in Appendix
\ref{appendix_b} to clarify some aspects of the derivation\footnote{The open worldline formalism 
is also discussed at length in
\cite{Fradkin:1966zz,Fainberg:1987jr,vanHolten:1995ds,Reuter:1996zm,Schubert:2001he,Ahmadiniaz:2020wlm}.}. This result  (Eq.~\eqref{greenfunction_worldline_momentumintegrated} of the Appendix) gives the amplitude for the propagation of a
spinning charge from $x_i$ to $x_f$ in the presence of
$A_\mu(x)$ to be (defining $\bar{D}_F^A(x_f,x_i)=D_F^A (x_f,x_i) \gamma_5)$,
\ba
\label{fradkin_gitman_result}
 &\bar{D}_F^A(x_f,x_i) = \frac{1}{N_5} \exp\bigg\{\bar{\gamma}_\lambda
  \frac{\partial}{\partial \theta_\lambda}\bigg\}\int_0^\infty d\varepsilon^0 
  \int d\chi^0 \int \mathcal{D}\varepsilon_{} \frac{\mathcal{D}\pi_{}}{2\pi}\int \mathcal{D}\chi_{}
  \mathcal{D}\nu_{}\int \mathcal{D}^4x_{} \int\mathcal{D}^5\psi_{}&
 \\
  &\times\exp\bigg\{-S_{R,0}+ig\int_0^1 d\tau \dot{x}_\mu(\tau)
    A_\mu\big(x_{}(\tau)\big)-i\frac{g}{2}\int_0^1 d\tau \varepsilon_{}(\tau)
    \psi_\mu(\tau)\psi_\nu(\tau)
    F_{\mu\nu}\big(x_{}(\tau)\big)\bigg\}\Bigg|_{\theta=0}\,.\nonumber&
\ea
The various terms and fields in the above expression demand detailed explanation, which is given below. (See also Appendix \ref{appendix_b} for further details.) The propagation of the real fermion is described by the super-pair $(x_\mu(\tau),\psi_\mu(\tau))$ of commuting and anti-commuting worldlines, respectively, with the dimensions 
of the fermion degrees of freedom specified by five Grassmannian fields 
labeled by 
$\lambda=1,2,3,4,5$, with the fifth necessary to account for the helicity-momentum constraint\footnote{For virtual particles, 
this extra fermion d.o.f. in the Green function can be omitted since one takes a trace over them.}. 

Importantly, one requires open boundary conditions of their path integrals, which satisfy 
\ba
  x_\mu(0)=x^{i}_{\mu}\,, \medspace\medspace\medspace\medspace\medspace\medspace\medspace\medspace\medspace\medspace x_\mu(1)=x^{f}_{\mu}\,, \medspace\medspace\medspace\medspace\medspace\medspace\medspace\medspace\medspace\medspace \psi_\lambda(1)=-\psi_\lambda(0)+2\,\theta_\lambda\,, \label{boundary_conditions_openworldline}
\ea
where $\theta_{\lambda}$ are five anti-commuting auxiliary sources, which are eventually set to zero.
We also define  
$\bar{\gamma}_{\mu}=-i\gamma_5\gamma_\mu$ and $\bar{\gamma}_5=\gamma_5$,
with $\gamma_5=\gamma_1\gamma_2\gamma_3\gamma_4$; they obey the identity 
$\{\bar{\gamma}_{\lambda},\bar{\gamma}_{\xi}\}=2\delta_{\lambda\xi}$.
These two objects anti-commute  ($\{\theta_{\lambda},\bar{\gamma}_{\lambda}\}=0$), which allows us to generate the correct tensor structure of the Green function in the fermion degrees of freedom.  The normalization
of the odd path integral in five dimensions is denoted  by $N_5$.

Further in Eq.~\eqref{fradkin_gitman_result},  $A_\mu$-independent terms are collected in the (super-gauge unfixed) free
worldline action of a real spinning charge
\ba
  &S_{R,0}=\frac{1}{4}\psi_\lambda(1)\psi_\lambda(0)+\int^1_0 d\tau\bigg\{i\pi(\tau)\dot{\varepsilon}(\tau)+ \nu_{}(\tau)\dot{\chi}(\tau)+ \varepsilon(\tau) m^2+\frac{\dot{x}_{\mu}^2(\tau)}{4\varepsilon(\tau)}
  \label{eq:energy-momentum}\\
  &+\frac{1}{4}\psi_\lambda(\tau)\dot{\psi}_\lambda(\tau)-\chi(\tau)\bigg(m\psi_5(\tau)+\frac{i}{2\epsilon(\tau)}\dot{x}^\mu(\tau)\psi_\mu(\tau)\bigg)\bigg\}\,.
  \label{open_worldline_free_action}
\ea
We leave manifest the time reparametrization
invariance of the worldlines as a gauge symmetry, keeping the 
\textit{einbeins} unfixed:
besides the super-pair of worldlines $\big(x_\mu(\tau),\psi_\mu(\tau)\big)$,
we introduced a new super-pair of commuting and anti-commuting \textit{einbeins} denoted $\varepsilon(\tau)$ and
$\chi(\tau)$, respectively, with trivial dynamics. Their conjugate partners are denoted $\pi(\tau)$ and $\nu(\tau)$. The first two terms in the curly brackets of Eq.~\eqref{eq:energy-momentum} fix the gauge in the reparametrization invariance of the worldlines. The next two terms in the curly brackets of Eq.~\eqref{eq:energy-momentum} enforce the energy-momentum constraint on the worldline, while the terms in the parenthesis in Eq.~\eqref{open_worldline_free_action} impose the helicity-momentum constraint~\cite{BastianelliLectures,Mueller:2017arw}.

The zero-modes of the new boson and fermion \textit{einbeins} are denoted as 
\ba
\varepsilon_{}^0=\varepsilon_{}(0), \medspace\medspace\medspace\medspace 
\chi_{}^0=\chi_{}(0), \medspace\medspace\medspace\medspace \tau\in[0,1]\,,
\ea
and correspond to conventional commuting and anti-commuting Schwinger
parameters, respectively, in Feynman diagram language. The terms
 $\pi(\tau)\dot{\varepsilon}(\tau)$ and
$\nu(\tau)\dot{\chi}(\tau)$ in Eq.~\eqref{open_worldline_free_action} act as super-gauge fixing terms. 

With all the terms specified as above, Equation \eqref{fradkin_gitman_result} is exact. It sums all possible worldline paths connecting the open boundary endpoints specified in 
Eq.~\eqref{boundary_conditions_openworldline} to
construct the dressed quantum amplitude for a spin-$1/2$ particle
to go from $x_i$ to $x_f$ in the presence of an Abelian background gauge field,  to all orders in the coupling. It also contains the semi-classical limit of the action of a charged spinning
particle in an Abelian background gauge field, proposed previously in \cite{Berezin:1976eg,Barducci:1976xq,Brink:1976sz}. 

For a single charge $r=1$, plugging Eq.~\eqref{fradkin_gitman_result} into Eq.~\eqref{functional_average} gives the dressed fermion propagator
\ba
  &\Big\langle \bar{D}_A^F(x_f^1,x_i^1) \Big\rangle_{A}=\frac{1}{N_5}\exp\bigg\{\bar{\gamma}_\lambda \frac{\partial}{\partial
      \theta_{\lambda}^1}\bigg\}\int_0^\infty d\varepsilon^0_1 \int d\chi^0_1 \int \mathcal{D}\varepsilon_1 \frac{\mathcal{D}\pi_1}{2\pi} \int \mathcal{D}\chi_1 \mathcal{D}\nu_1 \int \mathcal{D}x_1 \mathcal{D}\psi_1e^{-S_{R,0}^1}\nonumber\\
  &\times
   \frac{1}{\text{Z}}\int \mathcal{D}A\det\big(\slashed{D}+m\big)
  \exp\bigg\{-\frac{1}{4}\int d^4x F_{\mu\nu}^2-\frac{1}{2\zeta}\int
    d^4x\,(\partial_\mu
    A_\mu)^2+i\int
    d^4x  J_{R,\mu}^1 A_\mu \bigg\}\Bigg|_{\theta_{1}=0}\,.
    \label{functional_average_greenfunction_1}
\ea
with the real particle charged current density  defined (with the \textit{einbein} $\varepsilon(\tau)$ unfixed) to be 
\ba
J_{R,\mu}^1(x)=g\int^{1}_0 d\tau
\dot{x}_\mu^1(\tau)\delta^4\big(x-x_1(\tau)\big)-g\int^{1}_0 d\tau\varepsilon_1(\tau)
\psi_{\mu}^1(\tau)\psi_\nu^1(\tau)\frac{\partial}{\partial x_\nu
}\delta^4\big(x-x_1(\tau)\big)\,.
\label{current_external_fermion}
\ea
Further expressing the fermion loop contributions as functional traces in worldline form, and expanding in loops, precisely as in the previous section, we then get
\ba
  &\det\big(\slashed{D}+m\big)  \label{fermion_determinant_loopexpansion}
  =\sum_{\ell=0}^\infty \frac{(-1)^\ell}{\ell!} \exp\bigg\{\frac{1}{2}\Tr\int^\infty_0\frac{d\varepsilon_0}{\varepsilon_0}e^{-\varepsilon_0}\bigg\}\\&\times \prod_{i=1}^\ell  \int_0^\infty \frac{d\varepsilon_i^0}{2\varepsilon_i^0} \int \mathcal{D}\varepsilon_i\int \frac{\mathcal{D}\pi_i}{2\pi}\int_\mathrm{P} \mathcal{D}^4x_i \int_\mathrm{AP}\mathcal{D}^4\psi_i\exp\bigg\{-S_{V,0}^i+i\int d^4x J_{V,\mu}^iA_\mu\bigg\}\,,\nonumber
\ea
with $x_i^\mu(1)=x_i^\mu(0)$ (P) and
$\psi_i^\mu(0)=-\psi_i^\mu(1)$ (AP). The charged current densities of the virtual
fermions are defined identically
\ba
J_{V,\mu}^i(x)=g\int^{1}_0 d\tau
\dot{x}_\mu^i(\tau)\delta^4\big(x-x_i(\tau)\big)-g\int^{1}_0 d\tau \varepsilon_i(\tau)
\psi_{\mu}^i(\tau)\psi_{\nu}^i(\tau)\frac{\partial}{\partial x^\nu
}\delta^4\big(x-x_i(\tau)\big)\,,
\label{currents_loop_fermions}
\ea
and the corresponding free worldline action given by 
\ba
S_{V,0}^{i}=\int^1_0 d\tau \bigg\{&i\pi_i(\tau)\dot{\varepsilon}_i(\tau)+\varepsilon_i(\tau) m^2+\frac{\dot{x}_i^2(\tau)}{4\varepsilon_i(\tau)}-\frac{1}{4}\psi_\mu^i(\tau)\dot{\psi}_\mu^i(\tau)\bigg\}\,.
\ea
Substituting Eq.~\eqref{fermion_determinant_loopexpansion} into
Eq.~\eqref{functional_average_greenfunction_1}, we get
\ba
 & \Big\langle \bar{D}_A^F(x_f^1,x_i^1) \Big\rangle_{A}= \frac{1}{N_5}\exp\bigg\{\bar{\gamma}_\lambda \frac{\partial}{\partial
      \theta_\lambda^1}\bigg\}\int_0^\infty d\varepsilon^0_1 \int d\chi^0_1 \int  \mathcal{D}\varepsilon_1 \frac{\mathcal{D}\pi_1}{2\pi} \int \mathcal{D}\chi_1 \mathcal{D}\nu_1\int \mathcal{D}^4x_1\mathcal{D}^5\psi_1 \nonumber \\
 &\cdot  \sum_{\ell=0}^\infty\frac{(-1)^\ell}{\ell!} \exp\bigg\{\frac{1}{2}\Tr\int^\infty_0\frac{d\varepsilon_0}{\varepsilon_0}e^{-\varepsilon_0}\bigg\}\Bigg[\prod_{i=1}^\ell \int_0^\infty \frac{d\varepsilon_i^0}{2\varepsilon_i^0} \int \mathcal{D}\varepsilon_i\frac{\mathcal{D}\pi_i}{2\pi} \int_\mathrm{P} \mathcal{D}^4 x_i \int_\mathrm{AP}\mathcal{D}^4\psi_i\Bigg]e^{-S^{(1,\ell)}_0}\nonumber\\
&\cdot\frac{1}{\text{Z}}\int\mathcal{D}A  \exp\bigg\{-\frac{1}{2}\int d^4x \int d^4y A_\mu(x) \Big(D^{B}_{\mu\nu}\Big)^{-1}(x-y) A_\nu(y)+i\int d^4x A_\mu(x)J_\mu^{(1,\ell)}(x)\bigg\}\bigg|_{\theta_1=0}\,.
\label{eq:open-path-integral}
\ea
Note that the gauge functional average differs from the one in the vacuum-vacuum
amplitude in Eq.~\eqref{QED_partition_function_loopexpansion3} only due to the presence of the charged current of the 
real particle. Since currents and actions are additive for each virtual and real 
particle present, we defined
\ba
S_0^{(r,\ell)}=\sum_{n=1}^{r}S_{R,0}^n+\sum_{i=1}^\ell S_{V,0}^i\,, \medspace\medspace\medspace J_\mu^{(r,\ell)}(x) = \sum_{n=1}^r J_{R,\mu}^{n}(x) + \sum_{i=1}^\ell J_{V,\mu}^{i}(x)\,. \label{total_current_real_virtual}
\ea
Upon integrating over the gauge field, as previously,  we finally obtain the dressed propagator for the 
propagation of a real spinning charge from $x_i^1$ to $x_f^1$, which  can be expressed as  
\ba
\Big\langle \bar{D}_A^F(x_f^1,x_i^1) \Big\rangle_{A}&= \frac{\text{Z}_{\text{MW}}}{\text{Z}} \exp\bigg\{\bar{\gamma}_\lambda \frac{\partial}{\partial
      \theta_\lambda^1}\bigg\}\sum_{\ell=0}^\infty \frac{(-1)^\ell}{\ell!}  \text{W}^{(1,\ell)}\Big(x_f^1,x_i^1,\theta^1\Big)\bigg|_{\theta^1=0}\,,
      \label{2_point_function_result_2}
\ea
and where, in analogy with the vacuum-vacuum amplitude, we can define 
a more general Wilson line containing $\ell$ virtual fermion loops as
\ba 
\mathrm{W}^{(1,\ell)}(x_f^1,x_i^1,\theta^1)=\bigg\langle \exp\bigg\{-\frac{1}{2}\int d^4x\int d^4y J_\mu^{(1,\ell)}(x)D_{\mu\nu}^B(x-y) J_\nu^{(1,\ell)}(y)\bigg\}\bigg\rangle\,.
\ea 
The structure of Wilson lines and loops are that of the normalized expectation value $\langle \mathcal{O}^{(r,\ell)}\rangle$
of many-body functionals $\mathcal{O}^{(r,\ell)}$ of the set of $r$ and $\ell$ super-pairs of fermionic and bosonic worldlines $\{x_\mu(\tau),\psi_\mu(\tau)\}$, with boundary conditions in Eq.~\eqref{boundary_conditions_openworldline} for the $r$ real charges and P and AP for the $\ell$ virtual charges,
\ba 
&\Big\langle \mathcal{O}^{(r,\ell)}\big[\{x_\mu(\tau),\psi_\mu(\tau)\}\big]\Big\rangle = \prod_{n=1}^r \bigg\{\frac{1}{N_5}\int_0^\infty
d\varepsilon_n^0 \int d\chi_n^0\int \mathcal{D}\varepsilon_n\frac{\mathcal{D}\pi_n}{2\pi} \int \mathcal{D}\chi_n \mathcal{D}\nu_n\int \mathcal{D}^4x_n\mathcal{D}^5\psi_n\bigg\}
\nonumber\\
&\times \exp\bigg\{\frac{1}{2}\Tr\int^\infty_0\frac{d\varepsilon_0}{\varepsilon_0}e^{-\varepsilon_0}\bigg\} \prod_{i=1}^\ell\bigg\{ \int_0^\infty \frac{d\varepsilon_i^0}{2\varepsilon_i^0}\int\mathcal{D}\varepsilon_i\frac{\mathcal{D}\pi_i}{2\pi}\int_\mathrm{P}
\mathcal{D}^4x_i \int_\mathrm{AP}\mathcal{D}^4\psi_i\bigg\}\nonumber\\
&\times\exp\bigg\{-S_0^{(r,\ell)}\bigg\}\mathcal{O}^{(r,\ell)}[\{x_\mu(\tau),\psi_\mu(\tau)\}\big]\,.\label{langle_o_rangle_open}
\ea 
Eq.~\eqref{2_point_function_result_2} then expresses the functional average over fields in the l.h.s., as a 0+1 dimensional field theory in the r.h.s., with normalized expectation values $\mathrm{W}^{(1,\ell)}(x_f,x_i,\theta)$
given by the path integrals in Eq.~\eqref{langle_o_rangle_open}.
Note that in the above expression the sub-indices $i$ and $n$ denote different and independent (virtual and real, respectively) worldlines.

We can now write down the precise form of the $2r$ point function, encoding the amplitude of propagation of a system of $r$ real particles\footnote{Here we only consider 
particles ($N_o=N_i$). The general case including antiparticles will be discussed shortly.} from $\{x_\mu^i\}$ at $\tau=0$ to $\{x_\mu^f\}$ at $\tau=1$
\ba 
& \frac{1}{\mathrm{Z}[0,0]} \frac{\delta^{2r}\mathrm{Z}[\bar{\eta},\eta]}{\delta \eta(x_i^{N_i})\cdots\delta\eta(x_i^{1})\delta\bar{\eta}(x_f^1)\cdots\delta\bar{\eta}(x_f^{N_o})}=\sum_{\text{perm}}\epsilon_{N_o\ldots 1} \bigg\langle \prod_{n=1}^{r}D_F^A(x_f^n,x_i^n)\bigg\rangle_A\,,
\ea 
where $\epsilon_{N_o\ldots 1}$ is the totally anti-symmetric symbol and the sum runs over all permutations of the final points of the dressed Euclidean Green functions, encoding the parity of the \textit{out} many-body wavefunction under permutations of final single particle states. Following the same steps as for the $r=1$ case we get 
\ba 
&\bigg\langle\prod_{n=1}^{r}\bar{D}_F^A(x_f^n,x_i^n)\bigg\rangle_A =\frac{\text{Z}_{\text{MW}}}{\text{Z}}  \prod_{n=1}^{r}\bigg\{\exp\bigg\{\bar{\gamma}_\lambda \frac{\partial}{\partial
      \theta_\lambda^n}\bigg\}\bigg\} \nonumber\\
     & \times\sum_{\ell=0}^\infty \frac{(-1)^\ell}{\ell!}  \text{W}^{(r,\ell)}\Big(x_f^{r},x_i^{r},\theta^r,\cdots, x_f^{1},x_i^1,\theta^1 \Big)\bigg|_{\theta^n=0}\,,\label{2r_point_function_result}
\ea 
where
\ba 
\mathrm{W}^{(r,\ell)}\Big(x_f^{r},x_i^{r},\theta^r,\cdots, x_f^{1},x_i^1,\theta^1 \Big)=\bigg\langle \exp\bigg\{-\frac{1}{2}\int d^4x\int d^4y J_\mu^{(r,\ell)}(x)D_{\mu\nu}^B(x-y) J_\nu^{(r,\ell)}(y)\bigg\}\bigg\rangle  
\label{w_r_l_definition}
\ea 
This functional contains the exponentiation of all tree-level photon 
exchanges in the self-interactions of the charged fermions as well as of the virtual photon exchanges between them. It further includes the 
fermion loops arising from the couplings of the real particles
with the polarized virtual fermions of the sea and  photon exchanges between the
disconnected loops of the sea. These last are removed
order to order in perturbation theory by $(\text{Z}/\text{Z}_{\text{MW}})^{-1}$ in 
Eq.~\eqref{2r_point_function_result}, calculated in Section 2.

With the input from the above expression, Eq.~\eqref{2r_point_function_result} is exact just as was the case for  Eq.~\eqref{fradkin_gitman_result}. It provides the 2\,$r$-point correlator of $r$ spin-$1/2$ fields between
$x_i^n$ to $x_f^n$ to all orders in perturbation theory. It can be interpreted as a sum over
worldline contours of the interactions of $r$ pointlike spinning charges in the field created
by themselves and those induced by $\ell$ virtual particles at each loop order. 

Note that the tree-level $\ell=0$ case of a single field $r=1$ in Eq.~\eqref{2_point_function_result_2} should be understood as the generalization (with full spin corrections, back-reaction, helicity-momentum and energy-momentum constraints, and well-defined worldline expectation values) 
of a Wilson loop operator between $x_i^1$ to $x_f^1$. For $\ell>1$, we note that the 
the exponentiation of genuine fermion loops, as shown, cannot be obtained in the conventional Wilson loop picture.

We will now provide the explicit form of the {Euclidean interaction functional of QED in general $d$ dimensions, arbitrary gauge parameter $\zeta$ and general \textit{einbein} $\varepsilon(\tau)$. 
Using Eq.~\eqref{total_current_real_virtual}, we can reexpress the argument of the exponential in Eq.~\eqref{w_r_l_definition} as 
\ba
\label{eq:Wilson-line-prod}
-\frac{1}{2}\int d^4x
  \int d^4y J_\mu^{(r,\ell)}(x)D_{\mu\nu}^B(x-y)J_\nu^{(r,\ell)}(y) = \sum_{a,b=1}^{r+\ell} S_{ab}\,,
\ea
where $S_{ab}$ denotes the contribution to the action of the pair of worldlines $a$ and $b$ in the many-body system comprising the complete set of interactions in the amplitude. We can write it as
\ba
\label{eq:Wilson-line_log-sum}
S_{ab}=S_{ab}^\text{BB}+S_{ab}^\text{BF}+S_{ab}^\text{FB}+S_{ab}^\text{FF}  \,,  
\ea
where the labels (B) and (F) denote the boson and fermion currents, respectively,  
of $a$ and $b$ attached to the photon. Dimensionally regularizing 
Eq.~\eqref{photon_propagator_coordinate}, absorbing the mass dimensions of the coupling in $\mu$ and further 
using Eq.~\eqref{auxiliary_currents_jh}, we obtain the following individual contributions:
\ba
&S_{ab}^\text{BB}=-\frac{g^2\mu^{4-d}}{8\pi^{\frac{d}{2}}}\Gamma\bigg(\frac{d-2}{2}\bigg)\bigg(\frac{1+\zeta}{2}\bigg)\int^1_0d\tau_a\int^1_0d\tau_b \frac{\dot{x}^a_\mu \dot{x}^b_\mu}{\Big[\big(x^a_\mu-x^b_\mu\big)^2\Big]^{\frac{d}{2}-1}}\nonumber\\
&-\frac{g^2\mu^{4-d}}{8\pi^{\frac{d}{2}}}\Gamma\bigg(\frac{d}{2}\bigg)(1-\zeta)\int^1_0d\tau_a\int^1_0 d\tau_b \frac{\dot{x}_\mu^a(x_\mu^a-x_\mu^b)\dot{x}_\nu^b(x_\nu^a-x_\nu^b)}{\Big[\big(x^a_\mu-x^b_\mu\big)^2\Big]^{\frac{d}{2}}}\,,
\label{S_ab_BB}
\ea
which is the classical interaction appearing in a Wilson loop between two 
scalar charges. We use as a shorthanded $x^{a}_\mu\equiv x_\mu^a(\tau_{a})$, $\psi^a_\mu\equiv \psi^a_\mu(\tau_{a})$ and $\epsilon_{a}\equiv \epsilon_a(\tau_{a})$.  We observe that the non-diagonal gauge-dependent terms in the second line of Eq.~\eqref{S_ab_BB} can be cast as a total double derivative with respect to particle $a$ and $b$ proper times $\tau_a$ and $\tau_b$. Thus in the worldline formalism,  gauge transformations correspond to the addition of boundary terms to the action \cite{Schubert:2001he}, but only in the scalar interactions, as we will see next.

The coupling of the scalar current of a charge $a$, 
in the magnetic and boosted electric field created  by the spin 
precession of $b$, generates
\ba
\label{S_ab_BF}
&S_{ab}^\text{BF}=-\frac{g^2\mu^{4-d}}{4\pi^{\frac{d}{2}}}\Gamma\bigg(\frac{d}{2}\bigg)\int^1_0d\tau_a\int^1_0 d\tau_b\varepsilon_b\frac{\dot{x}_\mu^a \psi_\mu^b\psi_\nu^b \big(x_\nu^a-x_\nu^b\big)}{\Big[\big(x_\mu^a-x_\mu^b\big)^2\Big]^{\frac{d}{2}}}\nonumber\\
&-\frac{g^2\mu^{4-d}}{4\pi^{\frac{d}{2}}}\Gamma\bigg(\frac{d}{2}+1\bigg)(1-\zeta)\int^1_0 d\tau_a \int^1_0 d\tau_b \varepsilon_b \frac{\dot{x}_\mu^a(x_\mu^a-x_\mu^b)\psi_\nu^b(x_\nu^a-x_\nu^b)\psi_\rho^b(x_\rho^a-x_\rho^b)}{\Big[\big(x_\mu^a-x_\mu^b\big)^2\Big]^{\frac{d}{2}+1}}\,,
\ea
and its mirror image (setting $a\leftrightarrow b$), coming from the coupling of the spin precession of $a$ in the field created 
by the scalar current of $b$,
\ba
\label{S_ab_FB}
&S_{ab}^\text{FB}=+\frac{g^2\mu^{4-d}}{4\pi^{\frac{d}{2}}}\Gamma\bigg(\frac{d}{2}\bigg)\int^1_0d\tau_a\varepsilon_a\int^1_0 d\tau_b\frac{\dot{x}_\mu^b \psi_\mu^a\psi_\nu^a \big(x_\nu^a-x_\nu^b\big)}{\Big[\big(x_\mu^a-x_\mu^b\big)^2\Big]^{\frac{d}{2}}}\\
&+\frac{g^2\mu^{4-d}}{4\pi^{\frac{d}{2}}}\Gamma\bigg(\frac{d}{2}+1\bigg)(1-\zeta)\int^1_0 d\tau_a\varepsilon_a \int^1_0 d\tau_b  \frac{\dot{x}_\mu^b(x_\mu^a-x_\mu^b)\psi_\nu^a(x_\nu^a-x_\nu^b)\psi_\rho^a(x_\rho^a-x_\rho^b)}{\Big[\big(x_\mu^a-x_\mu^b\big)^2\Big]^{\frac{d}{2}+1}}\nonumber
\,.
\ea

The final term is the pure fermion-fermion contribution corresponding to the interaction induced by the precession of the spin of one worldline on that of the other:
\ba
\label{S_ab_FF}
&S_{ab}^\text{FF}=-\frac{g^2\mu^{4-d}}{4\pi^{\frac{d}{2}}}\Gamma\bigg(\frac{d}{2}\bigg)\int^1_0d\tau_a\varepsilon_a\int^1_0 d\tau_b\varepsilon_b\frac{\psi_\mu^a\psi_\nu^a\psi_\mu^b\psi_\nu^b }{\Big[\big(x_\mu^a-x_\mu^b\big)^2\Big]^{\frac{d}{2}}}\\
&-\frac{g^2\mu^{4-d}}{2\pi^{\frac{d}{2}}}\Gamma\bigg(\frac{d}{2}+1\bigg)\int^1_0 d\tau_a\varepsilon_a\int^1_0d\tau_b\varepsilon_b \frac{\psi_\mu^a \psi_\nu^a(x_\nu^a-x_\nu^b)\psi_\rho^b(x_\rho^a-x_\rho^b)\psi_\mu^b}{\Big[\big(x_\mu^a-x_\mu^b\big)^2\Big]^{\frac{d}{2}+1}}\nonumber\\
&+\frac{g^2\mu^{4-d}}{2\pi^\frac{d}{2}}\Gamma\bigg(\frac{d}{2}+2\bigg)(1-\zeta)\int^1_0 d\tau_a\varepsilon_a\int^1_0 d\tau_b\varepsilon_b \frac{\Big[\psi_\mu^a(x_\mu^a-x_\mu^b)\Big]^2\Big[\psi_\nu^b(x_\nu^a-x_\nu^b)\Big]^2}{\Big[\big(x_\mu^a-x_\mu^b\big)^2\Big]^{\frac{d}{2}+2}}\,.\nonumber
\ea
Note that due to the global supersymmetry of the QED worldline Lagrangian (whose explicit form can be found in Eq.~\eqref{lagrangian_greenfunction_wordline_momentumintegrated} of Appendix~\ref{appendix_b}), the action between any two charges $S_{ab}$ in QED can be generated through the application of the 0+1-dimensional $\mathcal{N}=1$ SUSY algebra to the scalar QED term $S_{ab}^\text{BB}$ \cite{Rajeev:1985ii,Polyakov:1987ez}. For $d=4$,  after fixing Feynman gauge $\zeta=1$ and the \textit{einbein}, these individual contributions can be put together in a compact expression as previously shown in Eq.~\eqref{w_l_definition_3} for the vacuum-vacuum amplitude. This can be seen by substituting Eqs.~\eqref{S_ab_BB}-\eqref{S_ab_FF} into Eq.~\eqref{eq:Wilson-line-prod}, and thence into Eq.~\eqref{w_r_l_definition}.

Having obtained explicit expressions for all the elements necessary to compute  $\text{W}^{(r,\ell)}$ in  Eq.~\eqref{w_r_l_definition}, we will now relate this to to the Dyson S-matrix element in Eq.~\eqref{amplitude_external_particles}. We start by evaluating the $r=1$ case, describing the creation of a free particle at past infinity and its subsequent evolution via the QED Hamiltonian up to plus infinity. For a positive energy charge propagating from past infinity, one can represent the fermion dressed self-energy in terms of a Dyson S-matrix element, given by\footnote{The external lines are created by imposing the limits $x_f^0\to\infty$ and $x_i^0\to-\infty$ in the dressed $\mathrm{N}$-point Green functions, which is equivalent to the standard LSZ reduction~\cite{Weinberg:1995mt}. The consequences of this truncation  will be very relevant for the discussion in Section \ref{section_5} on the IR behavior of the S-matrix.}
\ba 
\mathcal{S}_{fi}&=\lim_{\substack{x_f^0\to+\infty\\ x_i^0\to -\infty}}\big\langle \medspace p_f,s_f;\medspace x_f^0\medspace \big|\medspace p_i,s_i\medspace ;\medspace x_i^0\medspace \big\rangle\nonumber\\
&= \lim_{\substack{x_f^0\to+\infty\\ x_i^0\to -\infty}} \int d^3\v{x}_f\int d^3\v{x}_i {\Psi_{f,\beta}^{(+)\dag}(x_f)} \Psi^{(+)}_{i,\alpha}(x_i) \gamma^0_{\gamma\alpha}
\frac{1}{\mathrm{Z}}\bigg(\frac{1}{i}\bigg)^2\frac{\delta^2 \mathrm{Z}[\eta,\bar{\eta}]}{\delta \eta_\gamma(x_i)\delta\bar{\eta}_\beta(x_f)}\bigg|_{\substack{\bar{\eta}=0\\\eta=0}},
\ea 
where $\alpha$, $\beta$ and $\gamma$ are spin indices, and we introduced the following shorthands for positive energy plane-wave functions:
\ba
\Psi_{f,\beta}^{(+)}(x_f)=u_\beta(p_f,s_f)e^{-ip_f\wc x_f}, \medspace\medspace\medspace  \Psi_{i,\alpha}^{(+)}(x_i)=u_\alpha(p_i,s_i)e^{-ip_i\wc x_i}\,.
\ea 
The QED generating functional $\mathrm{Z}[\bar{\eta},\eta]$ in physical time is given in Eq.~\eqref{QED_generating_functional} of Appendix \ref{appendix_b}. The S-matrix can be reexpressed as 
\ba 
\mathcal{S}_{fi}^{(1)}=\lim_{\substack{x_f^0\to+\infty\\ x_i^0\to -\infty}} i\int d^3\v{x}_f\int d^3\v{x}_i {\Psi_{f,\beta}^{(+)\dag}}(x_f)  \Big\langle \bar{D}_{\beta\gamma}^{F,A}(x_f,x_i)\Big\rangle_A \gamma^0_{\gamma\alpha} \Psi^{(+)}_{i,\alpha}(x_i),\label{s_fi_1_electron}
\ea 
which involves replacing the dressed Euclidean Green function
in Eq.~\eqref{2r_point_function_result} 
with its Minkowski counterpart $\bar{D}^F_A(x_f,x_i)$ in Eq.~\eqref{greenfunction_worldline_momentumintegrated_minkowski} of Appendix \ref{appendix_b} multiplied by a factor of $i$. The integration of the dressed Green function over the gauge field configurations produces
\ba
\Big\langle \bar{D}_A^F(x_f,x_i) \Big\rangle_{A}&= i\frac{\text{Z}_{\text{MW}}}{\text{Z}} \exp\bigg\{\bar{\gamma}_\lambda \frac{\partial}{\partial
      \theta_\lambda^1}\bigg\}\sum_{\ell=0}^\infty \frac{(-1)^\ell}{\ell!}  \text{W}^{(1,\ell)}\Big(x_f,x_i,\theta\Big)\bigg|_{\theta=0}\,,
      \label{2_point_function_result_minkowski}
\ea
with $\mathrm{W}^{(1,\ell)}(x_f,x_i,\theta)$ the Wick rotation to Minkowski time of the normalized worldline expectation value in Eq.~\eqref{w_r_l_definition}. Plugging  Eq.~\eqref{2_point_function_result_minkowski} into \eqref{s_fi_1_electron}, noticing $\gamma_5^2=-1$ in the Minkowski case,  and identifying $\bar{\gamma}_0=\gamma_5\gamma_0$, one gets the loop expansion 
\ba 
\mathcal{S}_{fi} =\sum_{\ell=0}^\infty \mathcal{S}_{fi}^{(1,\ell)}\,,
\ea 
analogous to the one in Eq.~\eqref{QED_partition_function_worldline} for the vacuum-vacuum amplitude, with 
the $\ell$-loop contribution given by
\ba 
\mathcal{S}_{fi}^{(1,\ell)}&=\frac{\mathrm{Z}_\text{MW}}{\mathrm{Z}} \frac{(-1)^\ell}{\ell!}
\lim_{\substack{x_f^0\to+\infty\nonumber\\ x_i^0\to -\infty}}\int d^3\v{x}_f\int d^3\v{x}_i {\Psi_f^{(+)\dag}}(x_f)   \exp\bigg\{\bar{\gamma}_\lambda \frac{\partial}{\partial
      \theta_\lambda}\bigg\} \bar{\gamma}_0 \Psi_i^{(+)}(x_i)   \mathrm{W}^{(1,\ell)}(x_f,x_i,\theta)\bigg|_{\theta=0}\,.
\ea 
The first term $\ell=0$ contains the tree-level self-energy graphs attached to a single charge to all orders in perturbation theory. The $\ell\neq 0$ terms encode self-energy contributions, also to all orders in perturbation theory, with a fixed number $\ell$ of virtual fermions. As discussed before, the disconnected vacuum polarization graphs are removed by the vacuum-vacuum terms in $\mathrm{Z}/\mathrm{Z}_\text{MW}$.

Similarly, for an anti-particle created at past infinity, introducing the shorthands for negative energy plane-wave solutions
\ba 
\Psi^{(-)}_{f,\beta}(\bar{x}_f)=v_\beta(\bar{p}_f,\bar{s}_f)e^{-i\bar{p}_f\wc \bar{x}_f}, \medspace\medspace\medspace \Psi^{(-)}_{i,\alpha}(\bar{x}_i)=v_\alpha(\bar{p}_i,\bar{s}_i)e^{-i\bar{p}_i\wc \bar{x}_i},
\ea 
the Dyson S-matrix element reads
\ba
&\mathcal{S}_{fi}^{(1)}=\lim_{\substack{x_f^0\to+\infty\\ x_i^0\to -\infty}}\big\langle \medspace \bar{p}_f,\bar{s}_f;\medspace x_f^0\medspace \big|\medspace \bar{p}_i,\bar{s}_i\medspace ;\medspace x_i^0\medspace \big\rangle\nonumber\\
&=\lim_{\substack{\bar{x}_f^0\to+\infty\\\bar{x}_i^0\to-\infty}} -i\int d^3\bar{\v{x}}_f \int d^3\bar{\v{x}}_i {\Psi_{i,\alpha}^{(-)\dag}}(\bar{x}_i)\Big\langle D_{\alpha\gamma}^{F,A}(\bar{x}_i,\bar{x}_f)\Big\rangle_A \gamma^0_{\gamma\beta} \Psi^{(-)}_{f,\beta}(\bar{x}_f)\,.
\ea
which can be thought of as the dressed Green function of a positive energy particle propagating backwards in time from $\bar{x}_f$ to $\bar{x}_i$, and multiplied by a factor of $-1$. Using Eq.~\eqref{2_point_function_result_minkowski}, it can be rewritten as the loop expansion
\ba
&\mathcal{S}_{fi}^{(1)}=\sum_{\ell=0}^\infty \mathcal{S}_{fi}^{(1,\ell)},\medspace\medspace\medspace  \mathcal{S}_{fi}^{(1,\ell)}=-\frac{\mathrm{Z}_\text{MW}}{\mathrm{Z}} \frac{(-1)^\ell}{\ell!}\nonumber\\
&\times\lim_{\substack{\bar{x}_f^0\to+\infty\\\bar{x}_i^0\to-\infty}} \int d^3\bar{\v{x}}_f \int d^3\bar{\v{x}}_i {\Psi_i^{(-)\dag}}(\bar{x}_i) \exp\bigg\{\bar{\gamma}_\lambda\frac{\partial}{\partial \theta_\lambda}\bigg\} \bar{\gamma}_0 \Psi_f^{(-)}(\bar{x}_f)\mathrm{W}^{(1,\ell)}(\bar{x}_i,\bar{x}_f,\theta)\bigg|_{\theta=0}.
\ea 

Turning now to the many-body scattering problem with $r$ real particles, the calculation of the S-matrix elements involves just the reduction of the $2r$-point Green functions to products of the ($r=1$) $2$-point functions above, as emphasized previously. We begin by computing explicitly the worldline Dyson S-matrix elements for the $r=2$ case,  depicted schematically in Fig. \ref{figure_2}.
\begin{figure}[ht]
    \centering
    \includegraphics[scale=0.6]{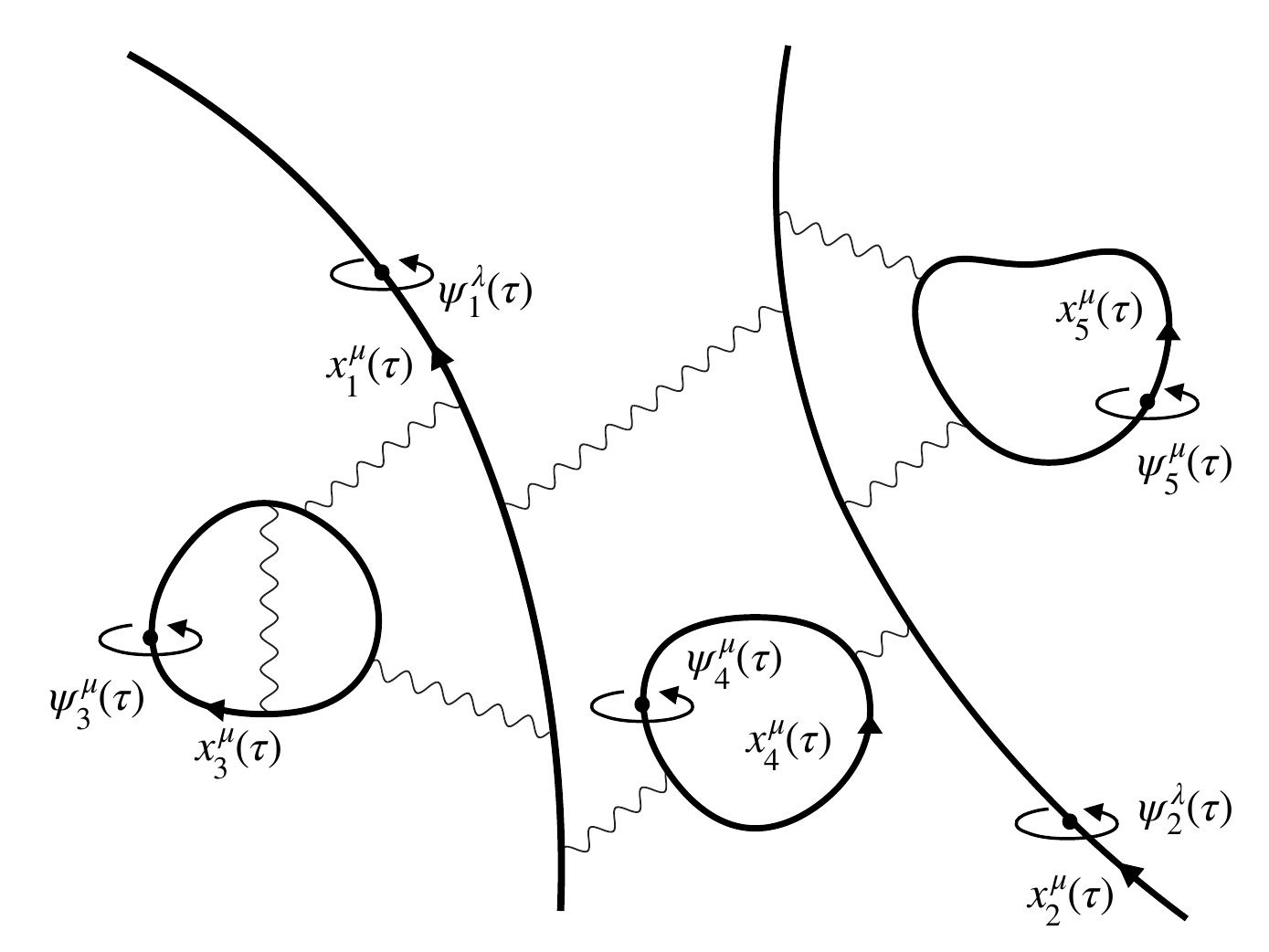}
    \caption{The $2\leftrightarrow 2$ scattering amplitude for $r=2$ real particles and $\ell=3$ virtual loops, corresponding to  $\mathcal{S}^{(2,3)}_{fi}$ in Eq.~\eqref{transition_n_particles}. The real charges $n=1,2$ are described by a super-pair $\{x^\mu_n(\tau),\psi^\lambda_n(\tau)\}$ of worldlines in propertime  with open boundary conditions; the virtual loops of charged wordlines $i=3,4,5$ by $\{x^\mu_i(\tau),\psi^\mu_i(\tau)\}$ with periodic and antiperiodic boundary conditions respectively. The real and virtual 0+1-dimensional pointlike charges interact in propertime on equal footing through their Lorentz forces.}
    \label{figure_2}
\end{figure}
In particular, M\"oller scattering ($e^- e^-\rightarrow e^- e^-$) between two fermions is given by the expression
\ba 
&\mathcal{S}_{fi}^{(2)}= \lim_{\substack{x_{f,2}^{0}\to +\infty\\ x_{i,2}^0\to-\infty}}  \lim_{\substack{x_{f,1}^{0}\to +\infty\\ x_{i,1}^0\to-\infty}} \int d^3\v{x}_f^1\int d^3\v{x}_i^1  {\Psi_{f_1,\beta_1}^{(+)^\dag}}(x_f^1)\gamma^0_{\gamma_1\alpha_1} \Psi_{i_1,\alpha_1}^{(+)}(x_i^1)\nonumber\\
&\times
\int d^3\v{x}_f^2\int d^3\v{x}_i^2  {\Psi_{f_2,\beta_2}^{(+)\dag}}(x_f^2)\gamma^0_{\gamma_2\alpha_2} \Psi^{(+)}_{i_2,\alpha_2}(x_i^2)
\frac{1}{\mathrm{Z}}\bigg(\frac{1}{i}\bigg)^4\frac{\delta^4\mathrm{Z}}{\delta \eta_{\gamma_2}(x_i^2)\delta\eta_{\gamma_1}(x_i^1)\delta\bar{\eta}_{\beta_1}(x_f^1)\delta\bar{\eta}_{\beta_2}(x_f^2)}\bigg|_{\substack{\bar{\eta}=0\\\eta=0}}\,.\nonumber
\ea 
Proceeding along the same lines and using Eq.~\eqref{2_point_function_result_minkowski}, this gives two contributions, related to the one possible permutation of the single particle states in a final state with two fermions,
\ba 
\mathcal{S}_{fi}^{(2)} = \mathcal{S}_{fi,1}^{(2)}+\mathcal{S}_{fi,2}^{(2)}=\sum_{\ell=0}^\infty \mathcal{S}_{fi,1}^{(2,\ell)}+\sum_{\ell=0}^\infty\mathcal{S}_{fi,2}^{(2,\ell)}\,,
\ea 
where the $\ell$-loop contributions are
\ba 
&\mathcal{S}_{fi,1}^{(2,\ell)}= \frac{\mathrm{Z}_\text{MW}}{\mathrm{Z}} \frac{(-1)^\ell}{\ell!} \lim_{\substack{x_{f,1}^{0}\to +\infty\\ x_{i,1}^0\to-\infty}} \int d^3\v{x}_f^1\int d^3\v{x}_i^1  {\Psi_{f_1}^{(+)\dag}}(x_f^1) \exp\bigg\{\bar{\gamma}_\lambda\frac{\partial}{\partial \theta_\lambda^1}\bigg\} \bar{\gamma}_0 \Psi_{i_1}^{(+)}(x_i^1) \label{s_fi_2_eminuseminus_eminuseminus_1}\\
& \times\lim_{\substack{x_{f,2}^{0}\to +\infty\\ x_{i,2}^0\to-\infty}}\int d^3\v{x}_f^2\int d^3\v{x}_i^2  {\Psi_{f_2}^{(+)\dag}}(x_f^2) \exp\bigg\{\bar{\gamma}_\lambda\frac{\partial}{\partial \theta_\lambda^2}\bigg\} \bar{\gamma}_0 \Psi_{i_2}^{(+)}(x_i^2)\,\mathrm{W}^{(2,\ell)}(x_f^1,x_i^1,\theta^1,x_f^2,x_i^2,\theta^2)\,\nonumber,
\ea 
and
\ba 
&\mathcal{S}_{fi,2}^{(2,\ell)}=  -\frac{\mathrm{Z}_\text{MW}}{\mathrm{Z}}  \frac{(-1)^\ell}{\ell!}\lim_{\substack{x_{f,1}^{0}\to +\infty\\ x_{i,2}^0\to-\infty}} \int d^3\v{x}_f^2\int d^3\v{x}_i^1 {\Psi_{f_2}^{(+)\dag}}(x_f^2) \exp\bigg\{\bar{\gamma}_\lambda\frac{\partial}{\partial \theta_\lambda^1}\bigg\} \bar{\gamma}_0 \Psi_{i_1}^{(+)}(x_i^1) \label{s_fi_2_eminuseminus_eminuseminus_2}\\
& \times  \lim_{\substack{x_{f,2}^{0}\to +\infty\\ x_{i,1}^0\to-\infty}} \int d^3\v{x}_f^1\int d^3\v{x}_i^2 {\Psi_{f_1}^{(+)\dag}}(x_f^1) \exp\bigg\{\bar{\gamma}_\lambda\frac{\partial}{\partial \theta_\lambda^2}\bigg\} \bar{\gamma}_0 \Psi_{i_2}^{(+)}(x_i^2)\mathrm{W}^{(2,\ell)}(x_f^2,x_i^1,\theta^1,x_f^1,x_i^2,\theta^2)\,.\nonumber
\ea 

Along the same lines,  Bhabha scattering case ($e^- e^+\rightarrow e^- e^+$) in the worldline formalism is given by the Dyson S-matrix element,
\ba 
&\mathcal{S}_{fi}^{(2)}=  \lim_{\substack{x_{f,1}^{0}\to +\infty\\ x_{i,1}^0\to-\infty}} \int d^3\v{x}_f^1\int d^3\v{x}_i^1  {\Psi_{f_1,\beta_1}^{(+)\dag}}(x_f^1)\gamma^0_{\gamma_1\alpha_1} \Psi^{(+)}_{i_1,\alpha_1}(x_i^1)\lim_{\substack{x_{f,2}^{0}\to +\infty\\ x_{i,2}^0\to-\infty}} \int d^3\bar{\v{x}}_f^2\int d^3\bar{\v{x}}_i^2 \\
& \times  \Psi^{(-)}_{f_2,\beta_2}(\bar{x}_f^2) \gamma^0 _{\gamma_2\beta_2}{\Psi_{i_2,\alpha_2}^{(-)\dag}}(\bar{x}_i^2)\frac{1}{\mathrm{Z}}\bigg(\frac{1}{i}\bigg)^4\frac{\delta^4\mathrm{Z}}{\delta \eta_{\alpha_2}(\bar{x}_i^2)\delta\eta_{\gamma_1}(x_i^1)\delta\bar{\eta}_{\beta_1}(x_f^1)\delta\bar{\eta}_{\gamma_2}(\bar{x}_f^2)}\bigg|_{\substack{\bar{\eta}=0\\\eta=0}}\,,\nonumber
\ea 
and contains two contributions  
\ba 
\mathcal{S}_{fi}^{(2)} = \mathcal{S}_{fi,1}^{(2)}+\mathcal{S}_{fi,2}^{(2)}=\sum_{\ell=0}^\infty \mathcal{S}_{fi,1}^{(2,\ell)}+\sum_{\ell=0}^\infty \mathcal{S}_{fi,2}^{(2,\ell)}\,.
\ea 
The first contribution yields the amplitude for the scattering of the particle and anti-particle pair in the \textit{in} state to the \textit{out} state, to all orders in perturbation theory:
\ba 
&\mathcal{S}_{fi,1}^{(2,\ell)} = -\frac{\mathrm{Z}_\text{MW}}{\mathrm{Z}}  \frac{(-1)^\ell}{\ell!}\lim_{\substack{x_{f,1}^{0}\to +\infty\\ x_{i,1}^0\to-\infty}} \int d^3\v{x}_f^1 \int d^3\v{x}_i^1 {\Psi_{f_1}^{(+)\dag}}(x_f^1)\exp\bigg\{\bar{\gamma}_\lambda\frac{\partial}{\partial \theta_\lambda^1}\bigg\}\bar{\gamma}_0\Psi_{i_1}^{(+)}(x_i^1)   \nonumber\\
&\times \lim_{\substack{\bar{x}_{f,2}^{0}\to +\infty\\ \bar{x}_{i,2}^0\to-\infty}} \int d^3\bar{\v{x}}_f^2 \int d^3\bar{\v{x}}_i^2{\Psi_{i_2}^{(-)\dag}}(\bar{x}_i^2)\exp\bigg\{\bar{\gamma}_\lambda\frac{\partial}{\partial \theta_\lambda^2}\bigg\}\bar{\gamma}_0\Psi_{f_2}^{(-)}(\bar{x}_f^2)  \mathrm{W}^{(2,\ell)}(x_f^1,x_i^1,\theta^1,\bar{x}_i^2,\bar{x}_f^2,\theta^2)\,.
\ea 
The second contribution, the annihilation of the real pair in the \textit{in} state to produce a different real pair in the \textit{out} state, is given by
\ba 
&\mathcal{S}_{fi,2}^{(2,\ell)} =+ \frac{\mathrm{Z}_\text{MW}}{\mathrm{Z}} \frac{(-1)^\ell}{\ell!}\lim_{\substack{x_{f,1}^{0}\to +\infty\\ \bar{x}_{f,2}^0\to +\infty}} \int d^3\v{x}_f^1 \int d^3\bar{\v{x}}_f^2 {\Psi_{f_1}^{(+)\dag}}(x_f^1)\exp\bigg\{\bar{\gamma}_\lambda\frac{\partial}{\partial \theta_\lambda^1}\bigg\}\bar{\gamma}_0\Psi_{f_2}^{(-)}(\bar{x}_f^2)   \nonumber\\
&\times \lim_{\substack{\bar{x}_{i,2}^{0}\to -\infty\\ x_{i,1}^0\to-\infty}} \int d^3\bar{\v{x}}_i^2 \int d^3\v{x}_i^2 {\Psi_{i_2}^{(-)\dag}}(\bar{x}_i^2)\exp\bigg\{\bar{\gamma}_\lambda\frac{\partial}{\partial \theta_\lambda^1}\bigg\}\bar{\gamma}_0\Psi_{i_1}^{(+)}(x_i^1)  \mathrm{W}^{(2,\ell)}(\bar{x}_i^2,x_i^1,\theta^1,x_f^1,\bar{x}_f^2,\theta^2)\,.
\ea

These expressions can be  generalized to scattering
processes involving an arbitrary number of real charges
$r$. For example,  an \textit{in} state with initial $n_i$ positive energy particles and an \textit{out} state with $n_f=n_i=r$ positive energy particles, has the Dyson S-matrix element, 
\ba 
\mathcal{S}_{fi}^{(r)} = \sum_{\ell=0}^\infty \mathcal{S}_{fi}^{(r,\ell)}\,,
\ea 
where the $\ell$-loop contribution is given by
\ba
&\mathcal{S}_{fi}^{(r,\ell)} = \frac{\mathrm{Z}_\text{MW}}{\mathrm{Z}}\frac{(-1)^\ell}{\ell!} \prod_{n=1}^r \left\{\lim_{\substack{x_{f,n}^0\to+\infty\\x_{i,n}^0\to-\infty}}\int d^3\v{x}_f^n  \int d^3\v{x}_i^n {\Psi_{f_n}^{(+)\dag}}(x_f^n) \exp\bigg\{\bar{\gamma}_\lambda\frac{\partial}{\partial \theta_\lambda^n}\bigg\} \bar{\gamma}_0  \Psi_{i_n}^{(+)}(x_i^n) \right\}\nonumber\\
&\times\mathrm{W}^{(r,\ell)}(x_f^r,x_i^r,\theta^r,\ldots,x_f^1,x_i^1,\theta^1)+\text{final state permutations}\,.\label{transition_n_particles}
\ea 

Eq.~\eqref{transition_n_particles} 
is the central result of this section. It provides a closed form  expression for the matrix element for any $\alpha\to\beta$ transition between $r$ real fermions to all loop orders. It can be reinterpreted as the sum over first-quantized many-body worldline configurations of real and virtual pointlike  spinning charges whose numbers grow with each order in the loop expansion. We observe that it exponentiates the hard and soft interactions of a many-body system, including the spin degrees of freedom, and  defines  the expectation value corresponding to averages over worldline trajectories that are seen explicitly to obey energy-momentum and helicity-momentum constraints. 

In the next section, we will flesh out the perturbative framework  obtained that follows from the formal expressions for the amplitudes 
derived in this section and in Section \ref{section_2}. We will  specifically point to its advantages relative to the conventional Feynman diagram framework for computing amplitudes. 

\section{\label{section_4}QED perturbation theory without Feynman diagrams}
The central results of Sections \ref{section_2} and \ref{section_3},  Eqs.~\eqref{QED_partition_function_worldline} and \eqref{transition_n_particles}, respectively, are valid 
to all orders in perturbation theory, providing a first-quantized many-body description, summarized in the latter equation, of an arbitrary number of photon exchanges  between $\ell$ virtual and $r$ real worldlines. To provide a familiar context in which to interpret these findings, we will connect our results to the conventional language of  Feynman diagrams, and thereby examine in detail the perturbative expansion in the coupling of the worldline amplitudes found in Sections \ref{section_2} and \ref{section_3}. 
We note that by considering general multi-loop and multi-open worldlines we are here going beyond the well-known one-loop approach to QFT without Feynman diagrams pioneered by Bern and Kosower, \cite{Bern:1991aq} and further elucidated by Strassler~\cite{Strassler:1992zr}, albeit discussed here in the limited context of QED.
\subsection{\label{section_4_1}Multi-loop diagrams in the QED vacuum}
We begin by recalling Eq.~\eqref{w_l_definition}:
\ba
\frac{\text{Z}}{\text{Z}_\text{MW}}=\sum_{\ell=0}^\infty
\text{Z}^{(\ell)}=\sum_{\ell=0}^\infty \frac{(-1)^\ell}{\ell!}\mathrm{W}^{(\ell)}
\ea
with
\ba
 \text{W}^{(\ell)}= \bigg\langle \exp\bigg\{\frac{1}{2}\int \frac{d^4k}{(2\pi)^4}i\tilde{J}_\mu^{(\ell)}(-q)\tilde{D}^B_{\mu\nu}(q)i\tilde{J}_\nu^{(\ell)}(+q)\bigg\}\bigg\rangle \,, \label{w_l_definition_momentum}
\ea
where the photon propagator is the expression in Eq.~\eqref{photon_propagator_momentum} 
and the Fourier transforms of the $\ell$ virtual fermion currents in Eq.~\eqref{total_current} are 
\ba
\tilde{J}_\mu^{(\ell)}(q) = \sum_{i=1}^\ell \tilde{J}^{i}_\mu(q)=g\sum_{i=1}^\ell \int^1_0 d\tau
e^{+iq\wc x^i(\tau)} \Big(\dot{x}_\mu^i(\tau)+i\varepsilon_i^0q_\nu
\psi_\mu^i(\tau)\psi_\nu^i(\tau)\Big)\,.
\label{total_current_momentum}
\ea

Because the required normalized expectation value in Eq.~\eqref{w_l_definition_momentum} is separable, loop-by-loop, to the product of one-loop normalized expectation values, one can  alternately rewrite this expression as 
\ba 
\frac{\mathrm{Z}}{\mathrm{Z}_\mathrm{MW}} = \sum_{n,\ell=0}^\infty \mathrm{Z}_{(n)}^{(\ell)}\,,\,\,\,\,\,  
 \mathrm{Z}^{(\ell)}_{(n)}=\frac{(-1)^\ell}{\ell!}\frac{1}{2^n}\frac{1}{n!} \bigg\langle \bigg\{\sum_{i,j=1}^\ell S^{ij}\bigg\}^n\bigg\rangle\label{perturbative_expansion_z}\,,
\ea
whose building block is the interaction functional representing single photon exchange between a pair of virtual fermion loops $i$ and $j$, 
\ba 
S^{ij}=\int\frac{d^4q}{(2\pi)^4}i\tilde{J}_\mu^i(-q)\tilde{D}_{\mu\nu}^B(q)i\tilde{J}_\nu^j(+q)\,,
\label{s_ij_momentum}
\ea 
and $\text{Z}^{(\ell)}_{(n)}$ 
represents the vacuum sub-graphs with $n$ photon lines connecting in all possible ways $\ell$ fermion loops; in other words, $\mathrm{Z}^{(\ell)}_{(n)}$ is built from $\ell$-loop $n$-photon polarization tensors.

Thus the perturbation expansion in the coupling of the vacuum-vacuum amplitude can be expressed as 
\ba
\frac{1}{\text{Z}^{(0)}}\frac{\text{Z}}{\text{Z}_{\text{MW}}}=1+\underbrace{\bigg(\text{Z}^{(1)}_{(0)}+\text{Z}^{(1)}_{(1)}+\cdots\bigg)}_{\text{multi-photon},\medspace\text{1-fermion loop}}+\underbrace{\bigg(\text{Z}^{(2)}_{(0)}+\text{Z}^{(2)}_{(1)}+\cdots\bigg)}_{\text{multi-photon},\medspace\text{2-fermion loops}}+\underbrace{\bigg(\text{Z}^{(3)}_{(0)}+\text{Z}^{(3)}_{(1)}+\cdots\bigg)}_{\text{multi-photon},\medspace\text{3-fermion loops}}+\cdots
\label{schematic_perturbative_expansion_vacuum}\,,
\ea
Note that here we have factored out the zero-point energy
of the vacuum from all diagrams\footnote{We omit in what follows the phase common to all diagrams.},
\ba
\text{Z}^{(0)}=\exp\bigg\{\frac{1}{2}\Tr\int^\infty_0\frac{d\varepsilon_0}{\varepsilon_0}e^{-\varepsilon_0}\bigg\}\,.
\ea

To compute in full generality terms in this expansion at arbitrary order, we first observe that 
\ba 
&\bigg\{\sum^\ell_{i,j=1}S_{ij}\bigg\}^n = \bigg\{S_{11}+\bigg(\sum_{i,j=1}^\ell S_{ij}-S_{11}\bigg)\bigg\}^n
=\sum_{n_{11}=0}^n \frac{n!}{(n-n_{11})!n_{11}!}S_{11}^{n_{11}}\bigg\{\sum_{i,j=1}^\ell S_{ij}-S_{11}\bigg\}^{n-n_{11}}\,,
\ea 
where $n_{11}$ is the number of photon lines attached at both ends to the $i=1$ loop fermion.
Repeating the procedure recursively for the remaining term in brackets in the r.h.s., this gives
\ba
\bigg\{\sum^\ell_{i,j=1}S_{ij}\bigg\}^n  = n! \sum_{n_{11}+n_{12}+\cdots+n_{\ell\ell}=n}\bigg\{ \prod_{i,j=1}^\ell \frac{S_{ij}^{n_{ij}}}{n_{ij}!}\bigg\}\,,
\ea 
where $n_{ij}$  ($n_{ji}$) are the number of virtual photon lines flowing from the fermion loop $j$ to $i$ ($i$ to $j$).
Inserting this back into Eq.~\eqref{perturbative_expansion_z}, the $n!$ out in front cancels, and one gets
\ba
\mathrm{Z}^{(\ell)}_{(n)} = \frac{(-1)^\ell}{\ell!} \sum_{n_{11}+n_{12}+\cdots+n_{\ell\ell}=n} \bigg\langle \bigg\{\prod_{i,j=1}^\ell \frac{S_{ij}^{n_{ij}}}{2^{n_{ij}}n_{ij}!}\bigg\}\bigg\rangle\,.
\ea 
Using now Eq.~\eqref{s_ij_momentum}, collecting the currents of each virtual fermion $i$, and noticing the normalized expectation values are independent, the terms in the coupling expansion of the vacuum-vacuum amplitude of QED are given by
\ba
&\mathrm{Z}^{(\ell)}_{(n)} = \frac{1}{\ell!} \sum_{n_{11}+n_{12}+\cdots+n_{\ell\ell}=n}\prod_{i,j=1}^\ell \bigg\{ \frac{1}{2^{n_{ij}}}\frac{1}{n_{ij}!} \prod_{k=1}^{n_{ij}} \bigg\{\int\frac{d^4q_k^{ij}}{(2\pi)^4} \tilde{D}_{\mu_k^{ij}\nu_k^{ij}}^B(q_k^{ij})\bigg\}\bigg\}\nonumber\\
&\times \prod_{i=1}^\ell \bigg\{- \bigg\langle  \prod_{j=1}^\ell \prod_{k=1}^{n_{ij}}\Big(iJ_{\mu_k^{ij}}(-q_k^{ij})\Big)\prod_{k=1}^{n_{ji}}\Big(iJ_{\nu_k^{ji}}(+q_k^{ji})\Big)\bigg\rangle \bigg\} \,.\label{perturbative_expansion_z_result}
\ea 
Through these formal manipulations, we now see that the general $\ell$-loop, $n$-photon, amplitudes are reduced to computing a product of $\ell$ one-loop, $n_i^\text{th}=\sum_{j=1}^\ell (n_{ij}+n_{ji})$ rank, vacuum polarization tensors. 

In the worldline framework, these reduce to evaluating the normalized expectation value of a product of $n_i$ currents, specifically, the terms in brackets in the second line of Eq.~\eqref{perturbative_expansion_z_result}.
Here, each $k$ of the $n_{ij}$ photon lines of momentum $q_k^{ij}$ are attached at both ends to the $i$ and $j$ virtual fermions. The signs in the arguments of the currents $-q_k^{ij}$ and $+q^{ji}_k$ indicate, respectively, the absorption and emission of the momentum of the photon by the fermion line. Because there are $n_{ij}!$ ways of attaching the two ends of $n_{ij}$ virtual photon lines and $2$ momentum flow directions for each of the photons, the sub-graph is normalized by the factors $n_{ij}!$ and $2^{n_{ij}}$. We absorbed the $\ell$ $-1$ parity factors in each vacuum polarization tensor, and the $1/\ell!$ accounts for the $\ell!$ identical permutations. 
The detailed computation of the
worldline path integrals in these general $n^{\rm th}$ rank tensors is spelled out in  
Appendix \ref{appendix_c}.
\begin{figure}[ht]
  \centering
  \includegraphics[scale=0.25]{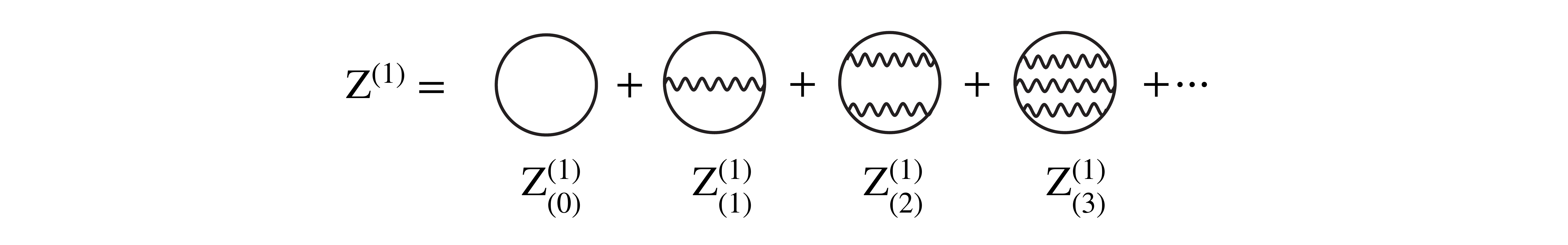}
  \caption{Perturbative expansion of the one-loop fermion contribution
    $\ell=1$ to the QED vacuum , shown here up to three loops $n=1,2,3,\ldots$ in
    the number of photons.}
  \label{fig_3}
\end{figure}

To understand how the perturbative expansion works, let us first consider the fermion one-loop graphs ($\ell=1$) depicted in Fig.~\ref{fig_3}.
Beginning with the simplest graph, setting $n=0$ in Eq.~\eqref{perturbative_expansion_z_result}, the noninteracting single free fermion loop diagram can be expressed as
\ba
\text{Z}^{(1)}_{(0)}=\Pi = -2(2\pi)^4\delta^{(4)}(0)\int_0^\infty\frac{d\varepsilon_0}{\varepsilon_0}\int\frac{d^4p}{(2\pi)^4}e^{-\varepsilon_0(m^2+p^2)},\label{z_1_0_result}
\ea
where the infinite volume phase  space factor $\text{V}=(2\pi)^4\delta^{(4)}(0)$ is a consequence of  translational  invariance, and where we used in the r.h.s. the result Eq.~\eqref{pi} derived in Appendix \ref{appendix_c}. The remaining integral in $\varepsilon_0$ is UV divergent. However, one must pair this amplitude of finding a free one-loop in the vacuum with the 
corresponding counterterm in the zero-point energy of the vacuum, whose series will produce
\ba
\text{Z}^{(0)}=\exp\bigg\{\frac{1}{2}\Tr\int^\infty_0 \frac{d\varepsilon_0}{\varepsilon_0}e^{-\varepsilon_0}\bigg\}=1+\frac{1}{2}\Tr\int^\infty_0 \frac{d\varepsilon_0}{\varepsilon_0}e^{-\varepsilon_0}+\cdots,\label{z_0_result}
\ea
The UV divergence $1/\varepsilon_0$ is then removed, leaving the mass-shell branch point
\ba
\label{eq:branch-point}
\frac{1}{2}\Tr \int_0^\infty \frac{d\varepsilon_0}{\varepsilon_0} e^{-\varepsilon_0}+\text{Z}_{(1)}^{(0)} = 2(2\pi)^4\delta^{(0)}(0)\int\frac{d^4p}{(2\pi)^4}\ln (p^2+m^2)\,.
\ea
In the absence of real 
photons, there are no diagrams involving tadpole subgraphs in the one-loop contributions $\ell=1$. They will appear however in multi-loop diagrams with $\ell>1$. Since  $\Pi_\mu(k)=0$, as seen from the derivation following Eq. 
\eqref{pi_mu} of Appendix \ref{appendix_c}, those containing tadpoles vanish as required. 

The first correction, in $\text{Z}_{(1)}$, is the amplitude for  finding the vacuum contribution from a single fermion loop with the internal exchange of a photon. Setting $n=1$ in the expression in  Eq.~\eqref{perturbative_expansion_z_result}, we get
\ba
&\text{Z}_{(1)}^{(1)}=\frac{1}{2}\int\frac{d^4q}{(2\pi)^4}\tilde{D}^B_{\mu\nu}(q)\Pi_{\mu\nu}(-q,+q)\label{z_1_1_result}\\
&=(2\pi)^4\delta^{(4)}(0)\, 2g^2\int\frac{d^4q}{(2\pi)^4}\int\frac{d^4p}{(2\pi)^4}\frac{\eta_{\mu\nu}}{q^2}\,\frac{\eta_{\mu\nu}(p^2+p\wc q+m^2)-p_\mu(p_\nu+q_\nu)-p_\nu(p_\mu+q_\mu)}{(p^2+m^2)((p+q)^2+m^2)},\nonumber
\ea
where we used Eq. \eqref{pi_munu} in Appendix \ref{appendix_c}.

Higher order contributions are systematically obtained by substituting  Eq.~\eqref{bern_kosower} from Appendix \ref{appendix_c} in  Eq.~\eqref{perturbative_expansion_z_result}.
For instance, the amplitude for finding the vacuum in a configuration with a single fermion loop exchanging two photons,
\ba
\text{Z}^{(1)}_{(2)}=\frac{1}{2\wc 4} \int\frac{d^4q_1}{(2\pi)^4}\int\frac{d^4q_2}{(2\pi)^4}\tilde{D}_{\mu\nu}^B(q_1)\tilde{D}_{\rho\sigma}^B(q_2)\Pi_{\mu\nu\rho\sigma}(-q_1,+q_1,-q_2,+q_2)\,,
\label{z_1_2_result}
\ea
where $\Pi_{\mu\nu\rho\sigma}(q_1,q_2,q_3,q_4)$ is the light-by-light scattering graph that
can be obtained from precisely the same master equation  (Eq.~\eqref{bern_kosower} in Appendix \ref{appendix_c}) as the vacuum polarization tensor. 
The one-loop diagrams Eqs.~\eqref{z_1_0_result}, \eqref{z_1_1_result} and \eqref{z_1_2_result} are depicted in Fig. \ref{fig_3}. Order-by-order in perturbation theory, the one-loop amplitudes obtained in Euclidean time in Eqs.~\eqref{eq:branch-point}- \eqref{z_1_2_result}, as well as any $\mathrm{Z}^{(\ell)}_{(n)}$, can be Wick rotated back to Minkowski times following the transformations given in Appendix \ref{appendix_a}.
\begin{figure}[ht]
  \centering
  \includegraphics[scale=0.25]{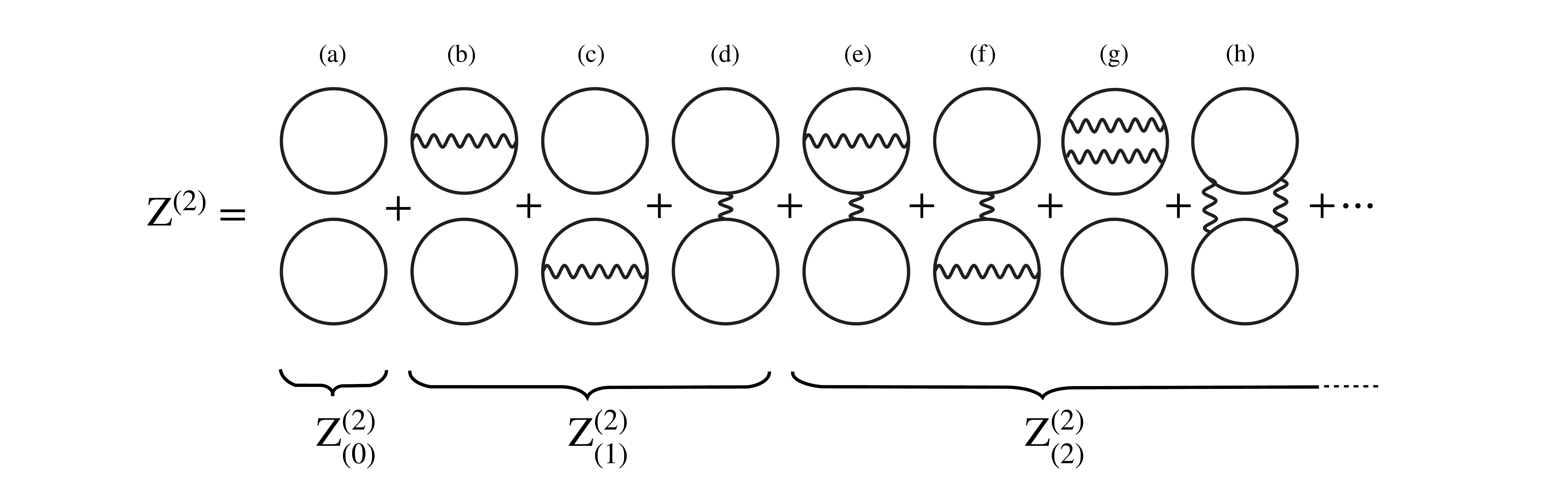}
  \caption{Perturbative expansion of the two-loop fermion
    configurations $\ell=2$ in the QED vacuum up to two photon loops
    $n=1,2,\ldots$.}
  \label{fig_4}
\end{figure}

The two-loop fermion configurations are obtained in the same vein. The vacuum contribution reads, using Eq.~\eqref{perturbative_expansion_z_result}, as 
\ba
\mathrm{Z}^{(2)}_{(0)}=\frac{1}{2!}(\Pi)^2= \frac{1}{2}\Big(\mathrm{Z}^{(1)}_{(0)}\Big)^2\,.
\ea
It corresponds to graph (a) in Fig.~\ref{fig_4}. 

More generally, all the free loop amplitudes satisfy $\mathrm{Z}^{(\ell)}_{(0)}=\big(\mathrm{Z}^{(1)}_{(0)}\big)^\ell/\ell!$; using  Eq.~\eqref{z_0_result}, they can be collected, resummed and UV regularized as
\ba
 \mathrm{Z}^{(0)} \bigg\{1+\sum_{\ell=1}^\infty \mathrm{Z}^{(\ell)}_{(0)}\bigg\}= \mathrm{Z}^{(0)}\exp\bigg\{\mathrm{Z}^{(1)}_{(0)}\bigg\} = \exp\bigg\{2(2\pi)^4\delta^{(4)}(0) \int\frac{d^4p}{(2\pi)^4}\ln\Big(p^2+m^2\Big)\bigg\}\,.
\ea
From Eq.~\eqref{perturbative_expansion_z_result}, for the two-loop graphs (graphs (b), (c) and (d)) with one photon exchange, 
we get
\ba
\mathrm{Z}^{(2)}_{(1)}= \frac{1}{2}\Pi\int\frac{d^4q}{(2\pi)^4}\tilde{D}_{\mu\nu}^B(q)\Pi_{\mu\nu}(-q,+q)+\frac{1}{2}\int\frac{d^4q}{(2\pi)^4}\tilde{D}_{\mu\nu}^B(q) \Pi_\mu(-q)\Pi_\nu(+q)\,.
\ea
The first term corresponds to graphs $(b)$
and $(c)$ in Fig. \ref{fig_4}, and the second term to graph
$(d)$. Since $i\Pi_\mu(q)=0$, this last diagram does not contribute, and we can express the result as 
\ba
\mathrm{Z}^{(2)}_{(1)}= \mathrm{Z}^{(1)}_{(0)}\mathrm{Z}^{(1)}_{(1)}\,,
\ea
where $\mathrm{Z}_{(1)}^{(1)}$ is defined in Eq.~\eqref{z_1_1_result}.

The vacuum-vacuum amplitude for two fermion loops with two photon internal lines, illustrated in Fig.~\ref{fig_4} graphs (e)-(h), can be read from 
Eq.~\eqref{perturbative_expansion_z_result} to be,
(dropping the diagrams with tadpole subgraphs $(e)$ and $(f)$ in Fig. \ref{fig_4})
\ba
  \mathrm{Z}_{(2)}^{(2)}&=\mathrm{Z}^{(1)}_{(0)}\mathrm{Z}^{(1)}_{(2)}+\frac{1}{2}\Big(\mathrm{Z}_{(1)}^{(1)}\Big)^2\\
  &+\frac{1}{4} \int\frac{d^4q_1}{(2\pi)^4}\int \frac{d^4q_2}{(2\pi)^4}\tilde{D}_{\mu\nu}^B(q_1)\tilde{D}_{\rho\sigma}^B(q_2)\Pi_{\mu\rho}(-q_1,-q_2)\Pi_{\nu\sigma}(q_1,q_2)\nonumber\,.
\ea
The first contribution corresponds to graph (g) in Fig.~\ref{fig_4}, the second term corresponds to the square of the second diagram in Fig.~\ref{fig_3}, and the final term is
depicted in graph (h) of Fig.~\ref{fig_4}. 

The above exercise is very suggestive of the efficacy  of the worldline framework for high order perturbative computations. 
Their advantage can be summarized as follows. 
Firstly, from the examples we considered,  we can deduce that all  multi-loop diagrams, to any level of complexity, can be universally generated from the master equation -- Eq. \eqref{perturbative_expansion_z_result}. The explicit, yet compact, elements for their computation given by Eq.~\eqref{bern_kosower} in Appendix  \ref{appendix_c}. In general $d$ dimensions, the results of applying this procedure to the calculation of amplitudes to fixed order are worldline diagrams that can be equivalently represented in terms of Schwinger/Feynman parameters, with their genuine UV loop divergences then regularized as in conventional perturbation theory. 

Secondly, while each Feynman graph in the latter corresponds to one particular permutation of the photon insertions in $\mathrm{Z}^{(\ell)}_{(n)}$, in contrast,  Eq. \eqref{perturbative_expansion_z_result} represents them all at once, automatically encoding the factorial growth of diagrams in perturbation theory. For instance,  $\mathrm{Z}_{(2)}^{(1)}$, shown in Fig.~\ref{fig_3}, is given by the integral in Eq.~\eqref{bern_kosower} of Appendix \ref{appendix_c}. In the usual perturbative calculation, it requires the evaluation of two Feynman graphs, corresponding to the dressing of the fermion loop with the two photons crossed or uncrossed, respectively. At large $n$ in perturbation theory, the situation becomes more complex, with the number of diagrams exploding factorially. 

A fundamental question then arises as to whether Eq.~\eqref{bern_kosower} grows factorially with the increasing number of diagrams in the large $n$ photon limit, or if cancellations do in fact occur between diagrams and reduce the scaling of Eq.~\eqref{bern_kosower} at large $n$ from the naive estimate of the number of diagrams involved. For instance, Cvitanovic and Kinoshita~\cite{Cvitanovic:1977dp} showed that in calculations of the higher order corrections to the electron anomalous magnetic moment~\cite{Cvitanovic:1977dp,Aoyama:2012qma,LAPORTA2017232} large cancellations occur at higher orders within certain gauge invariant sets of diagrams. Given that the number of diagrams inside these sets is very large ($\sim$ hundreds to sixth-order in perturbation theory for $g-2$) this led Cvitanovic to 
conjecture~\cite{Cvitanovic:1977dp} that $g-2$ may not satisfy an asymptotic expansion (at least in the quenched contributions with no virtual fermion loops ($\ell=0$)) as anticipated~\cite{Dyson:1952tj}, and that instead the series is rendered convergent by extensive cancellations between graphs in perturbation theory. 

A gauge invariant set of diagrams contributing to $g-2$ consists of all one-particle irreducible graphs with a fixed number $n$ of virtual photons dressing the electron vertex and no virtual fermion loops. For the further case of the dressings of a closed fermion loop, all the diagrams in a gauge set are then encoded in the single integral in Eq.~\eqref{bern_kosower} of Appendix \ref{appendix_c}. This ``quenched convergence" in the large $n$ photon limit has been addressed previously for loop diagrams within the worldline formulation of sQED and QED using Borel analysis~\cite{Dunne:2002qf,Dunne:2002qg,Dunne:2004xk,Huet:2010nt}.

Finally, we emphasize that in the conventional Wilson loop approach to gauge theories, genuine fermion loop contributions are not included within the framework itself, but are introduced ``by hand". As a specific example, the computation of the expectation value of the Wilson loop 
that appears in cusp anomalous dimensions (which we shall discuss further in Sec.~\ref{section_6}) in standard perturbative Wilson loop computations includes only the boson-boson contributions in Eq.~\eqref{eq:Wilson-line_log-sum} but not the boson-fermion and fermion-fermion contributions contained in the worldline computation. These have to be added separately, adding to the complexity of the computation. We will illustrate this point in detail in Paper II
by performing an explicit computation of the two-loop cusp anomalous dimension in the worldline approach and comparing it to the prior computation in standard perturbation theory~\cite{Korchemsky:1987wg}.
\subsection{\label{section_4_2}Resumming the loop expansion}
The multiplicity of virtual worldlines in
Eq.~\eqref{QED_partition_function_worldline} suggest that the 
sum in the number of closed loops $\ell$ can be performed formally for any amplitude. Equivalently, 
Eq.~\eqref{transition_n_particles} suggests
that these loop-resummed amplitudes can also be summed in the number
$r$ of real particles. This would then produce a ``field-free" worldline generating functional, depending on a  external 
boson source, which makes it suitable for the evaluation of 
amplitudes 
using semi-classical expansions.

To perform the sum in the number of loops $\ell$,
we introduce an external gauge boson source $B_\mu(x)$ that can be regarded as a dummy gauge background field, and define the functional operator
\ba
\mathcal{G}\bigg[\frac{\delta}{\delta B}\bigg]=\exp\bigg\{\frac{1}{2}\int d^4x \int d^4y \frac{\delta}{\delta B_\mu(x)} D_{\mu\nu}^B(x-y)\frac{\delta}{\delta B_\nu(y)} \bigg\}\,.
\label{G_functional_definition}
\ea
This generating functional connects virtual photon lines to all orders in perturbation theory.
We can thus rewrite the interaction term in Eq.~\eqref{w_l_definition} as
\ba
 & \exp\bigg\{\frac{1}{2}\int d^4x \int d^4y iJ_\mu^{(\ell)}(x) D_{\mu\nu}^B(x-y) iJ_\nu^{(\ell)}(y) \bigg\}=\mathcal{G}\bigg[\frac{\delta}{\delta B}\bigg]\exp\bigg\{i \int d^4x J_\mu^{(\ell)
} (x)B_\mu(x)\bigg\}\bigg|_{B=0}\,.
\ea
With this, the sum in Eq.~\eqref{QED_partition_function_worldline} over all vacuum-vacuum diagrams becomes,
\ba
 & \frac{\text{Z}}{\text{Z}_{\text{MW}}}=\mathcal{G}\bigg[\frac{\delta}{\delta B}\bigg]\exp\bigg\{\frac{1}{2}\Tr\int^\infty_0\frac{d\varepsilon_0}{\varepsilon_0}e^{-\varepsilon_0}\bigg\}
  \sum_{\ell=0}^\infty\frac{(-1)^\ell}{\ell!}\Bigg[\int_0^\infty \frac{d\varepsilon_0}{2\varepsilon_0}\int_\text{P} \mathcal{D}^4x \int_\text{AP} \mathcal{D}^4\psi\\
  &\times\exp\bigg\{-m^2\varepsilon_0-\frac{1}{4\varepsilon_0}\int^1_0d\tau\dot{x}_\mu^2(\tau)-\frac{1}{4}\int^{1}_0 d\tau \psi_\mu(\tau)\dot{\psi}_\mu(\tau)+i\int d^4x J_\mu(x)B_\mu(x)\Big)\bigg\}\Bigg]^\ell\Bigg|_{B=0}\nonumber\,,
\ea
where we used the fact that the currents can be decomposed into the sum of individual currents, as in Eq.~\eqref{total_current}.
Summing in $\ell$ and using Eq.~\eqref{auxiliary_currents_jh}, we can reexpress this result as 
\ba
  &\frac{\text{Z}}{\text{Z}_{\text{MW}}}= \mathcal{G}\bigg[\frac{\delta}{\delta B}\bigg]\exp\bigg\{\Gamma\big[B\big]\bigg\}
  \bigg|_{B=0}\,,
  \label{loop_resummation_vacuum}
\ea
where $\Gamma[B]$ is 
the one-loop ($\ell=1$) effective action 
in the presence of the background gauge field $B_\mu(x)$, 
\ba
&\Gamma[B]=\frac{1}{2}\Tr \int_0^\infty \frac{d\varepsilon_0}{\varepsilon_0}e^{-\varepsilon_0}-\frac{1}{2}\int_0^\infty \frac{d\varepsilon_0}{\varepsilon_0}\int_\text{P} \mathcal{D}^4x\int_\text{AP}  \mathcal{D}^4\psi\exp\bigg\{-m^2\varepsilon_0-\frac{1}{4\varepsilon_0}\int^1_0d\tau \dot{x}_\mu^2(\tau)\nonumber\\
&-\frac{1}{4}\int^1_0d\tau \psi_\mu(\tau)\dot{\psi}_\mu(\tau) 
  +ig\int^1_0 d\tau \dot{x}_\mu(\tau) B_\mu\big(x(\tau)\big)-i\frac{g\varepsilon_0}{2}\int^1_0 d\tau \psi_\mu(\tau)\psi_\nu(\tau) F_{\mu\nu}\big(x(\tau)\big)\bigg\}\,.\label{phi_b_definition}
\ea

One has thus reduced the multi (infinite)-loop problem of the QED vacuum to that of computing the one-loop worldline effective action in an arbitrary dummy background field. 
Indeed, since Eq.~\eqref{loop_resummation_vacuum} is valid for any choice of
$B_\mu(x)$, one could look for gauge configurations for which one could obtain a non-perturbative result for $\Gamma[B]$. Specifically, one can look configurations that minimize the action, allowing one to perform a semi-classical expansion around this value. For instance, 
for a constant electric field
$B_\mu(x)\propto x_\mu$,  Eq.~\eqref{phi_b_definition} 
can be  expanded in so-called worldline instanton configurations~\cite{Affleck:1981bma,Dunne:2005sx,Dunne:2006ff}. Note that this form of the ``dummy" background field gives the rate for $e^+ e^-$ pair production in strong external (not dummy) electric fields. 

The generating functional technique for amplitudes 
with real particles is analogous. We can similarly rewrite Eq.~\eqref{w_r_l_definition} as
\ba
&   \text{W}^{(r,\ell)}\Big(x_f^r,x_i^r,\theta^r,\cdots,x_f^1,x_i^1,\theta^1\Big)= \mathcal{G}\bigg[\frac{\delta}{\delta B}\bigg] \bigg\langle\exp\bigg\{+i\int d^4x J^{(r,\ell)}_\mu(x) B_\mu(x)\bigg\}\bigg\rangle\bigg|_{B=0}\,. \label{w_r_l_rewritten}
\ea
We consider first the case of a single real particle $r=1$. 
The gauge functional average of the dressed 2-point function in  Eq.~\eqref{2_point_function_result_minkowski} then becomes 
\ba
 & \Big\langle \bar{D}_A^F(x_f^1,x_i^1) \Big\rangle_{A}= i\frac{\text{Z}_{\text{MW}}}{\text{Z}}\mathcal{G}\bigg[\frac{\delta}{\delta B}\bigg]  \nonumber\\&\times\exp\bigg\{\bar{\gamma}^1_\lambda \frac{\partial}{\partial
      \theta_\lambda^1}\bigg\}\sum_{\ell=0}^\infty \frac{(-1)^\ell}{\ell!} \bigg\langle \exp\bigg\{+i\int d^4x J^{(1,\ell)}_\mu(x) B^\mu(x)\bigg\}\bigg\rangle\bigg|_{\substack{\theta_1=0 \\ B=0}}\,.
 \ea
Since the  currents and actions are additive sums of their single-particle counterparts,  the path integrals of the $r$ external and the $\ell$ virtual particles
can be separated. Using Eq.~\eqref{langle_o_rangle_open}, we get 
\ba
&\Big\langle \bar{D}_A^F(x_f,x_i) \Big\rangle_{A}= \frac{\text{Z}_{\text{MW}}}{\text{Z}}\mathcal{G}\bigg[\frac{\delta}{\delta B}\bigg] \bigg[\frac{i}{N_5}\exp\bigg\{\bar{\gamma}_\lambda \frac{\partial}{\partial
      \theta_\lambda^1}\bigg\}\int_0^\infty
d\varepsilon_1^0 \int d\chi_1^0
\int \mathcal{D}\varepsilon_1\frac{\mathcal{D}\pi_1}{2\pi} \int \mathcal{D}\chi_1 \mathcal{D}\nu_1\nonumber\\
     & \times\int \mathcal{D}x_1\mathcal{D}\psi_1\exp\bigg\{-iS_R+i\int d^4x J_{R,\mu}(x) B^\mu(x)\bigg\}\bigg|_{\theta_1=0}\bigg]\exp\bigg\{\frac{1}{2}\Tr\int^\infty_0\frac{d\varepsilon_0}{\varepsilon_0}e^{i\varepsilon_0}\bigg\}\nonumber\\
&\times\sum_{\ell=0}^\infty \frac{(-1)^\ell}{\ell!}\bigg[\int_0^\infty \frac{d\varepsilon_0}{2\varepsilon_0}\int\mathcal{D}\varepsilon\frac{\mathcal{D}\pi}{2\pi} \int_\text{P}
\mathcal{D}x \int_\text{AP}\mathcal{D}\psi\exp\bigg\{-iS_V+i\int d^4x J_{V,\mu}(x)B^\mu(x)\bigg\}\bigg]^\ell\bigg|_{B=0}
\nonumber\,.
\ea
The first quantity in brackets (in the first two lines) is the amplitude for a real fermion 
to propagate from $x_i^1$ to $x_f^1$ in $B_\mu(x)$. 
The quantity in brackets in  the third line above is the amplitude for a single virtual fermion loop in the presence of $B_\mu(x)$. Polarized virtual fermions from the vacuum and  real
particles are connected by the action of $\mathcal{G}$. Summing this expression to all orders in $\ell$,
\ba
 \Big\langle D_A^F(x_f^1,x_i^1)\gamma^5 \Big\rangle_{A}= \frac{\text{Z}_{\text{MW}}}{\text{Z}} \mathcal{G}\bigg[\frac{\delta}{\delta B}\bigg] D^F_B(x_f^1,x_i^1)\gamma^5 \exp\bigg\{\Gamma[B]\bigg\}\bigg|_{B=0}\,.
\ea
The functional average over gauge fields on the l.h.s. is  replaced on the r.h.s by the action of 
$\mathcal{G}$ on the product of a single real particle's dressed Green function $D_B^F(x_f^1,x_i^1)$ times a one-loop effective action $\Gamma[B]$, with $B$ being set finally to zero. 

To sum in $r$, we introduce two dummy anti-commuting 
sources $\bar{\eta}(x)$ and $\eta(x)$ and define the
quantity

\ba
& \frac{(-i)^r}{r!} \bigg\langle \prod_{n=1}^r  \int d^4x_f^n \int d^4x_i^n \bar{\eta}(x_f^n)    D_F^A(x_f^n,x_i^n)  \eta(x_i^n)\bigg\rangle_{A} \nonumber \\
  &=\frac{\text{Z}_\text{MW}}{\text{Z}} \mathcal{G}\bigg[\frac{\delta}{\delta B}\bigg]\exp\bigg\{\Gamma[B]\bigg\}\frac{(-i)^r}{r!}\prod_{n=1}^r \bigg\{\int d^4x_f^n \int d^4x_i^n  \bar{\eta}(x_f^n)D_F^B(x_f^n,x_i^n)\eta(x_i^n)\bigg\}\bigg|_{B=0}\,.
\ea
From Eqs.~\eqref{2_point_function} and \eqref{functional_average}, the l.h.s. is the definition of
the QED generating functional $\text{Z}[\bar{\eta},\eta]$ normalized by 
$\text{Z}=\text{Z}[0,0]$. 
On the r.h.s., we find the exponential of the transition amplitude for a real fermion  from the dummy state $\eta(x)$ to $\bar{\eta}(x)$ 
in the presence of $B_\mu(x)$. Summing this in $r$,
\ba
\frac{\text{Z}[\bar{\eta},\eta]}{\text{Z}_\text{MW}} = \mathcal{G}\bigg[\frac{\delta}{\delta B}\bigg]\exp\bigg\{\Gamma[B]\bigg\}\exp\bigg\{-i\int d^4x \int d^4y \bar{\eta}(x)D^F_B(x,y)\eta(y)\bigg\}\bigg|_{B=0} \label{loop_resummation_external}
\ea
For no real fermions $\bar{\eta}(x)=\eta(x)=0$, we recover Eq.~\eqref{loop_resummation_vacuum}. 

Equation \eqref{loop_resummation_external} contains the
factorization of three fundamental elements in QED: i) the exponentiated amplitude of a virtual fermion to describe a loop, given by $\Gamma[B]$, ii) the exponentiated amplitude for a 
a single fermion to propagate in an open configuration $D^F_B(x,y)$, and iii) the multiple insertions of the virtual photon line operator $\mathcal{G}[\delta/\delta B]$ connecting them. Note also that we factor out the infinite sea of disconnected photon loops $\text{Z}_\text{MW}$. 

In Section \ref{section_4_1}, we discussed in detail the efficiency of worldline calculations order-by-order in
perturbation theory. Some comments are now in order
regarding possible strategies to extend these calculations to all orders by using expressions like Eqs.~\eqref{loop_resummation_vacuum} and \eqref{loop_resummation_external} in this subsection.

 The  formal expressions we obtained as the principal results of Sections \ref{section_2} and \ref{section_3} for QED amplitudes 
 require the computation of many-body path integrals describing the 
  interactions between many point-like currents, with their multiplicity growing with the loop order $\ell$ in the expansion. Monte Carlo Euclidean techniques for the computation of these path integrals for scalar theories have been addressed in \cite{Gies:2001tj,Gies:2005sb}, and more recently in \cite{Franchino-Vinas:2019udt}. These have thus far only been in the quenched approximation.

Here we have instead reduced the previous quantum many-body problem to the one of solving the quantum dynamics of a single spin-1/2 particle in a given external gauge field, that is later set to zero. This has some advantages. The number of particles is fixed, one must only compute the path integral of a single fermion in an external field and without back-reaction, as it is the case of Eq.~\eqref{loop_resummation_vacuum}.
Further, the dummy field can be chosen conveniently so as to perform a semi-classical expansion of the worldline action around the corresponding instanton configuration. 

This worldline action is given by  Eq.~\eqref{semiclassical_expansion_action} of Appendix \ref{appendix_d}, where the semi-classical expansion is also discussed at length.  We see that replacing the fields sourced by the other particles in the loop expansion, $A_\mu^{x_j}(x)$ and $A_\mu^{\Psi_j}(x)$, by an arbitrary $B_\mu(x)$ reduces the complexity of the calculation.

\section{\label{section_5}  IR structure of QED worldline amplitudes}
To summarize where we are in the paper, we provided in Sections \ref{section_2} and \ref{section_3} compact expressions to compute  amplitudes in QED, to all loop orders in perturbation theory, in a formalism in which matter and gauge fields are integrated out explicitly and replaced by super-pairs of 0+1 dimensional worldlines that represent on equal footing real (external) and virtual (loop) particle propertime trajectories in position and spin.
In this first-quantized approach to QED, the quantum degrees of freedom of the theory are exactly accounted for as 0+1-dimensional boson and Grassmann path integrals over all possible worldline trajectories.

As noted, the computation of the required path integrals in these amplitudes is a difficult task. One approach is an  $\ell$-loop, $n$-photon, perturbative expansion; another is via a semi-classical expansion about so-called worldline instantons using Monte Carlo methods. The efficiency of each of these approaches, respectively, was discussed in Sections \ref{section_4_1} and \ref{section_4_2}.

We will now discuss in this framework the problem of the IR safety of the S-matrix outlined in the introduction to the paper. 
As we will show, the worldline formalism has significant advantages in tackling this problem.  The interactions between real and virtual worldlines, to all orders in perturbation theory, are given by the expressions in  Eqs.~\eqref{S_ab_BB}-\eqref{S_ab_FF}, 
representing in the IR the long range classical Coulomb dynamics of pairwise interactions of charged worldline super-pairs. This will  allow us to examine clearly the IR divergences of the scattering amplitudes, and construct IR safe amplitudes to all orders in perturbation theory.

\subsection{\label{section_5_1}Curing the IR divergences of the Dyson S-matrix}
To study the IR structure of QED, we first consider the Dyson S-matrix element for $r$ positive energy particles\footnote{The discussion here can be extended straightforwardly to scattering amplitudes involving particles and antiparticles.} in Eq.~\eqref{transition_n_particles}: 
\ba
&\mathcal{S}_{fi}^{(r)} = \frac{\mathrm{Z}_\text{MW}}{\mathrm{Z}} \prod_{n=1}^r \left\{\lim_{\substack{t_{f}^{n}\to+\infty\\t_{i}^{n}\to-\infty}}\int d^3\v{x}_f^n  \int d^3\v{x}_i^n u_{s_f^n}^{\dag}(p_f^n)e^{+ip_f^n\wc x_f^n} \exp\bigg\{\bar{\gamma}_\lambda\frac{\partial}{\partial \theta_\lambda^n}\bigg\} \bar{\gamma}_0  u_{s_i^n}(p_i^n)e^{-ip_i^n\wc x_i^n} \right\}\nonumber\\
&\times\sum_{\ell=0}^\infty\frac{(-1)^\ell}{\ell!}\mathrm{W}^{(r,\ell)}(x_f^r,x_i^r,\theta^r,\ldots,x_f^1,x_i^1,\theta^1)\bigg|_{\theta=0}\medspace\medspace\medspace+\medspace\medspace\medspace \text{final state permutations}\,.\label{s_fi}
\ea 
Recall that  $\mathrm{W}^{(r,\ell)}(x_f^r,x_i^r,\theta^r,\ldots,x_f^1,x_i^1,\theta^1)$ encodes the amplitude for the propagation of the $r$ particles from points placed on the celestial sphere at minus infinity to final points placed on the celestial sphere at plus infinity,
\ba 
x_f^n=(t_{f}^{n},\v{x}_{f}^n), \medspace\medspace\medspace x_i^n=(t_i^{n},\v{x}_i^n), \medspace\medspace\medspace t_f^{n}\to +\infty, \medspace\medspace\medspace t_i^{n}\to -\infty\,\medspace\medspace\medspace n=1,\ldots,r\,.\label{boundary_conditions_smatrix}
\ea 
It also contains $\ell$ virtual fermions attached to the $r$ real  particles, and has the structure of a normalized worldline expectation value of the path integrals in Eq.~\eqref{langle_o_rangle_open} over the $r$ worldline trajectories connecting the points at past and future infinity in Eq.~\eqref{boundary_conditions_smatrix},
\ba 
\mathrm{W}^{(r,\ell)}(x_f^r,x_i^r,\theta^r,\ldots,x_f^1,x_i^1,\theta^1)=\bigg\langle \exp\bigg\{-i\sum_{a,b=1}^{r+\ell} S^{ab}\bigg\}\bigg\rangle \,,
\label{w_r0}
\ea 
and over the $\ell$ closed worldline trajectories of the virtual fermions. Here $S^{ab}$ denotes the interaction of worldline $a$ with worldline $b$, and is given by the Wick rotation to Minkowski spacetime of the Lorentz forces in Eq.~\eqref{eq:Wilson-line_log-sum}. The indices $a,b=1,\ldots,r+\ell$ run over real and virtual particles. 

To explore the IR asymptotics of the S-matrix, and
to translate the previous results to the language of Feynman diagrams, we express the Fourier transform of $S^{ab}$ in Eq.~\eqref{eq:Wilson-line_log-sum} as ($d=4$ and $\zeta=1$)
\ba 
S^{ab}=-\frac{1}{2}\int\frac{d^4k}{(2\pi)^4}\tilde{J}_\mu^{a}(-k)\frac{g_{\mu\nu}}{k^2+i\epsilon}\tilde{J}_\nu^b(+k)\,,\label{S_ab_momentum}
\ea 
where $k$ is the 4-momentum of the virtual photon, and the currents in momentum space and Minkowski time are 
\ba 
 \tilde{J}_\mu^a(k)=g\int^1_0d\tau\Big( \dot{x}_\mu^a(\tau)+\varepsilon_0^ak^\nu\psi_\mu^a(\tau)\psi_\nu^a(\tau)\Big)e^{+ik\wc x_a(\tau)}.\label{currents_momentum_minkowski}
\ea 
The problem of the IR structure of the S-matrix is therefore reduced to that of examining the low energy behavior ($k\to 0$) of a theory of  worldline currents. 

In momentum space,
a contribution to $S^{ab}$ will be IR divergent when both currents entering Eq.~\eqref{S_ab_momentum} contain $1/k$ terms that survive when $k\to 0$. In this $k\rightarrow 0$ limit, the fermion component of the current in Eq.~\eqref{currents_momentum_minkowski} is sub-leading relative to the bosonic contribution and need not be considered further in this context. For the scalar component of the current, we can sample $N$ points of the worldline $x^\mu(\tau_\kappa)=x^\mu_\kappa$ for $\kappa=1,\ldots,N$, and connect pairs of points separated by a 4-distance $\delta x_\kappa^\mu =x_{\kappa+1}^\mu-x_\kappa^\mu$ with a straight line path for the propagation of the particle in a time $\delta \tau_\kappa=\tau_{\kappa+1}-\tau_\kappa$,
\ba 
x^\mu(\tau) =x_\kappa^\mu +\frac{\delta x_\kappa^\mu}{\delta \tau_\kappa}(\tau-\tau_\kappa), \,\,\,\,\, \tau\in (\tau_{\kappa+1},\tau_\kappa)\,,\label{worldline_discretized}
\ea 
with $\tau_1=0$ and $\tau_{N}=1$, as shown in Fig.~\ref{fig:worldline_piezewise} for the worldline of a particle starting at $x_1^\mu$ and ending at $x_N^\mu$. As emphasized before, the worldline has a closed path ($x_N^\mu=x_1^\mu$) for a virtual particle. For a real particle, these points are external and placed at plus and minus infinity, following Eq.~\eqref{boundary_conditions_smatrix}. In this discretized form, the path integral over worldline configurations in Eq.~ \eqref{langle_o_rangle_open} reduces to the sum over all possible positions of the intermediate points in the trajectory $x_\kappa^\mu$. 

\begin{figure}[ht]
    \centering
    \includegraphics[scale=0.3]{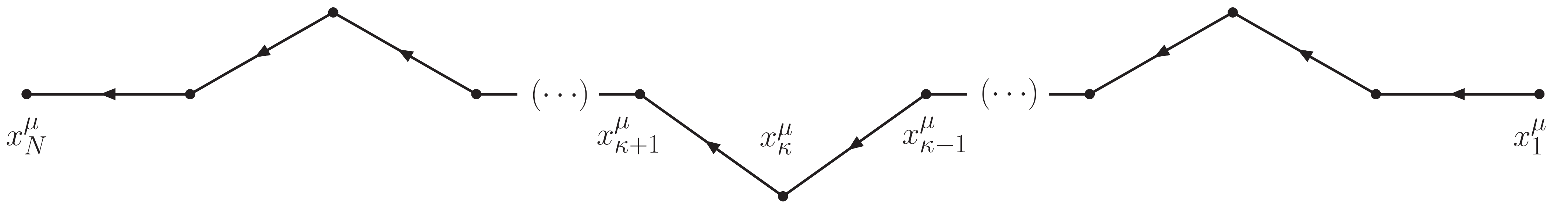}
    \caption{Discretized representation of a bosonic worldline $x_\mu(\tau)$ accounting for one possible configuration of a particle's trajectory in the path integral.}
    \label{fig:worldline_piezewise}
\end{figure}

In the soft limit $k\to 0$, the total current gives, using Eq.~\eqref{worldline_discretized},
\ba 
\tilde{J}^\mu(k)\simeq g\int^1_0 d\tau \frac{dx^\mu}{d\tau}e^{+ik\wc x(\tau)}=g\sum_{\kappa=1}^{N-1} \int^{\tau_{k+1}}_{\tau_k}d\tau \frac{dx^\mu}{d\tau}e^{+ik\wc x(\tau)}=\frac{g}{i}\sum_{\kappa=1}^{N-1}\frac{\delta x^\mu_\kappa}{k\wc \delta x_\kappa}\Big(e^{ik\wc x_{\kappa+1}}-e^{ik\wc x_{\kappa}}\Big)\,.\label{current_momentum_discretized}
\ea 
The form of the current above, within a discrete interval, holds the key to understanding the emergence of IR singularities in the Dyson  S-matrix and how to cure them.
For fixed $x_\kappa^\mu$ and $x_{\kappa+1}^\mu$, the term in the  parenthesis is of order $\mathcal{O}(k)$ and cancels the $1/k$ in the denominator when $k\to 0$. It follows that a worldline configuration with all points fixed - a particle confined to a finite 4-volume - does not introduce $1/k$ poles in $S^{ab}$ leading to IR divergences in the S-matrix.

However, if the real or virtual particle explores infinity in some step $x_\kappa^\mu\to \pm\infty$, which is a contribution accounted for by the path integral, as the one shown in Fig.~\ref{fig:worldline_piezewise_2}, the corresponding term $e^{ik\wc x_\kappa}$ in Eq.~\eqref{current_momentum_discretized} 
should be dropped due to the rapid oscillations in the phase in the asymptotic region. The remaining term in parenthesis in Eq.~\eqref{current_momentum_discretized} is then left unpaired, and contains a $1/k$ contribution in $k\to 0$ that introduces an IR divergence in $S^{ab}$. 
\begin{figure}[ht]
    \centering
    \includegraphics[scale=0.3]{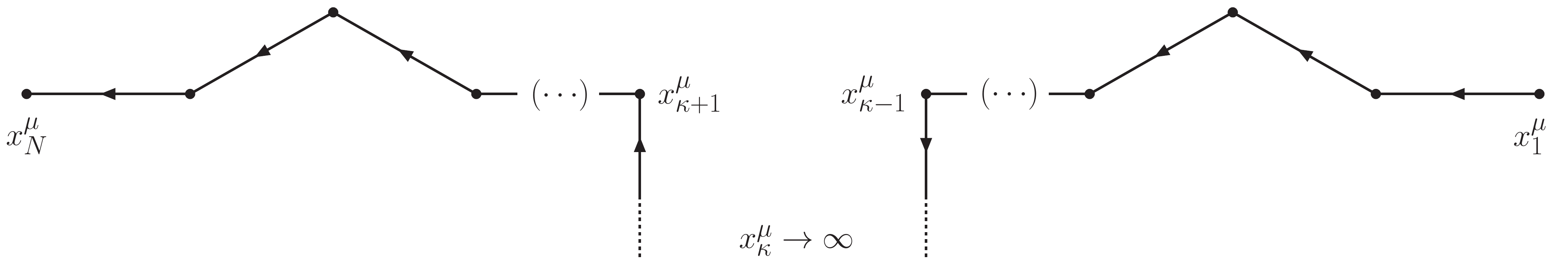}
    \caption{Discretized configuration of a bosonic worldline $x_\mu(\tau)$ with an internal point at infinity. A dashed line is drawn to indicate that the fermion line extends to infinity and becomes external as a result.}
    \label{fig:worldline_piezewise_2}
\end{figure}

To examine the $k\to 0$ behavior of these particular configurations,  we will distinguish between internal and external points placed at infinity. If the $\kappa$-th point at infinity is internal, like the one in Fig.~\ref{fig:worldline_piezewise_2}, it will appear twice in the sum in Eq.~\eqref{current_momentum_discretized}, which one can then  reorganize as
\ba 
&\tilde{J}^\mu(k)=\frac{g}{i}\sum_{\eta=\kappa+1}^{N-1}\frac{\delta x_\eta^\mu}{k\wc \delta x_\eta}\Big(e^{ik\wc x_{\eta+1}}-e^{ik\wc x_\eta}\Big)+\frac{g}{i}\frac{\delta x_\kappa^\mu}{k\wc \delta x_\kappa}\Big(e^{ik\wc x_{\kappa+1}}-e^{ik\wc x_\kappa}\Big)\nonumber\\
&+\frac{g}{i} \frac{\delta x_{\kappa-1}}{k\wc \delta x_{\kappa-1}}\Big(e^{ik\wc x_{\kappa}}-e^{ik\wc x_{\kappa-1}}\Big)+\frac{g}{i}\sum_{\eta=1}^{\kappa-2}\frac{\delta x_\eta^\mu}{k\wc \delta x_\eta}\Big(e^{ik\wc x_{\eta+1}}-e^{ik\wc x_\eta}\Big)\,.
\ea 
Noticing $\delta x_\kappa^\mu / k\wc \delta x_\kappa=\delta x^\mu_{\kappa-1}/k\wc \delta x_{\kappa-1}$ as $x_\kappa^\mu\to\infty$, and pairing the $\kappa+1$ and $\kappa-1$ terms left unpaired when $x_\kappa^\mu\to\infty$, this gives
\ba 
&\tilde{J}^\mu(k)
= \frac{g}{i}\sum_{\eta=\kappa+1}^{N-1}\frac{\delta x_\eta^\mu}{k\wc \delta x_\eta}\Big(e^{ik\wc x_{\eta+1}}-e^{ik\wc x_\eta}\Big)\nonumber\\
&+\frac{g}{i} \frac{\delta x_\kappa^\mu}{k\wc\delta x_\kappa}\Big(e^{ik\wc x_{\kappa+1}}-e^{ik\wc x_{\kappa-1}}\Big)+\frac{g}{i}\sum_{\eta=1}^{\kappa-2}\frac{\delta x_\eta^\mu}{k\wc \delta x_\eta}\Big(e^{ik\wc x_{\eta+1}}-e^{ik\wc x_\eta}\Big)\,,
\ea 
which has no $1/k$ terms when $k\to 0$.
Hence contributions within internal intervals of the currents can never introduce $1/k$ terms leading to IR divergences in $S^{ab}$, even if the particle trajectory propagates to infinity at intermediate times $\delta x_\kappa^\mu\to \infty$.

Since for virtual loop particles it is further verified that all points are internal ($x_{N}^\mu=x_1^\mu$), it follows from the previous observation that none of the $\ell$ virtual particles in the $\ell$-th order of the S-matrix loop expansion can introduce $1/k$ terms in $S^{ab}$ leading to IR divergences. We thus conclude for any virtual fermion $i$ participating in the process
\ba 
\label{finite-closed}
\lim_{k\to 0} \tilde{J}^\mu_i(k)= \text{const.}+\mathcal{O}(k),\,\,\,\,\,\,\,\, i=r+1,\ldots,r+\ell\,.
\ea 

The only remaining terms potentially introducing $1/k$ contributions in the $k\to 0$ limit to $S^{ab}$ are the elementary contributions in Eq.~\eqref{current_momentum_discretized} containing the external points $x^\mu_{N}=x^\mu_f$ and $x^\mu_1=x^\mu_i$ of the $r$ real particles. A worldline configuration for a real particle is shown in Fig.~\ref{fig:worldline_piezewise_3}.
The external points, unlike internal ones, appear only once in the sum in Eq.~\eqref{current_momentum_discretized} and are further placed at infinity by the boundary conditions in Eq.~\eqref{boundary_conditions_smatrix}. 
\begin{figure}[ht]
    \centering
    \includegraphics[scale=0.3]{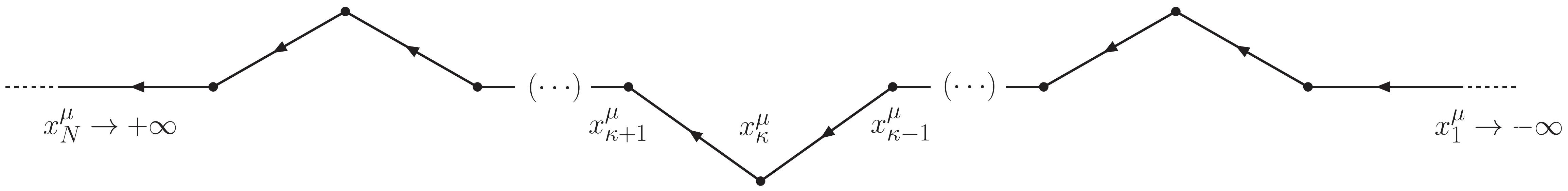}
    \caption{Discretized configuration of a bosonic worldline $x_\mu(\tau)$ with the external points placed at infinity. A dashed line is drawn to indicate that the fermion lines extend to infinity and become external as a result.}
    \label{fig:worldline_piezewise_3}
\end{figure}
Proceeding as previously, reorganizing the sum in Eq.~\eqref{current_momentum_discretized} we get
\ba 
&\tilde{J}^\mu(k)= \frac{g}{i}\frac{\delta x_{N-1}^\mu}{k\wc \delta x_{N-1}}\Big(e^{ik\wc x_N}-e^{ik\wc x_{N-1}}\Big)\label{realparticle_current_momentum_discretized}\\
&+\frac{g}{i}\sum_{\eta=2}^{N-2}\frac{\delta x_\eta^\mu}{k\wc \delta x_\eta}\Big(e^{ik\wc x_{\eta+1}}-e^{ik\wc x_\eta}\Big) +\frac{g}{i}\frac{\delta x_1^\mu}{k\wc \delta x_1} \Big(e^{ik\wc x_2}-e^{ik\wc x_1}\Big)\nonumber\,.
\ea 
In the $x_N^0\to\infty$ and $x_1^0\to\infty$ limits we then find
\ba 
\tilde{J}^\mu(k) = -\frac{g}{i}\frac{\delta x_{N-1}^\mu}{k\wc \delta x_{N-1}}e^{ik\wc x_{N-1}}+\frac{g}{i}\sum_{\eta=2}^{N-2}\frac{\delta x_\eta^\mu}{k\wc \delta x_\eta}\Big(e^{ik\wc x_{\eta+1}}-e^{ik\wc x_\eta}\Big) +\frac{g}{i}\frac{\delta x_1^\mu}{k\wc \delta x_1} e^{ik\wc x_2}\,.
\ea 
Then in the $k\to 0$ limit there are two non-vanishing $1/k$ contributions, encoding the attachment of an IR virtual photon line to the two external legs of the real particle,
\ba 
\lim_{k\to 0} \tilde{J}^\mu_n(k)= \frac{g}{i}\bigg\{\frac{\delta x_{1}^\mu}{k\wc \delta x_{1,n}}-\frac{\delta x_{N-1}^\mu}{k\wc \delta x_{N-1,n}} \bigg\}, \,\,\,\,\,\,\,\,\, n=1,\ldots,r\,,
\ea 
and where we neglected the phases in the $k\to0$ limit. Owing to the fact that the bosonic worldline currents are time reparametrization invariant, dividing and multiplying numerator and denominators above by the particle's proper time $s$ and mass $m$, respectively, produces
\ba
\lim_{x_{N}^0\to+\infty} m\frac{\delta{x}_{N-1}^\mu}{\delta s_{N-1}} = p_{f}^\mu, \,\,\,\, \lim_{x_{N}^0\to+\infty} m\frac{k\wc \delta{x}_{N-1}}{\delta s_{N-1}} = k\wc p_{f}^\mu\,,
\ea
and
\ba
\lim_{x_{1}^0\to-\infty} m\frac{\delta{x}_{1}^\mu}{\delta s_{1}} = p_i^\mu, \,\,\,\, \lim_{x_{1}^0\to-\infty} m\frac{k\wc \delta{x}_{1}}{\delta s_{1}} = k\wc p_i^\mu\,,
\ea
where $p_f^\mu$ and $p_i^\mu$ are the worldline momenta of the external legs. We can therefore rewrite
\ba 
\lim_{k\to 0} \tilde{J}^\mu_n(k)= \frac{g}{i}\bigg\{\frac{p_{i,n}^\mu}{k\wc p_{i,n}}-\frac{ p_{f,n}^\mu}{k\wc  p_{f,n} }\bigg\}, \,\,\,\,\,\,\,\,\, n=1,\ldots,r\,.\label{realparticle_current_momentum_discretized_2}
\ea 
Finally, the QED worldline interaction functional has the IR structure ($d=4$ and $\zeta=1$)
\ba 
&\sum_{a,b=1}^{r+\ell} S_{ab} = -\frac{g^2}{2}\sum_{n,m=1}^r\int_0^\Lambda\frac{d^4k}{(2\pi)^4} \frac{g_{\mu\nu}}{k^2+i\epsilon}\bigg\{\frac{p_{i,n}^\mu}{k\wc p_{i,n}}-\frac{p_{f,n}^\mu}{k\wc  p_{f,n}}\bigg\}\bigg\{\frac{p_{i,m}^\nu}{k\wc p_{i,m}}-\frac{ p_{f,m}^\nu}{k\wc  p_{f,m} }\bigg\}+\text{IR finite}\,,
\ea 
where the sum runs only over the real particles, contains only the external lines of these real particles, and $\Lambda$ denotes the IR region in which the previous soft approximations hold. The above statements are nothing but a worldline reformulation, to all orders in perturbation theory, of Low's soft theorem\footnote{Tests of Low's soft theorem by several experimental analyses to date show excellent agreement between Low's prediction and soft photon data in QED processes. On the other hand, a significant excess of soft photons is observed for a wide variety of processes involving hadron final states; these so-called ``anomalous soft photons", provide a phenomenological handle to study the interplay of QED and QCD sources of photon production. For a recent discussion, see \cite{Lebiedowicz:2021byo}.}\cite{Low:1958sn}.

We turn now to the cure of the IR problem stated above. In standard perturbation theory, the virtual IR divergences 
are cancelled in the cross-section by real IR photons attached to the squared S-matrix element. As we noted above, the origin of the virtual IR divergences can be directly traced back to 
dropping the asymptotic currents in Eq.~\eqref{realparticle_current_momentum_discretized}, when the $x_{N,n}^0=t_n^f\to\infty$ and $x_{1,n}^0=t_n^i\to-\infty$ limits are imposed by the standard S-matrix construction in Eq.~\eqref{boundary_conditions_smatrix}. However, as implicit in the discussion of Faddeev and Kulish \cite{Kulish:1970ut} (building on previous work by Dollard \cite{Dollard:1964}, Chung \cite{PhysRev.140.B1110} and Kibble \cite{Kibble:1968sfb,PhysRev.173.1527,PhysRev.174.1882,PhysRev.175.1624}), the order of limits matters. 

We can therefore, following Faddeev and Kulish, define as an alternative to the Dyson S-matrix, the equivalent worldline FK S-matrix:
\ba 
&\bar{\mathcal{S}}_{fi}^{(r)}= \frac{\mathrm{Z}_\text{MW}}{\mathrm{Z}} \prod_{n=1}^r \left\{\int d^3\v{x}_f^n  \int d^3\v{x}_i^n u_{s_f^n}^{\dag}(p_f^n)e^{+ip_f^n\wc x_f^n} \exp\bigg\{\bar{\gamma}_\lambda\frac{\partial}{\partial \theta_\lambda^n}\bigg\} \bar{\gamma}_0  u_{s_i^n}(p_i^n)e^{-ip_i^n\wc x_i^n} \right\}\nonumber\\
&\times\sum_{\ell=0}^\infty\frac{(-1)^\ell}{\ell!}\mathrm{W}^{(r,\ell)}(x_f^r,x_i^r,\theta^r,\ldots,x_f^1,x_i^1,\theta^1)\bigg|_{\theta=0}\medspace\medspace\medspace+\medspace\medspace\medspace \text{final state permutations}\,,\label{s_fi_fk}
\ea
with the limits $t_{f,i}^n\to \pm\infty$ taken \textit{only after} all the IR divergences of the 
diagrams generated by this new S-matrix are cancelled in the $k\to 0$ limit. $\bar{\mathcal{S}}_{fi}^{(r)}$ has the structure of a dressed 2r-point Green function with no truncated legs since time is kept finite until after $k\rightarrow 0$. It is therefore IR finite, by construction, to any order in perturbation theory. 
This is seen clearly by keeping the asymptotic charged currents that  consist of terms entering with phases $k\wc x_N$ and $k\wc x_1$ in Eq.~\eqref{realparticle_current_momentum_discretized},
\ba 
\tilde{J}^\mu_n(k)=\underbrace{\frac{g}{i}\frac{ p_{f,n}^\mu}{k\wc  p_{f,n} }e^{ik\wc x_N}}_{\text{Final asymptotic current}}-\underbrace{\frac{g}{i}\frac{ p_{f,n}^\mu}{k\wc  p_{f,n} }e^{ik\wc x_{N-1}} +(\cdots)+\frac{g}{i}\frac{p_{i,n}^\mu}{k\wc p_{i,n}}e^{ik\wc x_2}}_{\text{Current in the Dyson  S-matrix}}-\underbrace{\frac{g}{i}\frac{p_{i,n}^\mu}{k\wc p_{i,n}}e^{ik\wc x_1}}_{\text{Initial asymptotic current}}\,,
\ea 
instead of Eq.~\eqref{realparticle_current_momentum_discretized_2}. The three terms above are all of order $\mathcal{O}(1/k)$. However,  after taking the $k\to 0$ limit, they cancel exactly amongst each other and there are no $1/k$ contributions left in $S^{ab}$. We conclude therefore  that there are no IR divergences in the FK S-matrix $\bar{\mathcal{S}}_{fi}^{(r)}$, to all orders in perturbation theory.

\subsection{\label{section_5_2} An example: M\"oller scattering 
}
We will now discuss as a specific example of the previous general discussion, M\"oller scattering ($e^-e^-\to e^-e^-$), which is illustrated in Fig.~\ref{fig:2_2_scattering}. We will first analyze the IR divergences arising from the standard Dyson S-matrix, and subsequently contrast the resulting expressions with those of the FK S-Matrix.

\begin{figure}[ht]
    \centering
    \includegraphics[scale=0.3]{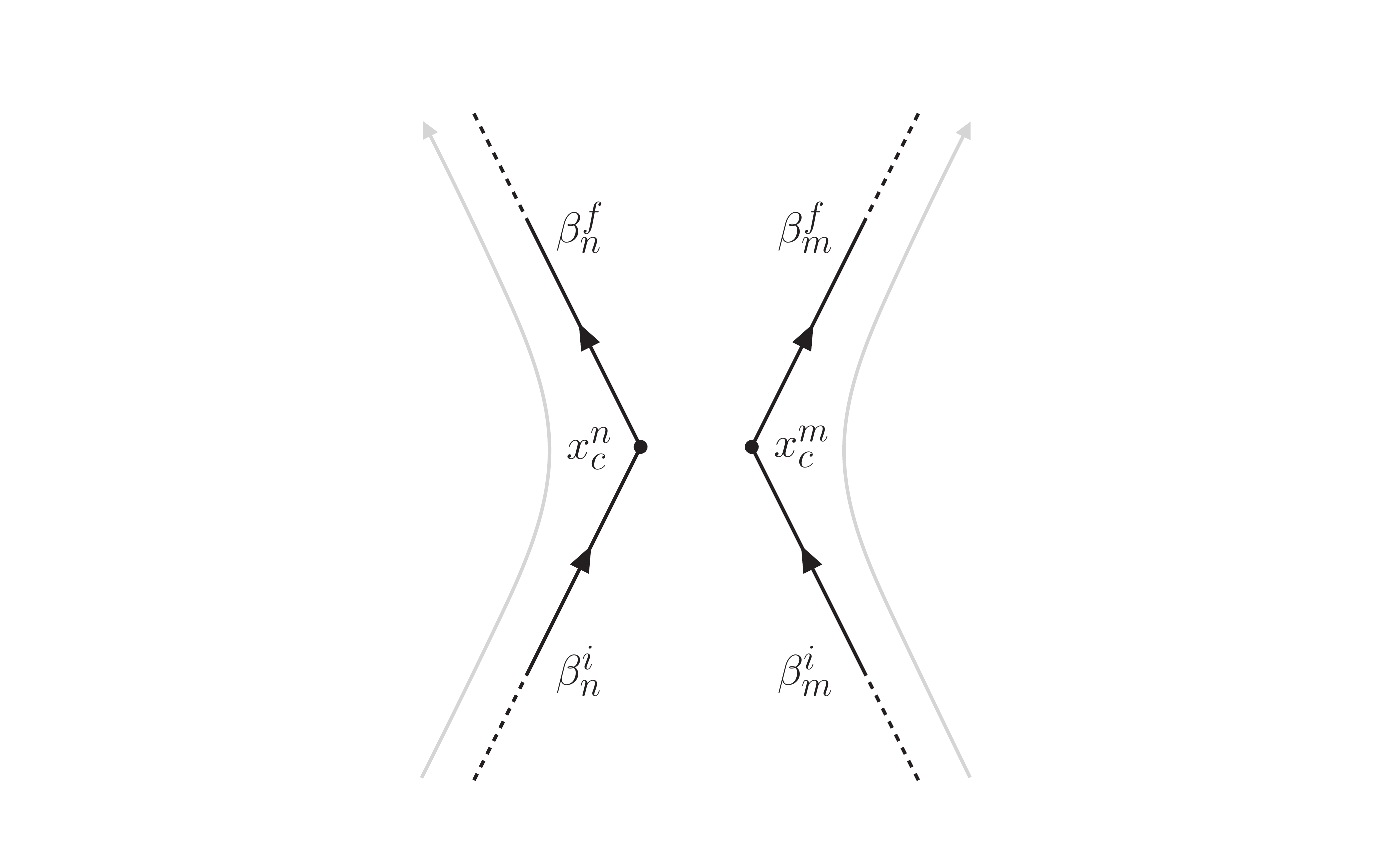}
    \caption{Worldlines of a pair of electrons $n,m=1,2$ with initial and final velocities $\beta_{i,f}^{1,2}$ without cusps, in grey, and two worldlines, in black, with cusps at $x_{n,m}^c$ and same initial and final velocities. In the soft photon limit they both have the same IR divergent contributions to the Dyson S-matrix.}
    \label{fig:2_2_scattering}
\end{figure}

We begin by writing the pair-wise worldline interactions as 
\ba
S^{ab}=S^{ab}_\text{IR}+S^{ab}_\text{UV}\,,
\ea 
where the soft and hard contributions are defined as
\ba 
S^{ab}_\text{IR}=-\frac{1}{2}\int_0^\Lambda\frac{d^4k}{(2\pi)^4}\tilde{J}^a_\mu(-k)\frac{g^{\mu\nu}}{k^2+i\epsilon}\tilde{J}^b_\nu(+k),\medspace\medspace\medspace 
S^{ab}_\text{UV}= -\frac{1}{2}\int_\Lambda^\infty\frac{d^4k}{(2\pi)^4}\tilde{J}^a_\mu(-k)\frac{g^{\mu\nu}}{k^2+i\epsilon}\tilde{J}^b_\nu(+k)\,,\label{S_nm_IR_def}
\ea 
where $\Lambda$ delimits the separation between the IR and UV regions.

We first consider the Dyson S-Matrix. $S^{ab}_\text{UV}$ is IR finite by construction. Following the previous discussion, the $1/k$ terms responsible for IR divergences in $S^{ab}_\text{IR}$ are introduced by the $t_{f}\to+\infty$ and $t_i\to-\infty$ legs of the scalar components of the worldline currents of the two real electrons. The virtual fermion contributions, as we discussed previously, are IR finite. Thus in the infrared, the sum over all real and virtual worldlines to any loop order $\ell$, can be approximated by a sum over  the real worldlines\footnote{We only consider here the leading $1/k$ contributions to the IR divergences. Keeping the virtual fermion loop contributions in the analysis that follows does not add further complications in the IR in this worldline picture. The underlying reason for this follows from the discussion around Eq.~\eqref{finite-closed} in Section \ref{section_5_1}. }: 
\ba 
\sum_{a,b=1}^{2+\ell}S^{ab}_\text{IR} \simeq  \sum_{n,m=1}^2 S^{nm}_\text{IR} \simeq  -\frac{1}{2}\sum_{n,m=1}^2 \int_0^\Lambda\frac{d^4k}{(2\pi)^4}\tilde{J}^n_{\mu,\text{IR}}(-k)\frac{g^{\mu\nu}}{k^2+i\epsilon}\tilde{J}^m_{\nu,\text{IR}}(+k)\,.
\ea 
Further, for sufficiently small $\Lambda$, virtual photons in 
$S^{nm}_\text{IR}$ attach to the asymptotic regions of the worldlines corresponding to $t_{f}\to+\infty$ and $t_i\to-\infty$. 
In these limits, the weakly interacting worldlines approach asymptotically the trajectories of free particles in $\tilde{J}^n_{\mu,\text{IR}}$ (or equivalently, $\tilde{J}^m_{\mu,\text{IR}}$), with initial and final velocities $\beta_i^n$ and $\beta_f^n$,
\ba 
&x_\mu^{n}(t) =x_{i,\mu}^{n}+\beta_{i,\mu}^{n}(t-t_i^{n})\,,\,\,\,\,\,\,\, t\in (t_i^{n},t^{n}_c)\,,\nonumber\\
&x_\mu^{n}(t) =x_{c,\mu}^{n}+\beta_{f,\mu}^{n}(t-t_c^{n})\,,\,\,\,\,\,\,\, t\in (t_c^{n},t^{n}_f)\,.\label{cusped_worldlines}
\ea
As shown in Fig.~\ref{fig:2_2_scattering}, any  arbitrary configuration of  worldlines at short distances, indicated by the cusps at $x_c^{n}$, are accounted for in  $S^{ab}_\text{UV}$. Using Eq.~\eqref{cusped_worldlines}, the  scalar currents in the IR region split into two parts,
\ba 
\tilde{J}^{n}_{\mu,\text{IR}}(k)= g\int^{t_c^{n}}_{t_i^{n}} dt\, \dot{x}_{\mu}^{n}(t)e^{+ik\wc x^{n}(t)-\epsilon (t^{n}_c-t)}+g\int^{t_f^{n}}_{t_c^{n}}dt\,\dot{x}_{\mu}^{n}(t)e^{+ik\wc x^{n}(t)-\epsilon(t-t_c^{n})}\,,
\ea 
where we introduced an $i\epsilon$ prescription to kill the incoming and outgoing currents at infinity. Performing the time integrals, and taking $t_i^{n}\to -\infty$,   $t_f^{n}\to+\infty$, this gives
\ba 
\tilde{J}^{n}_{\mu,\text{IR}}(k)= \frac{g}{i}\bigg\{\frac{\beta_{i,\mu}^{n}}{k\wc \beta_i^{n}-i\epsilon}-\frac{\beta_{f,\mu}^{n}}{k\wc \beta_f^{n}+i\epsilon}\bigg\}e^{+ik\wc x^n_c}\,.\label{j_n_ir}
\ea 
For sufficiently low $\Lambda$, the phase in the above expression, of order $1+\mathcal{O}(k)$, can be replaced by unity. The interactions in the IR region 
are then given by
\ba 
S^{nm}_\text{IR}\simeq\frac{1}{2}\int^\Lambda_0 \frac{d^4k}{(2\pi)^4}\frac{1}{k^2+i\epsilon}\bigg\{\frac{\beta_{i,\mu}^n}{-k\wc \beta_i^n-i\epsilon}-\frac{\beta_{f,\mu}^n}{-k\wc \beta_f^n+i\epsilon}\bigg\}\bigg\{\frac{\beta_{i,\mu}^m}{k\wc \beta_i^m-i\epsilon}-\frac{\beta_{f,\mu}^m}{k\wc \beta_f^m+i\epsilon}\bigg\}\,.\label{s_nm_IR}
\ea 
Because for weak long distance interactions (represented by the above expression in the asymptotics 
$\Lambda\rightarrow 0$) the  $n$ and $m$ worldline paths are well approximated by free  classical trajectories, they can be factorized out of the path integral in Eq.~\eqref{w_r0}. This gives  
\ba 
&\mathrm{W}^{(2,\ell)}(x_f^2,x_i^2,\theta^2,x_f^1,x_i^1,\theta^1)\simeq \exp\bigg\{-i\sum_{n,m=1}^2 S^{nm}_\text{IR}\bigg\}\bigg\langle  \exp\bigg\{-i\sum_{a,b=1}^{2+\ell}S^{ab}_\text{UV}\bigg\}\bigg\rangle\nonumber\\
&=\exp\bigg\{-i\sum_{n,m=1}^2 S^{nm}_\text{IR}\bigg\}\mathrm{W}^{(2,\ell)}_{\text{UV}}(x_f^2,x_i^2,\theta^2,x_f^1,x_i^1,\theta^1)\,,
\ea 
where we defined $\mathrm{W}^{(2,\ell)}_{\text{UV}}(\cdots)$ as the normalized worldline interaction functional of two real particles ($r=2$) with $\ell$ virtual fermions and an arbitrary number of virtual photons with energies above $\Lambda$.  Introducing this expression in the Dyson S-matrix in Eq.~\eqref{s_fi}, one obtains
\ba 
&\mathcal{S}_{fi}^{(2)}\simeq \exp\bigg\{ -i\sum_{n,m=1}^2 S^{nm}_\text{IR}\bigg\}\mathcal{S}_{fi,\text{UV}}^{(2)}\,,
\label{soft_hard_factorization}
\ea
Note that the S-matrix element for virtual photons with $k\geq\Lambda$ is IR finite by construction and given by
\ba 
&\mathcal{S}_{fi,\text{UV}}^{(2)}=\frac{\mathrm{Z}_\text{MW}}{\mathrm{Z}} \prod_{n=1}^2 \Bigg\{\lim_{\substack{t_{f}^{n}\to+\infty\\t_{i}^{n}\to-\infty}}\int d^3\v{x}_f^n e^{+ip_f^n\wc x_f^n} \int d^3\v{x}_i^ne^{-ip_i^n\wc x_i^n}  u_{s_f^n}^{\dag}(p_f^n)\exp\bigg\{\bar{\gamma}_\lambda\frac{\partial}{\partial \theta_\lambda^n}\bigg\} \bar{\gamma}_0  u_{s_i^n}(p_i^n) \Bigg\}\nonumber\\
&\times \sum_{\ell=0}^\infty\frac{(-1)^\ell}{\ell!}\mathrm{W}_\mathrm{UV}^{(2,\ell)}(x_f^2,x_i^2,\theta^2,x_f^1,x_i^1,\theta^1)\bigg|_{\theta=0}+(x_f^1\leftrightarrow x_f^2)\,.
\ea 

Eq. \eqref{soft_hard_factorization} corresponds to Weinberg's exponentiation of virtual Abelian IR divergences~\cite{PhysRev.140.B516}. 
It depends only on the directions of the charges at
infinity, is independent of their spin dimensions, and encodes their
interactions at late times with their self-generated classical gauge fields. This IR factor dresses the hard S-matrix $\mathcal{S}_{fi,\text{UV}}^{(2)}$ containing photons with $k\geq\Lambda$ outside the IR region. Note that this separation of scales is also assumed implicitly  in~\cite{PhysRev.140.B516}. 

\begin{figure}[ht]
  \centering
  \includegraphics[scale=0.3]{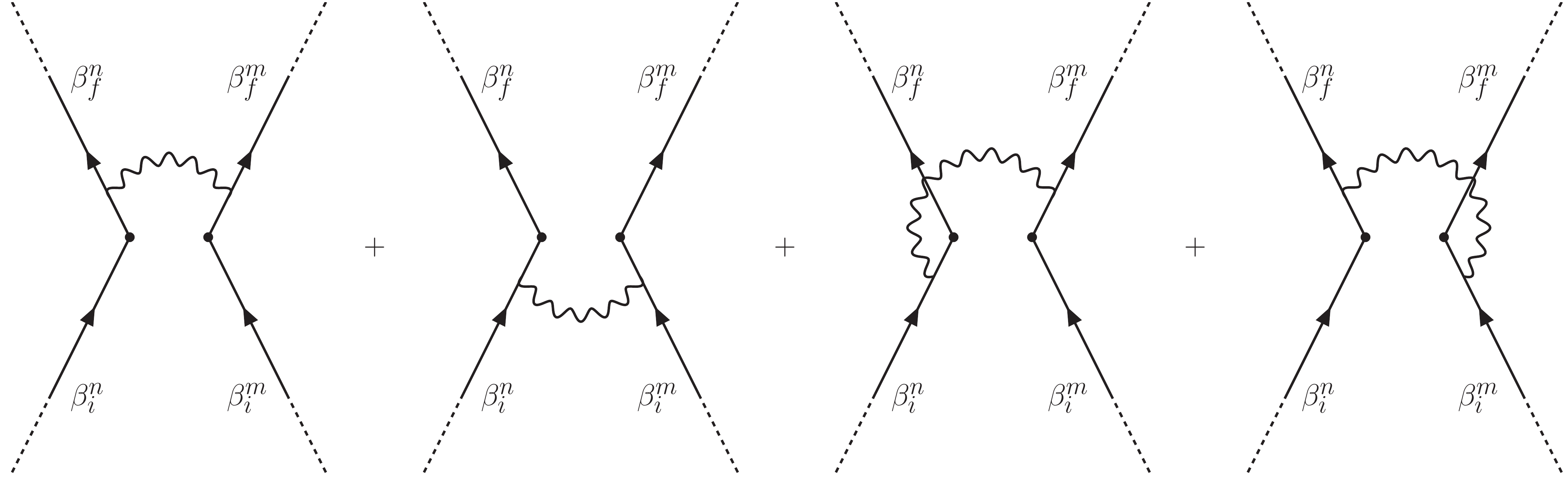}
  \caption{Diagrams 
    $S(\beta^n_f,x^n_c,\beta^m_f,x^m_c)$, $S(\beta^n_i,x^n_c,\beta^m_i,x^m_c)$, 
    $S(\beta^n_i,x^n_c,\beta^m_f,x^m_c)$, and $S(\beta^n_f,x^n_c,\beta^m_i,x^m_c)$ in
    Eq.~\eqref{s_nm_IR} representing all possible IR photon exchanges between two cusped worldlines.}
  \label{fig:s_nm_IR_diagrams}
\end{figure}

To evaluate the diagrams with these IR divergent dressings we keep the phases of the currents in Eq.~\eqref{j_n_ir}, depending on the cusp positions, and finally take the soft photon $k\to 0$ limits. The diagrams in Eq.~\eqref{s_nm_IR} are shown in Fig. \ref{fig:s_nm_IR_diagrams}
and denoted
\begin{align}
S^{nm}_{\text{IR}} =
S(\beta^n_i,x^n_c,\beta^m_i,x^m_c)+S(\beta^n_f,x^n_c,\beta^m_f,x^m_c)+S(\beta^n_i,x^n_c,\beta^m_f,x^m_c)+S(\beta^n_f,x^n_c,\beta^m_i,x^m_c)\,.\label{s_nm_ir_diagrams}
\end{align}
where
\ba 
S(\beta^n,x^n,\beta^m,x^m) = \frac{g^2}{2} \int \frac{d^4k}{(2\pi)^4}\,\frac{g_{\mu\nu}}{k^2+i\epsilon}\,\, \frac{\eta^n \beta_\mu^n}{-k\wc \beta^n-i\eta^n\epsilon}\,\,\frac{\eta^m\beta_\nu^m}{k\wc\beta^m-i\eta^m\epsilon}\,e^{-ik\wc (x^n-x^m)}\,,\label{s_betan_xn_betam_xm}
\ea 
and $\eta^n=+1$ or $\eta^n=-1$ for a charged current incoming to or outgoing from the point $x^n$.

The diagonal terms $n=m$ correspond to IR self-energy dressings of the S-matrix occurring at asymptotic times. The non-diagonal $n\neq m$, to virtual IR photons exchanged by the two electrons.
Consider the initial-initial photon exchange diagram, for which $\eta^n=+1$ and $\eta^m=+1$, and given  by
\begin{align}
S(\beta^n_i,x^n_c,\beta^m_i,x^m_c) = \frac{g^2}{2}\beta^n_i\wc\beta^m_i \int \frac{d^4k}{(2\pi)^4}\frac{1}{k^2+i\epsilon}\,\,\frac{1}{-k\wc\beta^n_i-i\epsilon}\,\,\frac{1}{k\wc\beta^m_i-i\epsilon}e^{-ik\wc(x^n_c-x^m_c)}\,.
\end{align}
\begin{figure}[ht]
    \centering
    \includegraphics[scale=0.4]{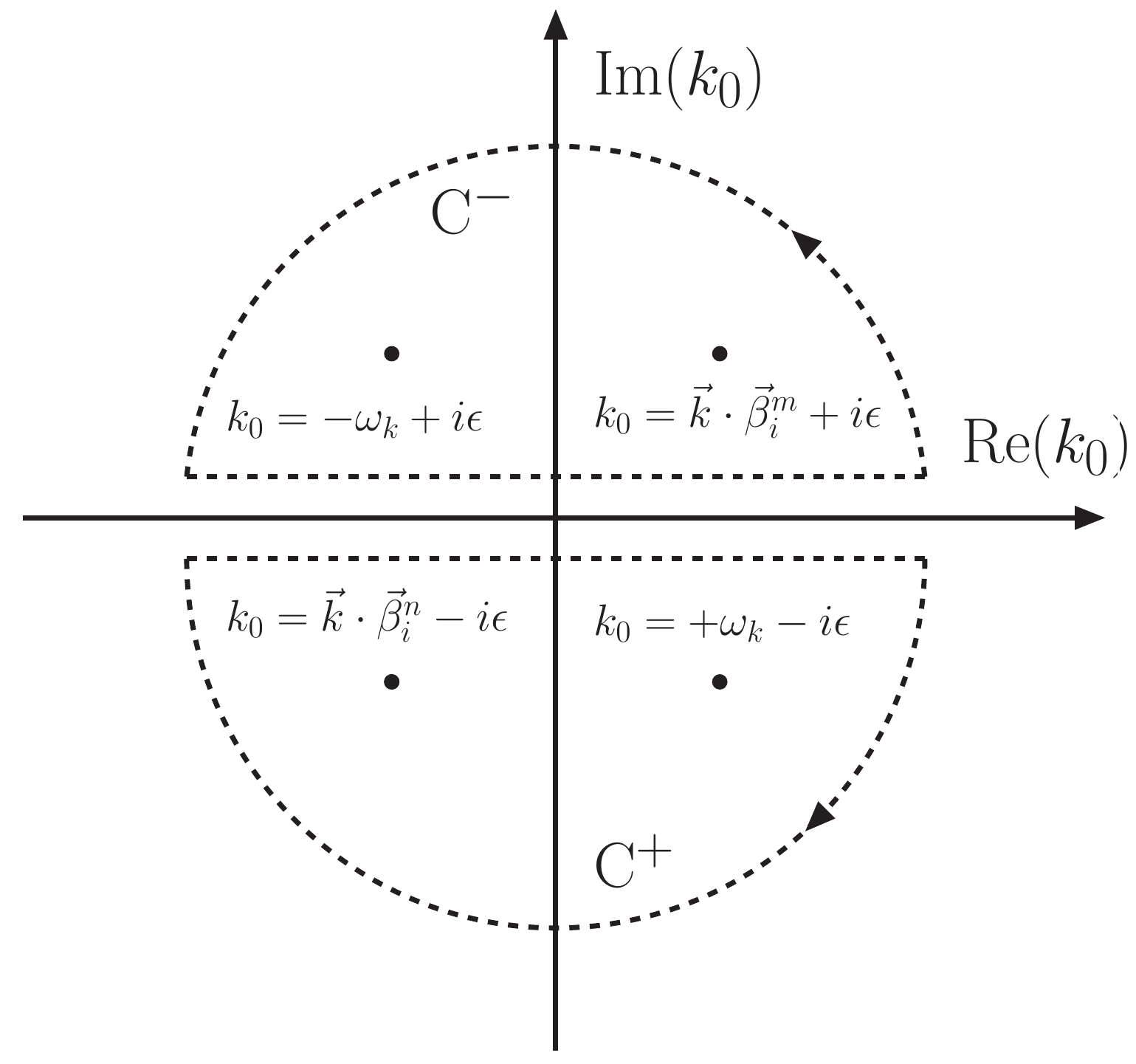}
    \caption{Poles and integration contour in the self-energy virtual photon graph attached to the initial leg, corresponding to one of the contributions in $S^{nn}_c$. }
    \label{fig:poles}
\end{figure}

Defining $\omega_k^2=\v{k}^2$, the integration in the virtual photon energy $k_0$,  as shown in Fig.~\ref{fig:poles}, has poles
at
\begin{align}
k_0=\omega_k-i\epsilon,  \medspace\medspace\medspace\medspace\medspace\medspace k_0=-\omega_k+i\epsilon, \medspace\medspace\medspace\medspace\medspace\medspace k_0=\v{k}\wc\v{\beta}^n_i-i\epsilon, \medspace\medspace\medspace\medspace\medspace\medspace k_0=\v{k}\wc\v{\beta}^m_i+i\epsilon\,.
\end{align}
The integration contour is determined by the sign of $t^n_c-t^m_c$. For $t^n_c>t^m_c$, the clockwise contour $\mathrm{C}^+$ in Fig.~\ref{fig:poles} should be chosen. The residue of the pole at $k_0=+\omega_k-i\epsilon$ encodes the interaction of the electron $n$ in the radiative mode of the gauge field of the electron $m$, created by the overall change in the \textit{in} and \textit{out} directions. The residue of the pole at $k_0=\v{k}\v{\beta}^n_i-i\epsilon$ corresponds to the Coulomb mode in the interaction of one electron in the Li\'{e}nard-Wiechert field of the other. This gives
\begin{align}
  &-iS(\beta^n_i,x^n_c,\beta^m_i,x^m_c)=\frac{g^2}{2}\int_\lambda^\Lambda \frac{d^3\v{k}}{(2\pi)^3}\frac{1}{2\omega_k} \frac{\beta^n_i}{k\wc\beta^n_i} \wc \frac{\beta^m_i}{k\wc\beta^m_i}e^{-ik\wc (x^n(t^n_c)-x^m(t^m_c))}\nonumber\\
  &+ig^2\lim_{\epsilon\to 0}\text{Im}\int_\lambda^\Lambda \frac{d^3\v{k}}{(2\pi)^3}\frac{1}{2\omega_k}\frac{\beta^n_i}{k\wc\beta^n_i}\wc \frac{\beta^m_i}{k\wc (\beta^m_i-\beta^n_i)+i\epsilon}e^{+ik\wc(x^n(t^m_c)-x^m(t^m_c))}\,,
  \label{s_p_p}
\end{align}
where we introduced an IR cut-off $\lambda$ to capture the logarithmic IR divergence in the resulting integrals.

For $t^n_c<t^m_c$, the anti-clockwise integration contour $\mathrm{C}^-$ should be chosen instead, with the residues corresponding now to the response of the electron $m$ in the radiative and Coulomb fields of $n$. In the IR $k\to 0$ limit, the long distance interactions accounted for in $S^{nm}_{\text{IR}}$ are however independent of the relative placement of the cusps of Fig.~\ref{fig:2_2_scattering}, and more generally, of the particular short distance configurations of the actual electron worldlines contributing to the path integral.  

Following the same procedure for the rest of diagrams, it can be shown that the IR divergent, long distance interactions between any two particles $n$ and $m$ in $S^{nm}_\text{IR}$ can be always split, following Weinberg \cite{PhysRev.140.B516} and Faddeev and Kulish \cite{Kulish:1970ut}, into a real and an imaginary contribution
\ba 
-iS^{nm}_\text{IR}=\mathrm{R}^{nm}+i\Phi^{nm}\,,
\ea 
where
\ba
 \mathrm{R}^{nm}=\frac{g^2}{2}\int_\lambda^\Lambda \frac{d^3\v{k}}{(2\pi)^3}\frac{1}{2\omega_k}\Bigg(\frac{\beta^m_f}{k\wc \beta^m_f}-\frac{\beta^m_i}{k\wc\beta^m_i}\Bigg)\wc\Bigg(\frac{\beta^n_f}{k\wc \beta^n_f}-\frac{\beta^n_i}{k\wc\beta^n_i}\Bigg)\,,
 \label{Rnm}
\ea  
and
\ba
  &\Phi^{nm}= \lim_{\epsilon\to 0}\bigg\{g^2\text{Im}\int_\lambda^\Lambda\frac{d^3\v{k}}{(2\pi)^3}\frac{1}{2\omega_k}\frac{\beta^m_f}{k\wc\beta^m_f}\wc \frac{\beta^n_f}{k\wc(\beta^n_f-\beta^m_f)+i\epsilon}\label{phi_nm}\\&+g^2\text{Im}\int_\lambda^\Lambda \frac{d^3\v{k}}{(2\pi)^3}\frac{1}{2\omega_k}\frac{\beta^n_i}{k\wc\beta^n_i}\wc \frac{\beta^m_i}{k\wc(\beta^m_i-\beta^n_i)+i\epsilon}\bigg\}\nonumber\,.
\ea
Performing then the remaining integrals, using  Eqs.~\eqref{I_bp_bq_result} and \eqref{J_bp_bq_result} of Appendix \ref{appendix_e}, this gives
\begin{align}
&-i\sum_{n,m=1}^2 S^{nm}_\text{IR}=\frac{g^2}{8\pi^2} \sum_{n',m'}\eta^{n'}\eta^{m'} \gamma^{n'm'}\coth\gamma^{n'm'}\log\frac{\Lambda}{\lambda}-i\frac{g^2}{8\pi}\sum_{n',m'}{}^{'} \eta^{n'}\eta^{m'} \coth\gamma^{n'm'}\log\frac{\Lambda}{\lambda}\,,\label{S_I_cusped_worldlines_result}
\end{align}
with the cusp angles between external legs of the charged-particle worldlines defined by
\begin{align}
\cosh \gamma^{nm}:= \frac{\beta^n\wc\beta^m}{\sqrt{(\beta^n)^2(\beta^m)^2}}\,. \label{relative_angle}
\end{align}
Recall further that $\eta=+1$ and $\eta=-1$ for an incoming and outgoing external charged-particle leg. The indices $n',m'$ in the r.h.s. indicate an \textit{in} or an \textit{out} leg of the real worldlines, and the pairs $(n',m')$ and $(m',n')$ are counted separately.
The prime in the imaginary part sum indicates that the sum runs only over the \textit{out}-\textit{out} and \textit{in}-\textit{in} pairings \cite{PhysRev.140.B516} since the Li\'{e}nard-Wiechert or Coulomb modes in all crossed diagrams, namely $S(\beta^n_i,x^n_c,\beta^m_f,x^m_c)$ and $S(\beta^n_f,x^n_c,\beta^m_i,x^m_c)$, vanish in the soft limit. 

We turn now to analyze the IR behavior of the FK S-matrix $\bar{\mathcal{S}}_{fi}^{(r)}$ in Eq.~\eqref{s_fi_fk}. The currents at long distances are well approximated again by the cusped worldlines of Fig.~\eqref{fig:2_2_scattering}. Using Eq.~\eqref{cusped_worldlines} and keeping $t_{f,i}^{n}$ finite, the currents give
\ba 
\tilde{J}^{n}_{\mu,\text{IR}}(k) = \underbrace{\frac{g}{i}\frac{\beta_{f,\mu}^{n}}{k\wc \beta_f^{n}+i\epsilon}e^{ik\wc x_f^{n}}}_\text{final asymptotic current}-\underbrace{\frac{g}{i}\bigg(\frac{p_{f,\mu}^{n}}{k\wc \beta_f^{n}+i\epsilon}-\frac{\beta^{n}_{i,\mu}}{k\wc \beta_i^{n}-i\epsilon}\bigg)e^{ik\wc x_c^{n}}}_\text{current at the cusp}-\underbrace{\frac{g}{i}\frac{\beta_{i,\mu}^{n}}{k\wc \beta_i^{n}-i\epsilon}e^{ik\wc x_i^{n}}}_\text{initial asymptotic current}\,.
\label{j_n_ir+as}
\ea

As observed previously in Section \ref{section_5_1}, in the standard approach~\cite{PhysRev.140.B516}, the first and the last terms in Eq.~\eqref{j_n_ir+as} are dropped due to the rapid oscillations in the phases occurring at asymptotic times. This would give back Eq.~\eqref{j_n_ir}. 
As implicit in Faddeev and Kulish's discussion on infrared safety \cite{Kulish:1970ut} the order of limits matters. For any fixed large 
$x^{f,i}$, the phases go to unity as $k\rightarrow 0$, and the asymptotic contributions cancel with that from the cusps in this limit. 

In the standard construction, these new terms correspond to the $(p\pm k)^2=m^2$ poles not accounted for by the LSZ truncation of the amplitude, order-by-order in perturbation theory. The asymptotic currents encode virtual photons pinched to the $x_f^\mu$ and $x_i^\mu$ points, and they can be regarded then as IR dressings of the \textit{in} and \textit{out} states in the gauge field to all orders in perturbation theory.

With the aid of the asymptotic currents in Eq.~\eqref{j_n_ir+as}, the new Lorentz-covariant worldline interaction functional includes photon exchanges at infinity that capture and remove IR singularities to any order in perturbation theory. This can be confirmed explicitly by plugging Eq.~\eqref{j_n_ir+as} into Eq.~\eqref{S_nm_IR_def}, and finding it to be free of IR singularities in $d=4$. Indeed, in the FK S-matrix $\bar{\mathcal{S}}_{fi}^{(r)}$, the IR dressings in Eq.~\eqref{s_nm_IR} become
\ba
\bar{S}_\text{IR}^{nm}= S^{nm}_\text{IR}+S^{nm}_{as}\label{s_nm_ir+as}\,,
\ea
where $S^{nm}_\text{IR}$ are the standard diagrams
in Eq.~\eqref{s_nm_ir_diagrams} and the new asymptotic diagrams $S^{nm}_{as}$ include a virtual photon attached in at least one of its ends to one of the asymptotic boundaries ($t_i\to-\infty$ and $t_f\to+\infty$)
\ba
&S_{as}^{nm}=S(\beta_n^f,x_n^f,\beta_m^f,x_m^f)+S(\beta_n^f,x_n^f,\beta_m^f,x_m^c)+S(\beta_n^f,x_n^f,\beta_m^i,x_m^c)+S(\beta_n^f,x_n^f,\beta_m^i,x_m^i)\nonumber\\
&+S(\beta_n^f,x_n^c,\beta_m^f,x_m^f)+S(\beta_n^i,x_n^c,\beta_m^f,x_m^f)+S(\beta_n^f,x_n^c,\beta_m^i,x_m^i)+S(\beta_n^i,x_n^c,\beta_m^i,x_m^i)\nonumber\\
&+S(\beta_n^i,x_n^i,\beta_m^f,x_m^f)+S(\beta_n^i,x_n^i,\beta_m^f,x_m^c)+S(\beta_n^i,x_n^i,\beta_m^i,x_m^c)+S(\beta_n^i,x_n^i,\beta_m^i,x_m^i)\,,\label{s_nm_as}
\ea
where each diagram $S(\beta_n,x_n,\beta_m,x_m)$ is also given 
 by Eq.~\eqref{s_betan_xn_betam_xm}. A subset of these asymptotic diagrams are displayed in Figs.~\ref{fig:asymptotic_diagrams_1} and \ref{fig:asymptotic_diagrams_2}.

\begin{figure}[ht]
  \centering
  \includegraphics[scale=0.3]{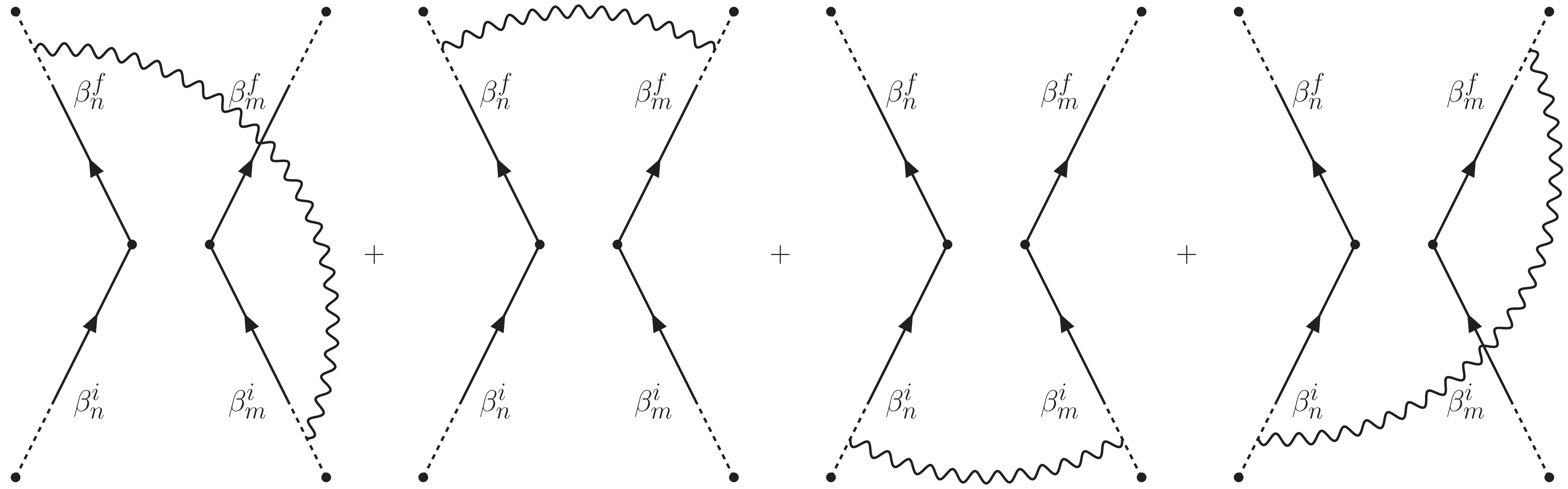}
  \caption{Subset of asymptotic diagrams in $S^{nm}_{as}$ of Eq.~\eqref{s_nm_as} corresponding to pure asymptotic interactions. The dots indicate the asymptotic points $x_f^{n,m}$ and $x_i^{n,m}$ within the charged currents. Only the \textit{out-out} and \textit{in-in} exchanges (represented by the middle two figures) contain imaginary Coulomb modes that cancel the corresponding imaginary IR divergences in $S^{nm}_\text{IR}$.}
    \label{fig:asymptotic_diagrams_1}
\end{figure}
\begin{figure}[ht]
  \centering
  \includegraphics[scale=0.3]{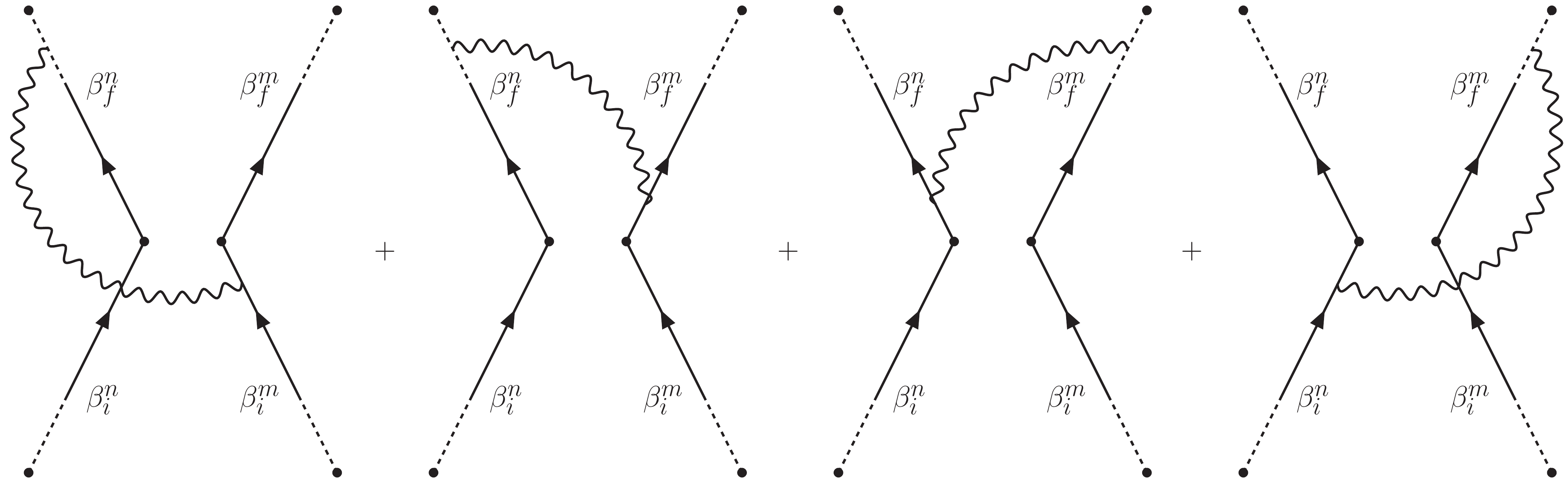}
  \caption{Subset of asymptotic diagrams in $S^{nm}_\text{as}$ in Eq.~\eqref{s_nm_as} representing interactions between the cusped and  asymptotic regions.}
  \label{fig:asymptotic_diagrams_2}
\end{figure}

The evaluation of the asymptotic diagrams, follows the same 
procedure as for the diagrams  discussed previously, and  also lead to real and imaginary contributions from the asymptotic regions, 
\ba 
-iS^{nm}_{as}= \mathrm{R}^{nm}_{as}+i\Phi^{nm}_{as}
\label{real-plus-imaginary-phases}
\ea 
with
\begin{align}
  &\mathrm{R}^{nm}_{as}=\frac{1}{2}g^2\int_\lambda^{\Lambda'}\frac{d^3\v{k}}{(2\pi)^3}\frac{1}{2\omega_k}\frac{\beta^n_f}{k\wc\beta^n_f}\wc \frac{\beta^m_i}{k\wc \beta^m_i}+\frac{1}{2}g^2\int_\lambda^{\Lambda'}\frac{d^3\v{k}}{(2\pi)^3}\frac{1}{2\omega_k}\frac{\beta^n_i}{k\wc\beta^n_i}\wc \frac{\beta^m_f}{k\wc \beta^m_f}\label{r_nm_asymptotic}\\
  &-\frac{1}{2}g^2\int_\lambda^{\Lambda'}\frac{d^3\v{k}}{(2\pi)^3}\frac{1}{2\omega_k}\frac{\beta^n_f}{k\wc\beta^n_f}\wc \frac{\beta^m_f}{k\wc \beta^m_f}-\frac{1}{2}g^2\int_\lambda^{\Lambda'}\frac{d^3\v{k}}{(2\pi)^3}\frac{1}{2\omega_k}\frac{\beta^n_i}{k\wc\beta^n_i}\wc \frac{\beta^m_i}{k\wc \beta^m_i}\,,\nonumber
\end{align}
where $\Lambda'$ denotes the IR region of the asymptotic interactions, and
\begin{align}
  &\Phi^{nm}_{as}=-\lim_{\epsilon\to 0}\bigg\{g^2\int^{\Lambda'}_{\lambda}\frac{d^3\v{k}}{(2\pi)^3}\frac{1}{2\omega_k}\frac{\beta^m_f}{k\wc\beta^m_f}\wc\frac{\beta^n_f}{k\wc (\beta^m_f-\beta^n_f)+i\epsilon}\nonumber\\ &+g^2\Im\int^{\Lambda'}_{\lambda}\frac{d^3\v{k}}{(2\pi)^3}\frac{1}{2\omega_k}\frac{\beta^n_i}{k\wc\beta^n_i}\wc\frac{\beta^m_i}{k\wc(\beta^m_i-\beta^n_i)+i\epsilon}\bigg\}\,. \label{phi_nm_asymptotic}
\end{align}

Performing the remaining integrals using Eqs.~\eqref{I_bp_bq_result} and~\eqref{J_bp_bq_result} in Appendix \ref{appendix_e}, the result of adding the diagrams of the asymptotic region $S^{nm}_{as}$ to the conventional IR divergent diagrams in $S^{nm}_\text{IR}$ is
\ba
-i\sum_{n,m=1}^2\bar{S}^{nm}_\text{IR}=\frac{g^2}{8\pi^2}\sum_{n'm'} \eta^{n'}\eta^{m'} \gamma^{n'm'}\coth \gamma^{n'm'}\log \frac{\Lambda}{\Lambda'} -i\frac{g^2}{8\pi}\sum_{n'm'}' \eta^{n'}\eta^{m'} \coth \gamma^{n'm'}\log\frac{\Lambda}{\Lambda'}\,.\label{cancellation_IR_divergences_allorders}
\ea
which is independent of the IR cut-off $\lambda$. Our result in Eq.~\eqref{cancellation_IR_divergences_allorders} therefore proves that the FK S-matrix $\bar{\mathcal{S}}_{fi}^{(r)}$ in QED is IR finite to all orders in perturbation theory. 

A few remarks on the cancellation of the IR divergences in Eq.~\eqref{cancellation_IR_divergences_allorders} are now in order.
Firstly, because the cancellation of these IR divergences in the conventional Dyson S-matrix approach occurs between virtual and real photon IR divergences in the cross-section, one might ask if virtual IR photon diagrams becoming IR finite when asymptotic interactions are included  means that real IR divergences they canceled previously now survive in the cross-section. This is fortunately not the case because diagrams with the emission of real photons are also accompanied by corresponding asymptotic diagrams, which similarly lead to IR finite cross-sections. Real photon emissions will be considered in detail in Paper II.

Secondly, in the Dyson S-matrix the imaginary Coulomb phase $i\Phi$ 
exists at the amplitude level, and becomes manifest in the calculation of the Dyson S-matrix element itself starting at next-to-leading order in the coupling $g$. It should therefore not be ignored in the context of our previous discussion of the IR finite FK S-matrix\footnote{Note that that this phase can also be recovered in the eikonal expansion of the S-matrix~\cite{Abarbanel:1969ek}. The resummation of ladder diagrams \cite{Cheng:1969eh} through Wilson lines, and the resummation of the resulting eikonal series thereafter for arbitrary momentum transfers away from zero, are of particular importance in the study of the Regge asymptotic behavior of QCD~\cite{Korchemskaya:1992je,Korchemskaya:1994qp} and gravity~\cite{Melville:2013qca}.}.

Further, we have included the $n=m$ terms in both in the conventional IR divergent Dyson and IR finite FK S-matrices. These Coulomb IR divergences in external lines (wherein the photon is attached to the same external charged-particle worldline) are rarely discussed because it is presumed that they are included in the counterterms corresponding to wavefunction renormalization of the external currents. However as noted by Weinberg~\cite{Weinberg:1995mt}, this is not the case because the presumed counterterms are themselves cancelled by counterterms arising from the internal lines and vertices. Therefore these $n=m$ IR Coulomb terms that are manifest in the worldline formalism should be kept. In the conventional approach, their IR divergences will cancel with the IR divergences of the real photons attached to the cross-section; in the FK approach, their IR divergences are removed by the corresponding counterterms, as is  manifest in the result obtained in  Eq.~\eqref{cancellation_IR_divergences_allorders}.

Finally, as also previously observed in \cite{Korchemsky:1987wg}, there is a subtle point regarding the expressions surviving the cancellation of the IR divergences in Eq.~\eqref{cancellation_IR_divergences_allorders}. 
The Coulomb phase is ill-defined for the cases in which $\gamma_{n'm'}=0$. For the case of the long distance interactions between different charged particles ($n'\neq m'$), this would correspond to the scenario in which the 
relative velocity of a charged-particle in the rest frame of the another, defined by the invariant relative rapidity
$v(\beta_{n'},\beta_{m'})=\sqrt{1-\beta_{n'}^2\beta_{m'}^2/(\beta_{n'}\wc\beta_{m'})^2}$, vanishes in the \textit{in} and/or \textit{out} state, reflecting a singularity of the scattering amplitude. For  photons attached to the same external line of a charged-particle, it is always true that $\gamma^{n'n'}=0$. This 
divergence corresponds to the interaction of a charge in uniform motion in its own Li\'enard-Wiechert field. In the Lorentz frame of the particle at rest, this is just the  divergence of a Coulomb field at short distances and will not impact scattering matrices at any finite relative velocity.

\subsection{\label{section_5_3}
Relating asymptotic worldlines to Faddeev-Kulish soft factors}
We will now relate our construction of IR finite amplitudes in the worldline formalism to the Faddeev-Kulish S-matrix in their original Hamiltonian approach \cite{Kulish:1970ut},  and to the recent discussion in the modern language of Wilson lines by Schwartz and Hannesdottir \cite{Hannesdottir:2019opa}. The fundamental issue noted in  \cite{Kulish:1970ut} is that to construct an IR finite S-matrix
operator, the evolution of the states in the far past and future have to account for contributions from long-range interactions that are neglected in the Dyson S-matrix. 

We begin with the discussion in \cite{Hannesdottir:2019opa}, where 
\begin{eqnarray}
&&\lim_{t_f\to+\infty}\mathrm{U}(t_f,t)|\Psi\rangle =\lim_{t_f\to+\infty} \mathrm{U}_{as}(t_f,t)|\Psi_{out}\rangle\,,\nonumber\\  
&&\lim_{t_i\to-\infty}\mathrm{U}(t_i,t)|\Psi\rangle = \lim_{t_i\to-\infty} \mathrm{U}_{as}(t_i,t)|\Psi_{in}\rangle\,,
\end{eqnarray}
where $i\partial_t \mathrm{U}_{as}(t,t') = \mathrm{H}_{as}(t)\mathrm{U}_{as}(t,t')$ and $\mathrm{H}_{as}(t)$ is the QED Hamiltonian in the asymptotic soft photon limit. In the traditional construction of the Dyson S-matrix, it is assumed that this asymptotic dynamics is well approximated by $\mathrm{U}_0(t,t')$, with $i\partial_t \mathrm{U}_0(t,t')=\mathrm{H}_0(t)\mathrm{U}_0(t,t')$, with  $\mathrm{H}_0(t)$ being the free QED Hamiltonian. The previous equation can be rewritten as 
\ba 
&|\Psi\rangle =\lim_{t_f\to+\infty} \mathrm{U}^\dag(t_f,t)\mathrm{U}_{as}(t_f,t)|\Psi_{out}\rangle ,\,\,\,\,\,\, |\Psi\rangle  =\lim_{t_i\to-\infty} \mathrm{U}^\dag (t_i,t)\mathrm{U}_{as}(t_i,t)|\Psi_{in}\rangle\,.
\ea 
The operator relating the \textit{out} and \textit{in} states defines the FK S-matrix: 
\ba 
|\Psi_{out}\rangle = \lim_{t_{f,i}\to\pm\infty} \mathrm{U}_{as}^\dag(t_f,t)\mathrm{U}(t_f,t_i)\mathrm{U}_{as}(t_i,t)|\Psi_{in}\rangle \equiv \bar{\mathcal{S}}|\Psi_{in}\rangle ,
\ea 
where the relations $\mathrm{U}^\dag(t_i,t)=\mathrm{U}(t,t_i)$ and $\mathrm{U}(t_f,t)\mathrm{U}(t,t_i)=\mathrm{U}(t_f,t_i)$ were employed.

Equivalently, for the Dyson S-matrix, one must replace above  $\mathrm{U}_{as}(t,t')$ with $\mathrm{U}_{0}(t,t')$:
\ba 
\mathcal{S}= \lim_{t_{f,i}\to\pm\infty} \mathrm{U}_{0}^\dag(t_f,t)\mathrm{U}(t_f,t_i)\mathrm{U}_{0}(t_i,t)
\,.
\ea 
Thus one finds the former can be expressed in terms of the latter as 
\ba 
\bar{\mathcal{S}}= \lim_{t_{f,i}\to\pm\infty} \mathrm{U}_{as}^\dag(t_f,t)\mathrm{U}_0(t_f,t)\mathcal{S} \mathrm{U}_0^\dag(t_i,t)\mathrm{U}_{as}(t_i,t)\,.
\label{S_Bar_S_Dyson}
\ea 

In the FK S-matrix (referred to as the ``Hard" S-matrix
in \cite{Hannesdottir:2019opa}), an \textit{in} state is dressed
by asymptotic interactions from $t$ to $t_i=-\infty$, subsequently 
evolves and scatters in the central region from $t_i=-\infty$ to $t$ and from $t$ to $t_f=+\infty$ as per the Dyson prescription, and returns to $t$ dressed analogously by final asymptotic 
interactions, with  $t=0$ chosen in \cite{Hannesdottir:2019opa}. 
Feynman rules for the leading IR divergent diagrams including contributions from the asymptotic regions, are shown explicitly to  
cancel with the IR divergences generated by the Feynman graphs corresponding to the Dyson S-matrix in the central region.

In the original  FK S-matrix discussion in \cite{Kulish:1970ut}, the asymptotic amplitudes likewise enter as conjugated to the ones in the central region, but with phases at $t_f\to +\infty$ and $t_i\to -\infty$ instead of $t=0$ in \cite{Hannesdottir:2019opa}. Since the role of the asymptotic interactions is to transform a free state at $t=0$ to a dressed \textit{in} and \textit{out} states at $t_f\to +\infty$ and $t_i\to -\infty$, respectively,  the matrix elements of the Hard S-matrix  \cite{Hannesdottir:2019opa} are  equivalent to the FK form in \cite{Kulish:1970ut}. 

The expression in Eq.~\eqref{S_Bar_S_Dyson} was further reexpressed by Faddeev and Kulish as 
\ba 
\bar{\mathcal{S}}=\exp\Big\{-\mathrm{R}(t_f)-i\Phi(t_f)\Big\}\mathcal{S}\exp\Big\{+i\Phi(t_i)+\mathrm{R}(t_i)\Big\}\,,\label{S_Bar_S_Dyson_2}
\ea 
where the real and imaginary parts of the IR divergences of the diagrams from the asymptotic regions are encoded in the action of the operators 
\ba 
\mathrm{R}(t)= e\sum_{\lambda} \int\frac{d^3\v{p}}{(2\pi)^3}\rho_p\int\frac{d^3\v{k}}{(2\pi)^3}\frac{1}{2\omega_k}\frac{p^\mu}{k\wc p}\bigg\{a^{\lambda,\dag}_k\epsilon^{\lambda,*}_{k,\mu}e^{ik\wc pt/p^0}-a_k^\lambda \epsilon^\lambda_{k,\mu}e^{-i k\wc p t/p^0}\bigg\}\,,
\ea 
and
\ba 
\Phi(t)=\frac{e^2}{8\pi^2}\int\frac{d^3\v{p}_1}{(2\pi)^3}\int\frac{d^3\v{p}_2}{(2\pi)^3}:\rho_{p_1}\rho_{p_2}:\frac{p_1\wc p_2}{\sqrt{(p_1\wc p_2)^2-m^4}}\ln |t|\,,
\ea 
on the \textit{in} and \textit{out} states.
Here $\rho_p=\sum_s(b_p^{s,\dag}b_p^s-d_p^{s,\dag}d_p^s)$ is the fermion number operator 
of the charges in the \textit{in} and \textit{out} states, $b_p^s$ ($d_p^s$) and $b_p^{s,\dag}$ ($d_p^{s,\dag}$), annihilation and creation operators respectively, of a fermion (antifermion) with  momentum $p$ and spin $s$,  $a_k^\lambda$ and $a_k^{\lambda,\dag}$ are annihilation and creation operators of a photon of momentum $k$ and polarization $\lambda$. 
The properties of these operators were considered in some detail in the original Faddeev and Kulish paper, where $\mathrm{R}$ is a coherent state operator spanning the Hilbert space of asymptotic charged particle states. However as noted in \cite{Kulish:1970ut}, and further emphasized in \cite{Hannesdottir:2019opa}, these states,  dressed with clouds of soft photons, can never be thought of as direct products of similarly dressed single particle states. As also noted previously, the phase $\Phi(t)$ is the relativistic generalization of the Coulomb (or Eikonal) phase\footnote{For a nice discussion of this phase in the parton model formulation of QED at high energies, we refer the reader to \cite{Bjorken:1970ah}.}, which is crucial in canceling the infinite phase factors that arise in the Dyson S-matrix.

In our worldline formulation of scattering amplitudes, we integrated out the gauge and matter fields explicitly, giving rise to non-local interaction functionals quadratic in the fermion currents, as in Eq.~\eqref{S_ab_momentum}.  This prevents us
from defining a FK S-matrix in the same way as in Eq.~\eqref{S_Bar_S_Dyson_2}, where the asymptotic dressings are linear in the gauge field, interactions are local, and currents can be factorized out from interactions in the central region to dress either the traditional S-matrix, or the \textit{in} and \textit{out} states, as desired.

In the operator formulation however, it is clear that the action of $\mathrm{R}(t)$ - more generally, the action of $\mathrm{R}(t)$ in conjunction with terms coming from the central region in the Dyson S-matrix $\mathcal{S}$, order-by-order in perturbation theory - will introduce the radiative modes of the asymptotic interactions that, once exponentiated, will lead precisely to the $\mathrm{R}^{nm}_{as}$ terms we found in Eq.~\eqref{r_nm_asymptotic}. Likewise, the action of $\Phi(t)$ on the asymptotic Coulomb modes, once exponentiated, will lead to the $\Phi^{nm}_{as}$ contributions we found in Eq.~\eqref{phi_nm_asymptotic}.

It is worth noting that each charged worldline in this framework 
is not separately dressed into an individual cloud, but this cloud depends on the rest of charges: the dressings involve photon exchanges between different charges in the asymptotic region, and also photon exchanges between the central and the asymptotic regions. As discussed previously, these features are crucial to completely cancelling the IR divergences in $\bar{\mathcal{S}}$. Thus the identification of charged worldlines with charged particles, while a useful mnemonic, is not fully robust. 

Our form of the IR finite FK S-matrix $\bar{\mathcal{S}}$ presents potential advantages. In the usual FK formulation~\cite{Kulish:1970ut,Hannesdottir:2019opa}, the S-matrix elements are only defined perturbatively, and the IR finiteness of amplitudes can only be proven order-by-order in perturbation theory. In the worldline approach, the amplitudes are defined to all orders in perturbation theory, with an arbitrary number of virtual photon and fermion loops attached. In particular, we showed that the exponentiation of IR divergences obtained from perturbation theory~\cite{PhysRev.140.B516} appears naturally, with the added intuition of  a theory of classical currents of nearly free first-quantized particles as they approach the far past and future. 
More importantly, the emergence of the IR dressings of the central and asymptotic regions as non-local worldline interaction functionals between classical electromagnetic currents is transparent. This allowed us to derive an all order proof of the IR finiteness of the FK S-matrix of QED. 

Another potential advantage presented by the worldline framework, as discussed in Section 4, is its efficiency in high order perturbative calculations. Each order in perturbation theory can be generated in the worldline from a single one-loop ($\ell=1$) master expression, which contains $n$-factorial combinations of Feynman graphs corresponding to different permutations of  photon insertions into the fermion lines.
We will discuss this further in the next section in the context of perturbative computations of the anomalous dimension of a cusped Wilson loop.

An interesting recent development is the interpretation of the  asymptotic interactions encoded in the FK S-matrix in the language of symmetries. It was pointed out very early by Weinberg~\cite{Weinberg:1964kqu} that the universal features of soft theorems in Abelian theories~\cite{PhysRev.140.B516} might arise directly from conservation laws of some fundamental symmetry. In his interpretation, the form of soft theorems is dictated from Lorentz invariance to simply enforce charge conservation and the principle of equivalence. More recently, Low’s theorem has been shown to be equivalent to the Ward identity \cite{Campiglia:2015qka,Kapec:2015ena,Kapec:2017tkm} of an asymptotic symmetry of the amplitudes generated by the group of large $U(1)$ gauge transformations acting over the \textit{in} and \textit{out} states with angle-dependent constants at null infinity~\cite{He:2014cra}. In gravity, Weinberg’s soft graviton theorem can be understood as the Ward identity~\cite{He:2014cra,Strominger:2013jfa} of the BMS group of asymptotically flat spacetime symmetries.\cite{Bondi:1962px,Sachs:1962wk}

The invariance of amplitudes under large gauge transformations implies that any scattering process is accompanied by a transition between degenerate vacua containing a differing number of IR photons. These shifts between different vacua have been found \cite{Kapec:2017tkm} to be implemented precisely by the same asymptotic interactions encoded in the dressings of the Dyson S-matrix proposed by Faddev and Kulish in \cite{Kulish:1970ut}. In the conventional construction of amplitudes through the Dyson S-matrix, the \textit{in} and \textit{out} states are assumed free charged-particle states, and then the vacuum of the theory is assumed to be unique. Because this would violate the conservation law, the Dyson S-matrix has to be identically zero. The emergence of IR divergences in the form of Low's soft photon theorem, order by order in perturbation theory, can be therefore interpreted as a necessary condition to avoid spoiling the BMS-charge conservation laws associated to the BMS-like symmetry of the theory.

As emphasized before, since we have integrated out the gauge field completely, the FK S-matrix in the worldline form of Eq.~\eqref{s_fi_fk} prevents us from disentangling the dressings in the gauge field of the charged-particle states in the \textit{in} and \textit{out} states at plus and minus infinity, from the interactions of these charges with the gauge field occurring at intermediate times in the central region. This indeed has to be the case, since the interaction of these charges to all orders in perturbation theory is encoded in a non-local interaction functional of all charged-particle currents. In this worldline framework then, the vacuum transitions mentioned previously are automatically implemented within the definition of the scattering amplitude itself, by means of the asymptotic currents corresponding to the terms with phases $k\wc x_N$ and $k\wc x_1$ in Eq.~\eqref{realparticle_current_momentum_discretized}. We plan to return to the discussion of asymptotic symmetries in the worldline context in future work. 

\section{\label{section_6} Cusp anomalous dimension from worldlines}

The fundamental objects in the worldline formulation of QED are normalized expectation values of non-local interaction functionals of charged currents created by super-pairs of scalar $x_\mu(\tau)$ and Grassmann $\psi_\mu(\tau)$ worldlines. These expectation values are expressed as generalized Wilson loops, containing the non-local dynamics of point-like super-pairs of particles.
Worldlines  therefore provide a natural framework to examine the renormalization properties of Wilson loops in QED\footnote{Early works on the renormalization of Wilson loops include \cite{Polyakov:1980ca,Arefeva:1980zd,Dotsenko:1979wb,PhysRevD.24.879,Brandt:1982gz,Korchemsky:1987wg}. For more recent impressive progress  in computing cusp anomalous dimensions to high orders in QED, QCD and $\mathcal{N}=4$ supersymmetric Yang-Mills theory, see \cite{PhysRevLett.126.021601,Grozin:2014hna,Catani:2019rvy,Henn:2019swt,Grozin:2015kna}.}. In particular, we will show 
how an efficient calculation of the cusp anomalous dimension can be carried out in the worldline formulation, with all the diagrams  generated, to any order in perturbation theory, from the master formula discussed in Section \ref{section_4}.

To dress a cusped worldline to all orders in perturbation theory, we consider the interactions of a single real charged-particle $(r=1)$ and an arbitrary number of virtual fermions $\ell$. This reduces to evaluating normalized worldline expectation values of interaction functionals of their respective currents, which are given by  Eq.~\eqref{total_current_real_virtual} to be
\ba 
J_\mu^{(1,\ell)}(x)=J_{R,\mu}^1(x)+\sum_{i=1}^\ell J_{V,\mu}^i(x)\,.
\ea 
The  asymptotic behaviour of a real charged-particle is dominated by its scalar degrees of freedom, and the evolution of its 
worldline $x_\mu(t)$ is well approximated by a  classical trajectory $x_\mu^\text{cl}(t)$ possessing a cusp. Following Eq.~\eqref{current_external_fermion}, this is 
\ba 
J_{R,\mu}^1(x)\simeq J_{\mu}^\text{cl}(x)= g\mu^{2-d/2}\int^{+\infty}_{-\infty}dt \dot{x}_\mu^\text{cl}(t)\delta^4(x-x_{cl}(t))\,,
\ea 
where $d=4-2\,\epsilon$ and we absorbed the mass dimension of the coupling in a parameter $\mu$ with mass dimension unity, and
\ba 
x_\mu^\text{cl}(t)=\beta_\mu^i\,t \,\,\,\,\text{for}\,\,\,\, t<0\,,\,\,\,\, x_\mu^\text{cl}(t)= \beta_\mu^f\,t\,\,\,\, \text{for}\,\,\,\, t>0\,,\label{single_cusp_worldline}
\ea 
where, as previously, $\beta_i$ and $\beta_f$ are the initial and final velocities. The cusp is placed at $x_\mu^\text{cl}(0)=0$ without loss of generality. 

The dressings of this scalar charged-particle's Wilson line in the Abelian gauge field are encoded, to all orders in perturbation theory, in the dressed 2-point function of Eq.~\eqref{2_point_function_result_minkowski}, with the spin degrees of freedom and the path integration over the real worldline configurations omitted. We define the amplitude $\mathcal{W}$ accordingly as
\ba 
\mathcal{W}=\frac{\mathrm{Z}_\text{MW}}{\mathrm{Z}}\sum_{\ell=0}^\infty \frac{(-1)^\ell}{\ell!}\mathrm{W}^{(1,\ell)}(x_f,x_i) \,,\label{mathcalW}
\ea 
where $x_f=x^\text{cl}(t_f)$ and $x_i=x^\text{cl}(t_i)$ as $t_{f,i}\to\pm\infty$, and the $\ell^{\rm th}$ contribution in the loop expansion is given by the normalized worldline expectation value taken only over the worldline configurations of the $\ell$ virtual fermions
\ba 
&\mathrm{W}^{(1,\ell)}(x_f,x_i)\nonumber\\
&= \bigg\langle \exp\bigg\{-\frac{i}{2}\int d^dx\int d^dy \Big(J_\mu^{(\ell)}(x)+J_\mu^\text{cl}(x)\Big)D_{\mu\nu}^B(x-y)\Big(J_\nu^{(\ell)}(y)+J_\nu^\text{cl}(y)\Big) \bigg\}\bigg\rangle\,.
\ea 
Further expanding Eq.~\eqref{mathcalW} in powers of $g$, precisely in the same way as discussed in Sect. \ref{section_4_1} for the vacuum-vacuum amplitude, gives the series
\ba 
&\mathcal{W}= \bigg( \sum_{\ell=0}^\infty \sum_{n=0}^\infty \mathrm{Z}^{(\ell)}_{(n)}\bigg)^{-1}\sum_{\ell=0}^\infty\sum_{n=0}^\infty  \frac{(-1)^\ell}{\ell!}\frac{1}{2^nn!} \label{mathcalW_expansion}\\
&\times \bigg\langle \prod_{i=1}^n\int d^dx_i\int d^dy_i  \Big(iJ_{\mu_i}^{(\ell)}(x_i)+iJ_{\mu_i}^\text{cl}(x_i) \Big)iD_B^{\mu_i\nu_i}(x_i-y_i) \Big(iJ_{\nu_i}^{(\ell)}(y_i)+iJ_{\nu_i}^\text{cl}(y_i)\Big)\bigg\rangle \nonumber\,.
\ea 
Here the disconnected fermion loop graphs are removed by the corresponding terms $\mathrm{Z}^{(\ell)}_{(n)}$ in the loop expansion of the vacuum-vacuum amplitude.
The resulting $n^\text{th}$ and $\ell^\text{th}$ term in the power expansion is of order $\alpha^n$ and contains $n$ virtual photons and $\ell$ virtual fermions attached to the cusped charged-particle worldline. 

To one-loop order ($n=1$ and $\ell=0$), Eq.~\eqref{mathcalW_expansion} gives
\ba 
\mathcal{W}=1 -\frac{g^2}{8\pi^2}(\mu^2\pi)^{\epsilon}\Gamma(1-\epsilon) \lim_{\delta\to\infty}\int_{-\delta}^{+\delta} dt_1 \int^{+\delta}_{-\delta} dt_2 \frac{\dot{x}^\text{cl}(t_1)\wc\dot{x}^\text{cl}(t_2)}{(x^\text{cl}(t_1)-x^\text{cl}(t_2))^{2-2\epsilon}}+\cdots\,,\label{Z_jcusp_Z_oneloop}
\ea 
where we used the photon propagator in Eq.~\eqref{photon_propagator_coordinate} in $d=4-2\epsilon$ dimensions and Minkowski signature,
\ba 
D^{\mu\nu}_B(x)=
\frac{ig^{\mu\nu}}{4\pi^{d/2}}\Gamma\left(\frac{d-2}{2}\right)\frac{1}{(ix)^{d-2}}\,.
\ea 
This one-loop contribution contains UV divergences 
 from virtual photons attached to equal points in the worldline $x^\text{cl}(t_1)=x^\text{cl}(t_2)$, and IR divergences coming from the virtual photons attached to the far past and future $\delta\to\infty$. 
\begin{figure}[ht]
    \centering
    \includegraphics[scale=0.27]{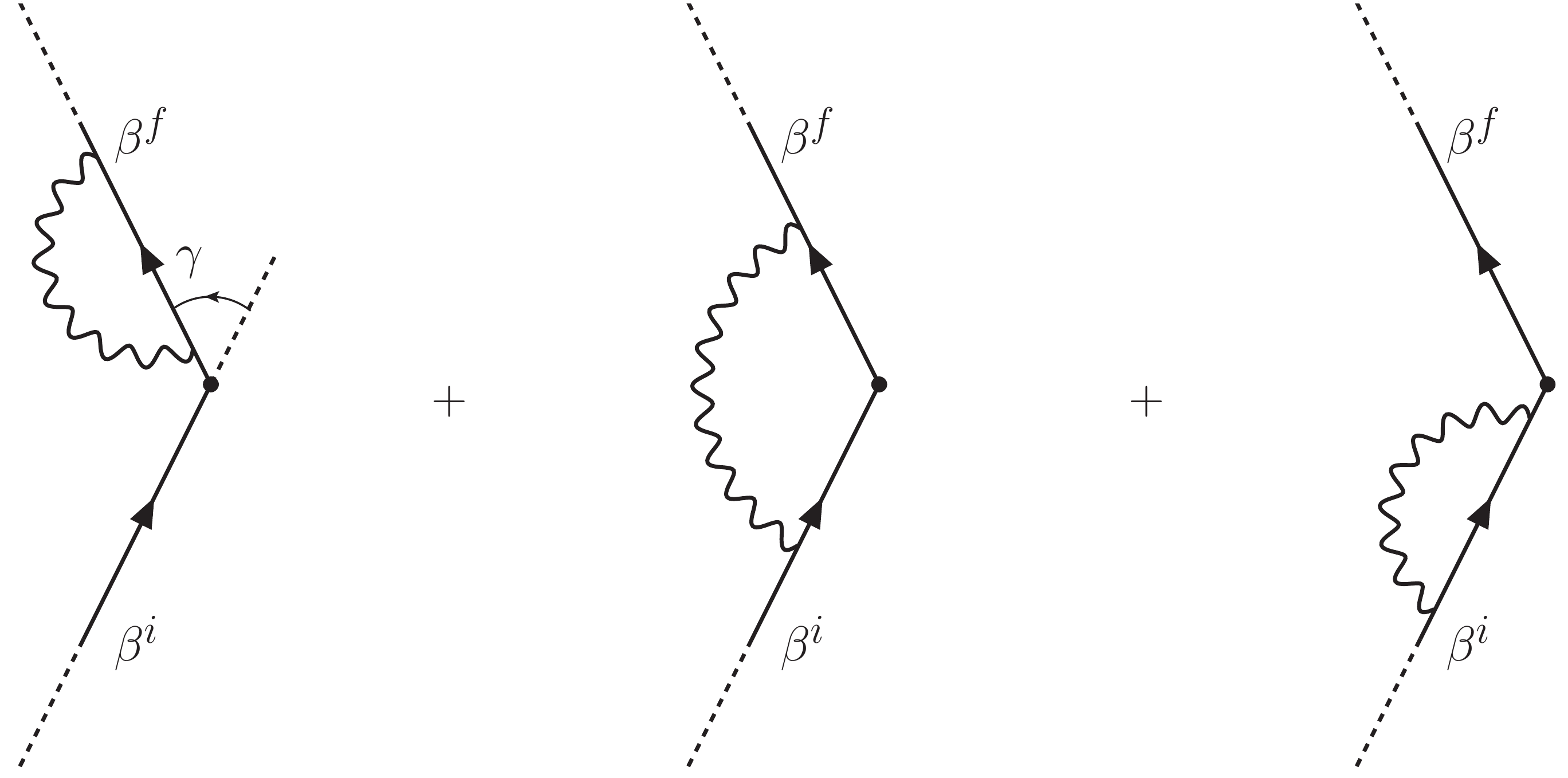}
    \caption{One-loop diagrams ($n=1$ and $\ell=0$) in Eq.~\eqref{Z_jcusp_Z_oneloop} contributing to the dressing of the cusped current $J^\text{cl}_\mu$. Note that the $(\beta_f,\beta_i)$ and $(\beta_i,\beta_f)$ contributions in 
    Eq.~\eqref{eq:four-contributions} are counted separately.}
    \label{fig:figure_cusp_1}
\end{figure}

Using Eq.~\eqref{single_cusp_worldline} the integral can be split into four contributions, their corresponding diagrams shown in Fig.~\ref{fig:figure_cusp_1}, giving 
\ba 
\mathcal{W} =1 -\frac{g^2}{8\pi^2}(\mu^2\pi)^{\epsilon}\Gamma(1-\epsilon) \Big\{I(\beta_f,\beta_i)+I(\beta_i,\beta_f)-I(\beta_f,-\beta_f)-I(-\beta_i,\beta_i)\Big\}+\cdots\,,
\ea 
where the required integrals $I(\beta_f,\beta_i)$ are defined as
\ba
\label{eq:four-contributions}
I(\beta_f,\beta_i)=\lim_{\delta\to\infty}\int_0^\delta dt_1\int_{0}^\delta dt_2 \frac{\beta_f\wc \beta_i}{(\beta_ft_1+\beta_it_2)^2}=\int_0^{\infty} dt \frac{1}{t^{1-2\epsilon}} \int_0^1 ds &\frac{\beta_f \wc
    \beta_i}{(\beta_fs+(1-s)\beta_i)^{2}}\,.
\ea
The Schwinger parameters $t=t_1+t_2$ and
$s=t_1/t$ are introduced in the second equality. The scale-less $t$ integral receives both the $t\to 0$
(UV) divergence and the $t\to\infty$ (IR) divergence from photons
attached to equal and very distant points in the worldline,
respectively, in Eq.~\eqref{Z_jcusp_Z_oneloop}. 
To capture the IR divergence, we regulate the integral over $t$ as 
\begin{align}
\int_0^{\infty}\frac{dt}{t^{1-2\epsilon}}e^{-\lambda t}=\frac{1}{\lambda^{2\epsilon}}\Gamma(2\epsilon)\,.\label{time_integral}
\end{align}
The $s$ integral remains finite. Setting $\epsilon=0$ then yields the geometrical factor
\begin{align}
  \int_0^1 ds &\frac{\beta_f \wc
    \beta_i}{(\beta_fs+(1-s)\beta_i)^{2}}=\frac{\beta_f\wc\beta_i}{\sqrt{(\beta_f\wc\beta_i)^2-\beta_f^2\beta_i^2}}\text{Atanh}\left(\frac{\sqrt{(\beta_f\wc\beta_i)^2-\beta_f^2\beta_i^2}}{\beta_f\wc\beta_i}\right)\,,
    \label{angle_dependent_factor_cusp}
\end{align}
where one identifies the 4-dimensional 
cusp angle between the final $f$ and initial $i$ directions as
$\gamma_{fi}$, as shown previously Fig.~\ref{fig:figure_cusp_1}, 
\begin{align}
\cosh \gamma_{fi}= \frac{\beta_f\wc\beta_i}{\sqrt{(\beta_f)^2(\beta_i^n)^2}}\to \text{tanh}\gamma_{fi} = \frac{\sqrt{(\beta_f\wc \beta_i)^2-\beta_f^2\beta_i^2}}{\beta_f\wc \beta_i}\,,\label{cusp_angle_definition}
\end{align}
allowing us to write  Eq. \eqref{angle_dependent_factor_cusp} as 
\begin{align}
\int_0^1 ds &\frac{\beta_f \wc
    \beta_i}{(\beta_fs+(1-s)\beta_i)^{2}}=\gamma_{fi}\,\text{coth}\gamma_{fi}\,.
\end{align}

The diagrams with initial-initial
and final-final photon loops are given by the expressions  $-I(-\beta_i,\beta_i)$
and $-I(\beta_f,-\beta_f)$, corresponding to the substitutions $\gamma_{ff}=0$ and $\gamma_{ii}=0$ in the equation above\footnote{We omit here the divergent imaginary contribution that is related to shifts of the cusp angle $\gamma_{fi}\pm i\pi$.}. Collecting all the terms in $\epsilon$ in Eq.~\eqref{Z_jcusp_Z_oneloop}, and expanding around $\epsilon=0$, gives 
\begin{align}
\left(\frac{\mu^2\pi}{\lambda^2}\right)^{\epsilon}\Gamma(2\epsilon)=\frac{1}{2\epsilon}+\frac{1}{2}\log\left(\frac{\mu^2\pi}{\lambda^2}\right)+\gamma_E + (\cdots)\,,
\end{align}
where $\gamma_E$ is the Euler-Mascheroni constant. Plugging all these results into Eq.~\eqref{Z_jcusp_Z_oneloop} gives, to one-loop,
\begin{align}
 \mathcal{W} =1-\frac{g^2}{4\pi^2}\left(\frac{1}{2\epsilon}+\frac{1}{2}\log\left(\frac{\mu^2 \pi}{\lambda^2}e^{\gamma_E}\right)\right)\big(\gamma_{fi}\text{coth}\gamma_{fi}-1\big)\label{Z_jcusp_Z_oneloop_result}+\cdots\,,
\end{align}
Eq.~\eqref{Z_jcusp_Z_oneloop_result} contains both the cusp IR $\sim\log(1/\lambda)$ and UV $\sim 1/\epsilon$ divergences. The former corresponds precisely to the self-energy contributions ($n=m$) in the IR dressings of a scattering amplitude found in Eq.~\eqref{S_I_cusped_worldlines_result}. It can therefore be cancelled following the Faddeev and Kulish procedure discussed in Section \ref{section_5_2}, leading to Eq.~\eqref{cancellation_IR_divergences_allorders}.

To renormalize the UV divergences in $\mathcal{W}$, we follow the procedure in \cite{Brandt:1982gz}.
In Eq.~\eqref{Z_jcusp_Z_oneloop_result}, the prefactor of the UV divergence
$1/\epsilon$ has a dependence on the cusp angle related to the short-time integration in the vicinity of the cusp. The required renormalization factor $\mathcal{Z}(\gamma_{fi};g,\epsilon)$ is then
also angle dependent. This suggests, following \cite{Brandt:1982gz}, 
\ba 
\mathcal{W}_R(\gamma_{fi};g_R,\mu/\lambda)=\mathcal{Z}(\gamma_{fi};g,\epsilon)\mathcal{W}(\gamma_{fi};g,\mu/\lambda,\epsilon)\bigg|_{g=g(g_R,\mu,\epsilon)}
\ea 
where $g_R$ is the renormalized coupling and the renormalized amplitude $\mathcal{W}_R$ (encoding the dressings of the classical cusped scalar worldline) satisfies the renormalization group equation, (RGE) 
\ba 
\bigg\{\mu\frac{\partial}{\partial \mu}+\beta(g_R)\frac{\partial}{\partial g_R}+\Gamma(\gamma,g_R)\bigg\}\mathcal{W}_R = 0\,,\,\,\,\,\, \Gamma(\gamma_{fi},g_R) = -\frac{d \ln \mathcal{Z}}{d \ln \mu}\,,
\ea 
with the anomalous dimension $\Gamma(\gamma_{fi},g_R)$ introduced by the cusp. The QED $\beta$-function is defined by ($d=4-2\epsilon$) 
\ba 
\mu\frac{d\alpha}{d\mu} = -2\alpha\bigg\{\epsilon +\frac{\alpha}{4\pi}\beta_0+\bigg(\frac{\alpha}{4\pi}\bigg)^2\beta_1+\cdots\bigg\}\,,
\ea 
where $\alpha=g_R^2/4\pi$. Expanding $\mathcal{Z}$ in powers of $g_R$ and matching, to subtract to one-loop order, the $1/2\epsilon$ UV divergence of $\mathcal{W}$ gives
\ba 
&\mathcal{Z}(\gamma_{fi};g,\epsilon)=1+\sum_{n=1}^\infty g_R^{2n}\mathcal{Z}_n=1+\frac{\alpha}{\pi}\frac{1}{2\epsilon}(\gamma_{fi}\coth \gamma_{fi}-1)+\cdots\,.
&
\ea 
Hence the one-loop cusp anomalous dimension is then 
\ba 
\Gamma(\gamma_{fi},g_R)=\frac{\alpha}{\pi}(\gamma_{fi}\coth\gamma_{fi}-1)+\mathcal{O}(\alpha^2)\,.
\ea 

The previous computation can be extended to 
higher orders. We first  note that all tree-level dressings $(\ell=0)$ in Eq.~\eqref{mathcalW_expansion} exponentiate, as anticipated, by setting $\ell=0$ in Eq.~\eqref{mathcalW} and avoiding
therefore the power expansion in Eq.~\eqref{mathcalW_expansion}. One thus obtains,
\ba 
\mathcal{W}=\exp\bigg\{-\frac{g^2}{4\pi^2}\left(\frac{1}{2\epsilon}+\frac{1}{2}\log\left(\frac{\mu^2 \pi}{\lambda^2}e^{\gamma_E}\right)\right)\big(\gamma_{fi} \text{coth}\gamma_{fi}-1\big)\bigg\}\label{mathcalW_treelevelexponentiation}+\cdots\,,
\ea 
where the ellipses indicate that the $\ell\neq 0$ contributions coming from the attachment of virtual fermion loops remain to be included.  The corresponding graphs for these in $\mathcal{W}$ are  generated in exactly the same way, as the $\ell=0$ term by the compact expression in Eq.~\eqref{mathcalW_expansion}. The building blocks required of this computation, for arbitrary $n$ and $\ell$, are given by the multi-loop worldline polarization tensors of Eq.~\eqref{bern_kosower} in Appendix \ref{appendix_c}. 

For instance, at two-loop order ($n=2$ and $\ell=1$), the first
finite correction to the tree level graphs in Eq.~\eqref{mathcalW_treelevelexponentiation} 
using Eq.~\eqref{mathcalW_expansion} 
contains as its building block the expression
\ba 
\frac{1}{2}\int\frac{d^dk_1}{(2\pi)^d}iD^{\mu_1\nu_1}(k_1)\int\frac{d^dk}{(2\pi)^d}iD^{\mu_2\nu_2}(k_2)iJ_{\mu_1}^\text{cl}(+k_1)i\Pi_{\mu_1\nu_1}(+k_1,-k_2)iJ_{\mu_2}^\text{cl}(-k_2)\,,\label{two_loop_cuspanomalousdimension}
\ea 
where the polarization tensor $\Pi_{\mu\nu}(k_1,k_2)$ is given by Eq.~\eqref{bern_kosower}, as shown in Appendix \ref{appendix_c}.
As we discussed in Sec. \ref{section_4_1}, automatically encoded in a single and compact expression are multi-loop diagrams with many virtual fermions contributing to the cusp anomalous dimension at higher orders. 
In contrast, in the standard computations of the two-loop cusp anomalous dimension~\cite{Korchemsky:1985xu}, the exact form of a contribution like the one in Eq.~\eqref{two_loop_cuspanomalousdimension} has to be  obtained separately from the corresponding Feynman graphs that in the UV limit include the spin structure of fermion loops. The efficient computation of higher order loop contributions to the cusp anomalous dimension in the worldline formulation of QED will be addressed in detail in Paper II.

\section{\label{section_7}Conclusions and outlook}
We developed in this paper a reformulation of QED as a fully covariant, first-quantized, many-body theory of 0+1 dimensional super-pairs of bosonic and Grassmann worldlines. In particular, the QED path integral can be expressed as a power series expansion in generalized Wilson loops, where the Wilson loop at a given $\ell$-th order (corresponding to $\ell$ virtual worldlines describing loops in propertime) contains a path integral over all possible paths of pointlike boson and fermion super-pairs. The dynamics of the $\ell$ virtual fermions in this path integral is encoded in the exponentiated photon exchanges between the $\ell$ loop currents that are functionals of the boson and fermion super-pairs. 

We further extended our formalism to construct scattering amplitudes for charged wordlines interacting via the exchange of arbitrary numbers of virtual photons and charged fermion loops. One can similarly express the amplitude for the scattering of $r$ worldlines to $\ell$-th loop order in terms of a generalized Wilson line/loop structure corresponding to the path integral over first-quantized, many-body worldline configurations of the $r$ real worldlines and $\ell$ virtual worldlines. As previously, the dynamics of their interactions is contained in the exponentiated photon exchanges between the real and virtual currents, with additional energy-momentum and helicity-momentum constraints imposed on the real worldlines.

We outlined the usefulness of this formalism for high order perturbative computations by showing that terms to arbitrary order in a loop expansion of the vacuum-vacuum amplitude can be generated from a compact expression for multi-loop polarization tensors.  This master formula can further be applied to efficiently perform higher order calculations of any given QED process, automatically encoding the factorial growth of diagrams in conventional perturbation theory. We also demonstrated that all-order worldline results can be extracted from the generating functional for the QED one-loop worldline effective action in an background field. This suggests that the non-perturbative structure of worldline amplitudes (an example being Schwinger pair production in strong electromagnetic fields) can alternatively be explored by the use of semi-classical expansions in appropriate background fields. 

We also showed that the worldline formalism provides a powerful framework to explore the infrared structure of QED. 
By examining the classical long-range interactions of the charged-particle worldline currents, we showed that virtual IR divergences and Abelian exponentiation theorems naturally emerge in the worldline expressions as soft dressings of the Dyson S-matrix. We showed that these soft dressings that set the Dyson S-matrix to zero can be subtracted by interaction terms with the exchange of soft photons between charged currents, with at least of one them in the asymptotic region of very early or late times governed by the Coulomb dynamics of classical currents.  This observation allowed us to construct a Faddeev-Kulish worldline S-matrix including these asymptotic contributions and prove it to be free of IR singularities to all orders in perturbation theory. We illustrated our general argument for the more specialized case of M\"{o}ller scattering and discussed the mapping of our worldline results to the original Faddeev-Kulish results and more recent discussions in the literature.

We finally addressed the renormalization of cusped Wilson lines which naturally correspond to worldline cusps in our framework. We showed how the one-loop cusp anomalous dimension is recovered in this framework and outlined how this formalism enables efficient computation of cusp anomalous dimensions to higher orders. 

Indeed in Paper II, we will discuss an explicit computation of the QED two loop cusp anomalous dimension and compare and contrast this with previous computations in the literature. We will also discuss in Paper II the extension of the computation of scattering amplitudes  with all order virtual exchanges discussed here to that of amplitudes involving the emission of real photons. This will complete our proof of the infrared safety of the worldline Faddeev-Kulish S-matrix. 

It would also be interesting to understand if the infrared behavior of the worldline Faddeev-Kulish S-matrix can be recast in the universal language of asymptotic symmetries of gauge theories. In particular, since QED can be exactly rewritten in this first-quantized albeit non-local formalism, how does one interpret  the non-local FK dressings as Goldstone bosons of spontaneously broken asymptotic symmetries? We plan to pursue such  questions in follow-up work.

A further application of our work is to explore whether worldline Monte Carlo methods can be employed specifically to compute the $\ell$-th loop, n photon polarization tensors that we identified as building blocks of perturbative computations in this framework. If so, this would have immediately applications in precision QED tests for a wide range of observables. 

Further important extensions of our work are to QCD and gravity, where semi-classical worldline formulations have been known for decades. In the case of QCD, we noted that the powerful string amplitude and worldline formalisms have been shown to be equivalent. However it is not known whether the procedure followed in this paper can be extended to QCD with the necessary systematic approximations. This problem has a correspondence in the well known language of non-Abelian exponentiation. A specific goal  would be to extend our methods to compute high order cusp anomalous dimensions in QED to 
QCD. Another would be to apply these ideas to the Regge regime, where worldline computations in the presence of semi-classical background fields is promising both in precision computations and in achieving deeper insight into infrared phenomena such as color memory~\cite{Pate:2017vwa,Ball:2018prg}. With regard to the former, as we noted briefly, plans are underway to study puzzling apparent violations of Low's theorem at the LHC.

Finally, looking towards the longer term, the fact that QED can be formulated in full generality as a many-body theory of first-quantized worldlines may have powerful implications for quantum computing. Since Grassmann worldlines are natural qubits, it would be interesting to explore whether the supersymmetry between Grassmann and boson super-pairs can be exploited to simplify digitization of  the large Hilbert space occupied by bosonic worldlines.

\section*{Acknowledgements}
R. V. would like to thank Peter Braun-Munzinger, Stefan Fl\"{o}rchinger, Hofie Hannesdottir, Otto Nachtmann, Johanna Stachel, Andy Strominger, and especially  Monica Pate and Ana-Maria Raclariu, for discussions that influenced the work presented in this paper. We would also like to thank Niklas Mueller for previous  collaborative work on worldlines, as well as Fiorenzo Bastianelli, James Edwards, Christian Schubert and the ``worldliners" seminar group for useful discussions on this topic. 

X. F. is supported by a Fulbright Visitor Scholarship,  
grant ED481B-2019-040 of Conseller\'ia de Cultura, Educaci\'on e Universidade of Xunta de Galicia (Spain) and in part by the U.S. Department of Energy, Office of Science, Office of Nuclear Physics, under contract No. DE- SC0012704. 

R. V. is supported by the U.S. Department of Energy, Office of Science, Office of Nuclear Physics, under contract No. DE- SC0012704. He would also like to acknowledge support by the DFG Collaborative Research Centre SFB 1225 (ISOQUANT) at Heidelberg University.

The possible applications of the framework outlined here to quantum information science were discussed by R. V. with Steve Girvin, Rob Pisarski and Nathan Wiebe, with partial support provided by the U.S. Department of Energy, Office of Science National Quantum Information Science Research Center's Co-design Center for Quantum Advantage (${\rm C}^2$QA) at BNL under contract number DE-SC0012704. 

A. T. is supported by the U.S. Department of Energy, Office of Science, Office of Nuclear Physics under Award Number DE-SC0004286.

\appendix

\section{Conventions}
\label{appendix_a}
The QED generating functional in $d=3+1$ Minkowski spacetime is defined as
\ba 
&\mathrm{Z}[\mathcal{J},\bar{\eta},\eta]=\int \mathcal{D}A_M \mathcal{D}\bar{\Psi}_M\mathcal{D}\Psi_M\exp\bigg\{-\frac{i}{4}\int d^4x_M F_{\mu\nu}^MF^{\mu\nu}_M-\frac{i}{2\zeta}\int d^4x_M \Big(\partial_\mu^M A^\mu_M)^2\label{QED_generating_functional}\\
&+i\int d^4x_M \mathcal{J}_\mu A^\mu_M+i\int d^4x_M\bar{\Psi}_M(i\slashed{D}_M-m)\Psi_M +i\int d^4x_M \bar{\eta}\Psi_M +i\int d^4x_M\bar{\Psi}_M\eta\bigg\}\nonumber\,,
\ea 
with $\mu=0,1,2,3$. Here the label $M$ represents fields in Minkowski spacetime. Further $\slashed{D}_M=\gamma_\mu^M D^\mu_M$  $D_\mu^M=\partial_\mu^M-igA_\mu^M$,  and $g_{\mu\nu}=\text{diag}(+1,-1,-1,-1)$. Defining $[\hat{p}_\mu^M,\hat{x}_\nu^M]=ig_{\mu\nu}$,  
\begin{align}
\hat{p}_\mu^M = i\partial_\mu^M = i\partial/\partial x^\mu_M, \medspace\medspace\medspace  \langle x | p \rangle = e^{-ip_\mu^M x^\mu_M}\,.\label{cannonical_commutation}
\end{align}
The Dirac matrices are defined as
\ba 
\{ \gamma_\mu^M,\gamma_\nu^M\}=2g_{\mu\nu}, \medspace\medspace\medspace \gamma_5^M=\gamma_0^M\gamma_1^M\gamma_2^M\gamma_3^M, \medspace\medspace\medspace (\gamma_5^M)^2=-1,\medspace\medspace\medspace \{\gamma_5^M,\gamma_\mu^M\}=0\,.
\ea 
The analytic continuation of  Eq.~\eqref{QED_generating_functional} 
to $d=4$ Euclidean space is defined by the Wick rotation to imaginary times
\ba
x_0^M=-ix_4^E, \medspace\medspace\medspace x_i^M=x_i^E\,.\label{wick_rotation_coordinates}
\ea 
The cannonical commutation relation in Eq.~\eqref{cannonical_commutation} and gauge invariance determines the rest:
\ba
 \medspace\medspace\medspace p_0^M=+ip_4^E, \medspace\medspace\medspace p_i^M=-p_i^E, \medspace\medspace\medspace A_0^M=+iA_0^E, \medspace\medspace\medspace A_i^M=-A_i^E\medspace\medspace\medspace \mathcal{J}_0^M=-i\mathcal{J}_4^E, \medspace\medspace\medspace \mathcal{J}_i^M=\mathcal{J}_i^E\,,\label{wick_rotation_momentumfields}
\ea
with $\eta_{\mu\nu}=\text{diag}(+1,+1,+1,+1)$, $\mu=1,2,3,4$. It follows that, from the relations above,
\ba 
&x_\mu^M x^\mu_M = -x_\mu^E x_\mu^E= -x_E^2, \medspace\medspace\medspace p_\mu^M p^\mu_M=-p_\mu^Ep_\mu^E=-p_E^2, \medspace\medspace\medspace x_\mu^M p^\mu_M =+ x^E_\mu  p_\mu^E= +x_E\wc p_E\,,
\ea 
and
\ba
F_{\mu\nu}^M F^{\mu\nu}_M = +F_{\mu\nu}^EF_{\mu\nu}^E, \medspace\medspace\medspace \partial_\mu^M A^\mu_M = -\partial_\mu^E A_\mu^E\,.\nonumber
\ea 
with $F_{ij}^M=F_{ij}^E$ and $F_{0j}^M=-iF_{4j}^E$ and $F_{\mu\nu}^E$ an antisymmetric tensor.

While not affected by the Wick rotation, it will be convenient for us to define Hermitian Euclidean Dirac matrices $(\gamma_\mu^E)^\dag=\gamma_\mu^E$ and $(\gamma_5^E)^\dag=\gamma_5^E$ - so that  $(\gamma_\mu^E)^2=1$,  $(\gamma_5^E)^2=1$ and the squares of Dirac operators are positive operators, where $\slashed{p}^E=p_\mu^E\gamma_\mu^E$. This requires
\ba
\gamma_0^M=\gamma_4^E, \medspace\medspace\medspace \gamma_i^M=+i\gamma_i^E,\medspace\medspace\medspace\{\gamma_\mu^E,\gamma_\nu^E\}=2\,\eta_{\mu\nu}, \medspace\medspace\medspace \gamma_5^E=\gamma_4^E\gamma_1^E\gamma_2^E\gamma_3^E, \medspace\medspace\medspace \{\gamma_5^E,\gamma_\mu^E\}=0\,.\label{wick_rotation_gammamatrices}
\ea 
For the purposes of the discussion in the main text, it will be also convenient for us to introduce the Hermitian matrices
\ba
\bar{\gamma}_\mu^E=+i\gamma_5^E\gamma_\mu^E, \medspace\medspace \medspace (\bar{\gamma}_\mu^E)^\dag=\bar{\gamma}_\mu^E\,.\label{bar_gamma_definition}
\ea 
One can verify that
$\{\gamma_5^E,\bar{\gamma}_\mu^E\}=0$ and $\{\bar{\gamma}_\mu^E,\bar{\gamma}_\nu^E\}=2\eta_{\mu\nu}$. Since $(\gamma_5^E)^2=1$, the algebra can be completed by
\ba 
\{\bar{\gamma}_\lambda,\bar{\gamma}_\kappa\}=2\,\eta_{\lambda\kappa} \medspace\medspace \medspace\medspace\medspace \medspace \lambda=\mu,5\,.\label{bar_gamma_anticommutator}
\ea
The Euclidean representation of spinors, and their Wick rotataion  is subtle; for a nice discussion, see \cite{vanNieuwenhuizen:1996tv}. Since 
we are primarily interested in Minkowski scattering amplitudes, 
we will employ for spin the definitions 
\ba
\sigma_{\mu\nu}^M = \frac{i}{2}[\gamma_\mu^M,\gamma_\nu^M], \medspace\medspace\medspace \sigma_{\mu\nu}^E= \frac{i}{2}[\gamma_\mu^E,\gamma_\nu^E], \medspace\medspace\medspace \bar{\sigma}_{\mu\nu}^E=\frac{i}{2}[\bar{\gamma}_\mu^E,\bar{\gamma}_\nu^E]= \sigma_{\mu\nu}^E\,,
\ea 
It follows that
\ba
i\slashed{D}^M = -D_\mu^E \gamma_\mu^E = -\slashed{D}^E ,\medspace\medspace\medspace \sigma_{\mu\nu}^MF^{\mu\nu}_M = -\sigma_{\mu\nu}^E F_{\mu\nu}^E\,.
\ea

In the rest of the paper, we will drop the superscripts in fields and coordinates, unless
explicitly required.
All amplitudes in physical time or imaginary time can be recovered by the replacements in
Eq.~\eqref{wick_rotation_coordinates}, Eq.~\eqref{wick_rotation_momentumfields} and Eq.~\eqref{wick_rotation_gammamatrices}. In particular,
the starting point for our discussion, the Euclidean vacuum-vacuum amplitude in Eq.~\eqref{QED_partition_function} can be obtained in this way from Eq.~\eqref{QED_generating_functional} 
and setting the external sources to zero.

\section{Worldline path integral for the dressed fermion propagator}
\label{appendix_b}
We will present here the derivation of the dressed spin-$1/2$ propagator in the worldline formalism. This approach can be  straightforwardly extended to higher-point correlators. We will follow the logic of the previous  computation by Fradkin and Gitman 
\cite{Fradkin:1991ci} but shall further clarify some subtle features as well as adapt the derivation to our conventions. 

We begin with the equations of motion of the Euclidean dressed fermion propagator in the gauge background field, 
\ba
(\slashed{D}+m)D_F^A(x,y)=(\partial_\mu\gamma_\mu-igA_\mu\gamma_\mu+m)D_F^A(x,y)=\delta^{(4)}(x-y)\,,
\label{eq:prop_F1}
\ea
and seek to represent $D_F^A(x,y)$ as a sum over worldline trajectories connecting
$x$ and $y$. Following \cite{Fradkin:1991ci}, to
explicitly account for the helicity-momentum constraint that the worldlines must satisfy, one considers
alternatively $\bar{D}_F^A(x,y)=D_F^A(x,y)\gamma^5$, with
$\gamma^5=\gamma^0\gamma^1\gamma^2\gamma^3$.  
Multiplying both sides of Eq.~\eqref{eq:prop_F1} with $i\gamma^5$, this modified propagator satisfies
\ba
  -(i\gamma_5\slashed{D} +i\gamma
  _5m)i\bar{D}_F^A(x,y)=\delta^{(4)}(x-y)\,.
  \label{greenfunction_gamma5}
\ea
It will be also convenient for us to define the Hermitian matrices $\bar{\gamma}_\mu=+i\gamma_5\gamma_\mu$ and complete the algebra $\{\bar{\gamma}_\lambda, \bar{\gamma}_\kappa\}=2\,\eta_{\lambda\kappa}$ with $\lambda=\mu,5$ and $\bar{\gamma}_5=\gamma_5$.  (Eqs.~\eqref{bar_gamma_definition} and \eqref{bar_gamma_anticommutator} of Appendix \ref{appendix_a}.)
Inverting Eq.~\eqref{greenfunction_gamma5}, the problem reduces to computing 
$\bar{D}_F^A(x,y)=\langle x | \bar{D}_F^A | y \rangle$ with the operator defined as 
\ba
\bar{D}_F^A =
\frac{i}{\bar{\gamma}_\mu D_\mu+i\gamma_5m}=\int^{\infty}_0d\varepsilon\, e^{\varepsilon(\bar{\gamma}_\mu D_\mu+i\gamma_5m)^2}\int d\chi\, e^{i\chi (\bar{\gamma}_\mu D_\mu+i\gamma_5m)}
\label{s_f_superpropertimes}
\ea
The operator has been decomposed into a bosonic piece $(\bar{\gamma}_\mu D_\mu+i\gamma_5m)^2$ and a fermionic piece  $\bar{\gamma}_\mu D_\mu+i\gamma_5m$. 
The former can be reexpressed as the integral on the 
r.h.s in terms of the  commuting propertime variable
$\varepsilon$. The latter can be represented, as shown, by the Grassmann integral in the 
anti-commuting propertime variable $\chi$. Identifying
$\hat{p}_\mu=i\partial_\mu$, the bosonic contribution to the propagator yields
\ba
-(\bar{\gamma}_\mu D_\mu+i\gamma_5m)^2=m^2+\big(\hat{p}_\mu+gA_\mu(\hat{x})\big)^2+i\frac{g}{2}\bar{\gamma}_\mu\bar{\gamma}_\nu F_{\mu\nu}(\hat{x})\,.
\ea
Further, since the quantity above is bosonic, we can interchange the order of integrands and write the dressed propagator as
\ba
\bar{D}^A_F(x_f,x_i)=\langle x_f|\bar{D}^A_F|x_i\rangle = \int_0^\infty d\varepsilon \int d\chi\, \langle x_f|e^{-\hat{H}}|x_i\rangle \,,
\label{eq:fermion-prop-path_integral}
\ea
where the Hamiltonian operator including the propertime super-pair $(\varepsilon,\chi)$ is given by 
\ba
&\hat{H}=H(\hat{p}_\mu,\hat{x}_\mu,\bar{\gamma}_n,\varepsilon,\chi)=\varepsilon\bigg\{m^2+\big(\hat{p}_\mu+gA_\mu(\hat{x})\big)^2+i\frac{g}{2}F_{\mu\nu}(\hat{x})\bar{\gamma}_\mu\bar{\gamma}_\nu\bigg\}\nonumber\\
  &+\chi\Big\{\bar{\gamma}_\mu\big(\hat{p}_\mu+gA_\mu(\hat{x})\big)-\gamma_5m\Big\}\,.
  \label{worldline_hamiltonian_openworldline}
\ea
The first bracket represents the mass-shell constraint satisfied by the worldlines, which becomes apparent after anti-symmetrizing $i\bar{\gamma}_\mu\bar{\gamma}_\nu$ into the spin tensor $\bar{\sigma}_{\mu\nu}$, while the second bracket corresponds to the helicity-momentum constraint\footnote{For a recent discussion of this constraint for 
massless fermions, see \cite{Mueller:2019gjj} and references therein.}.

To obtain the path integral representation of the dressed propagator,  one needs to discretize this matrix element in $n$ steps. We will however first generalize the parameters $\varepsilon$ and $\chi$ to commuting
$\varepsilon(\tau)$ and anti-commuting $\chi(\tau)$ functions of the worldline parameter
$\tau$, on equal footing with $x_\mu(\tau)$ and $p_\mu(\tau)$. This implies that
$\hat{\varepsilon}$ and $\hat{\chi}$ can be represented by operators (albeit with trivial dynamics) acting on  states, just as for $\hat{x}_\mu$
and $\hat{p}_\mu$. We define $\hat{\pi}$ as the operator conjugate to
$\hat{\varepsilon}$, and $\hat{\nu}$ as the corresponding conjugate of 
$\hat{\chi}$. With these modifications, we obtain
\ba
  [\hat{p}_\mu,\hat{x}_\nu]=ig_{\mu\nu} ,\medspace [\hat{\pi},\hat{\varepsilon}]=i, \medspace \{\hat{\nu},\hat{\chi}\}=0,\medspace \langle x|p\rangle
= e^{-ip_\mu x^\mu}, \medspace \langle \varepsilon|\pi\rangle =
e^{-i\pi\varepsilon},\medspace \langle \chi|\nu\rangle = e^{-\chi\nu}\,.
\label{commutation_relations}
\ea 
This then allows us to write the expectation value as 
\ba
 & \langle x_n|e^{-\hat{H}}|x_0\rangle = \left\{\prod_{k=1}^{n-1}\int d^4x_k\right\}\left\{ \prod_{k=0}^{n-1}\int \frac{d^4p_k}{(2\pi)^4}\right\}\left\{\prod_{k=0}^{n-1}\int d\varepsilon_k\int \frac{d\pi_k}{(2\pi)}\right\}\left\{\prod_{k=0}^{n-1}\int d\chi_k \int d\nu_k\right\} \nonumber\\
  &\times\left\{\prod_{k=0}^{n-1}\langle x_{k+1},\varepsilon_{k+1},\chi_{k+1}|p_{k},\pi_k,\nu_k\rangle \langle p_{k},\pi_k,\nu_k|e^{-H(\hat{p},\hat{x},\bar{\gamma},\hat{\varepsilon},\hat{\chi})/n}|x_{k},\varepsilon_k,\chi_k\rangle \right\}\,,
\ea
with the boundary conditions $\varepsilon_n=\varepsilon$ and $\chi_n=\chi$.
Upon the trivial action of the Hamiltonian on the states, and using Eq.~\eqref{commutation_relations}, we get
\ba
 & \langle x_n|e^{-\hat{H}}|x_0\rangle = \left\{\prod_{k=1}^{n-1}\int d^4x_k\right\}\left\{ \prod_{k=0}^{n-1}\int \frac{d^4p_k}{(2\pi)^4}\right\}\left\{\prod_{k=0}^{n-1}\int d\varepsilon_k\int \frac{d\pi_k}{(2\pi)}\right\}\left\{\prod_{k=0}^{n-1}\int d\chi_k \int d\nu_k\right\}\nonumber\\
&\times  \text{T}\exp\bigg\{-i\sum_{k=0}^{n-1}\bigg(\pi_k\frac{\delta\varepsilon_{k}}{\delta \tau_k}\delta \tau_k+i\nu_k\frac{\delta\chi_{k}}{\delta \tau_k}\delta \tau_k+p_\mu^k\frac{\delta x_\mu^k}{\delta \tau_k}\bigg)-\sum_{k=0}^{n-1}H(p_k,x_k,\bar{\gamma},\varepsilon_k,x_k)\delta \tau_k\bigg\}\,,
\label{discretized_openworldline_2}
\ea
with $\delta a_k=a_{k+1}-a_k$,
and we divided and multiplied by $1/n=\delta \tau_k$ to take the continuum limit. 
By integrating in $\pi_k$ and $\nu_k$ we obtain $n$ commuting and $n$ anti-commuting
Dirac deltas that set all the $\varepsilon_k=\varepsilon$ and
$\chi_k=\chi$ to the constant conventional Schwinger parameters, which can be understood to be the zero-modes of these \textit{einbeins}. Taking $\delta
\tau_k\to 0$, the gauge-unfixed path integral in Eq.~\eqref{eq:fermion-prop-path_integral} is given by 
\ba
  &\bar{D}_F^A(x_f,x_i)=\int_0^{\infty} d\varepsilon \int d\chi \int_{x(0)=x_i}^{x(1)=x_f} \mathcal{D}x\int \frac{\mathcal{D}p}{(2\pi)^4}\int \mathcal{D}\varepsilon\frac{\mathcal{D}\pi}{2\pi} \int \mathcal{D}\chi \mathcal{D}\nu \,\text{T}\exp\bigg\{\int^1_0 d\tau \nu(\tau)\dot{\chi}(\tau)\nonumber\\
 &-i\int^1_0 d\tau\pi(\tau) \dot{\varepsilon}(\tau) -i\int^1_0 d\tau p_\mu(\tau)\dot{x}_\mu(\tau) -\int_0^1 d\tau H\big(p_\mu(\tau),x_\mu(\tau),\bar{\gamma}_\lambda,\varepsilon(\tau),\chi(\tau)\big)\bigg\}\,.
 \label{continuous_openworldline1}
\ea
This exponentiation is still incomplete since the path integral
involves the time ordered exponential of an operator $\hat{H}$
containing Dirac matrices. To get rid of the time ordering 
and exponentiate  spin dynamics, Fradkin and Gitman suggested  introducing five auxiliary anti-commuting sources
$\{\rho_\lambda(\tau),\rho_\kappa(\tau)\}=0$, which also anti-commute with $\bar{\gamma}_\lambda$. Then Eq.~\eqref{continuous_openworldline1} can be rewritten as
\ba
&\bar{D}_F^A(x_f,x_i)=\int_0^{\infty} d\varepsilon \int d\chi \int_{x(0)=x_i}^{x(1)=x_f} \mathcal{D}x\int \frac{\mathcal{D}p}{(2\pi)^4}\int \mathcal{D}\varepsilon\frac{\mathcal{D}\pi}{2\pi} \int \mathcal{D}\chi \mathcal{D}\nu \exp\bigg\{\int^1_0 d\tau \nu(\tau)\dot{\chi}(\tau)\nonumber\\
 &-i\int^1_0 d\tau \pi(\tau)\dot{\varepsilon}(\tau)-i\int^1_0 d\tau p_\mu(\tau)\dot{x}_\mu(\tau)\bigg\}\nonumber\\
 &\times\exp\bigg\{-\int_0^1 d\tau H\bigg(p_\mu(\tau),x_\mu(\tau),\frac{\delta}{\delta \rho_\lambda(\tau)},\varepsilon(\tau),\chi(\tau)\bigg)\bigg\}\mathrm{T}\exp\bigg\{\int^1_0 d\tau \rho_\lambda(\tau)\bar{\gamma}_\lambda(\tau)\bigg\}\,.
 \label{continuous_openworldline2}
\ea
Now using the Baker-Campbell-Hausdorff relation recursively, we get
the following Magnus expansion:
\ba
\text{T}\,e^{\big\{\int^1_0 d\tau\rho_\lambda(\tau)\bar{\gamma}_\lambda\big\}}
=e^{\big\{\int^1_0 d\tau\rho_\lambda(\tau)\bar{\gamma}_\lambda+\frac{1}{2}\int^1_0 d\tau_1\int^{\tau_1}_0 d\tau_2 \big[\rho_\lambda(\tau_2)\bar{\gamma}_\lambda,\rho_\kappa(\tau_1)\bar{\gamma}_\kappa\big]+\cdots\big\}}\,,
\label{magnus}
\ea
where the ellipses denote a series of higher order nested commutators. However since
\ba
\big[\rho_\lambda(\tau_2)\bar{\gamma}_\lambda,\rho_\kappa(\tau_1)\bar{\gamma}_\kappa\big]=-2\rho_\lambda(\tau_2)\rho_\lambda(\tau_1)\,,
\ea
forms a bosonic subalgebra, it commutes with $\rho_\lambda(\tau)\bar{\gamma}_\lambda$. The higher order nested commutators in Eq.~\eqref{magnus} therefore vanish and 
we get
\ba
 \text{T}\exp\bigg\{\int^1_0 d\tau\rho_\lambda(\tau)\bar{\gamma}_\lambda\bigg\}=\exp\bigg\{\int^1_0d\tau \rho_\lambda(\tau)\bar{\gamma}_\lambda-\int^1_0d\tau_1\int^{\tau_1}_0d\tau_2\, \rho_\lambda(\tau_2)\rho_\lambda(\tau_1)\bigg\}\,.
 \label{magnus2}
\ea

Symmetrizing the double integral to make $\tau_2$ run from $0$ to $1$, and
using the fact that $\{\rho_\lambda(\tau_2),\rho_\lambda(\tau_1)\}=0$, we get for the second term on the r.h.s. of Eq.~\eqref{magnus2}:
\ba
  &\int^1_0 d\tau_1\int^{\tau_1}_0 d\tau_2\,\rho_\lambda(\tau_2)\rho_\lambda(\tau_1) =\frac{1}{2}\int^1_0d\tau_1\int^1_0d\tau_2\bigg[\theta(\tau_2-\tau_1)-\theta(\tau_1-\tau_2)\bigg]\rho_\lambda(\tau_2)\rho_\lambda(\tau_1)\nonumber\\
  &=\frac{1}{2}\int^1_0d\tau_1\int^1_0 d\tau_2\, \text{sign}(\tau_2-\tau_1)\rho_\lambda(\tau_2)\rho_\lambda(\tau_1)\,.
\ea
Noting that $\langle \tau_2 |
(d/d\tau)^{-1}|\tau_1 \rangle=\frac{1}{2}\text{sign}(\tau_2-\tau_1)$, the above scalar product of $\rho_\lambda(t)$ can be cast as into a path integral over an anti-commuting worldline $\zeta_\lambda(t)$
\ba
  &\exp\bigg\{-\frac{1}{2}\int^1_0d\tau_1\int^1_0 d\tau_2\, \text{sign}(\tau_2-\tau_1)\rho_\lambda(\tau_2)\rho_\lambda(\tau_1)\bigg\}\\
  &=\frac{1}{N_5}\int \mathcal{D}^5\zeta(\tau)\exp\bigg\{-\frac{1}{4}\int_0^1 d\tau \zeta_\lambda(\tau)\dot{\zeta}_\lambda(\tau)+\int_0^1 d\tau\rho_\lambda(\tau)\zeta_\lambda(\tau)\bigg\}\,,
  \nonumber
\ea
with the normalization\footnote{This ill-defined factor in front of the Green function is dropped in \cite{Fradkin:1991ci}. We shall keep it since it does not depend on the interaction and is eliminated in perturbation theory as an overall normalization.} given by the path integral in absence of
sources ($\rho_\lambda(\tau)=0$):
\ba
N_5=\int \mathcal{D}^5\zeta(t) \exp\bigg\{-\frac{1}{4}\int_0^1 d\tau \zeta_\lambda(\tau) \dot{\zeta}_\lambda(\tau)\bigg\}\,.
\ea
Anti-periodic boundary conditions in the Grassmanian path integral are assumed
($\zeta_\lambda(1)=-\zeta_\lambda(0)$) to make it invariant under 
shifts of the worldlines. The other term on the r.h.s. of
Eq.~\eqref{magnus2} 
can be written as
\ba
\exp\bigg\{\int^1_0 d\tau \rho_\lambda(\tau)\bar{\gamma}_\lambda\bigg\}= \exp\bigg\{\bar{\gamma}_\lambda\frac{\partial}{\partial \theta_\lambda}\bigg\}\exp\bigg\{\int^1_0 d\tau \rho_\lambda(\tau)\theta_\lambda\bigg\}\bigg|_{\theta_\lambda=0}\,,
\ea
where $\theta_\lambda$, analogously to the $\rho_\lambda$, are five anti-commuting auxiliary constant sources anti-commuting with the
$\bar{\gamma}_\lambda$ matrices too. Hence after these manipulations, we obtain for the time ordered
exponential Eq.~\eqref{magnus2} (which appears in the dressed propagator in Eq.~\eqref{continuous_openworldline2}) the result
\ba
  &\text{T}\exp\bigg\{\int^1_0 d\tau \rho_\lambda(\tau)\bar{\gamma}_\lambda\bigg\}= \frac{1}{N_5}\exp\bigg\{\bar{\gamma}_\lambda\frac{\partial}{\partial \theta_\lambda}\bigg\}\label{magnus3}\\
  &\times \int \mathcal{D}^5\zeta(\tau)\exp\bigg\{-\frac{1}{4}\int_0^1 d\tau \zeta_\lambda(\tau)\dot{\zeta}_\lambda(\tau) +\int^1_0 d\tau \rho_\lambda(\tau)\big(\zeta_\lambda(\tau)+\theta_\lambda\big)\bigg\}\bigg|_{\theta_\lambda=0}\,.\nonumber
\ea
The form of the interaction term suggests that we can shift the path of integration by
$\psi_\lambda(\tau)=\zeta_\lambda(\tau)+\theta_\lambda$. Since
$\zeta_\lambda(1)=-\zeta_\lambda(0)$, we then get $\psi_\lambda(1)+\psi_\lambda(0)=2\theta_\lambda$. Using this, and the commutation relations, we get
\ba
-\frac{1}{4}\int^1_0 d\tau\zeta_\lambda(\tau)\dot{\zeta}_\lambda(\tau)  = -\frac{1}{4}\psi_\lambda(1)\psi_\lambda(0)-\frac{1}{4}\int^1_0 d\tau \psi_\lambda(\tau)\dot{\psi}_\lambda(\tau)\,.
\ea
Hence in these new variables, Eq.~\eqref{magnus3} becomes
\ba
  &\text{T}\exp\bigg\{\int^1_0 d\tau \rho_\lambda(\tau)\bar{\gamma}_\lambda \bigg\}=\frac{1}{N_5}\exp\bigg\{\bar{\gamma}_\lambda\frac{\partial}{\partial \theta_\lambda}\bigg\}\nonumber \\
  &\times\int \mathcal{D}^5\psi(\tau)\exp\bigg\{-\frac{1}{4}\psi_\lambda(1)\psi_\lambda(0)-\frac{1}{4}\int_0^1 d\tau \psi_\lambda(\tau)\dot{\psi}_\lambda(\tau)+\int_0^1 d\tau\rho_\lambda(\tau)\psi_\lambda(\tau)\bigg\}\bigg|_{\theta_\lambda=0}\,.
  \label{magnus4}
\ea
Finally, inserting Eq.~\eqref{magnus4} into Eq.~\eqref{continuous_openworldline2}, we obtain our result for the dressed propagator, 
\ba
&\bar{D}_F^A(x_f,x_i)=\frac{1}{N_5}\exp\bigg\{\bar{\gamma}_\lambda\frac{\partial}{\partial \theta_\lambda}\bigg\} \int_0^{\infty} d\varepsilon_0 \int d\chi_0 \int \mathcal{D}\varepsilon\frac{\mathcal{D}\pi}{2\pi} \int \mathcal{D}\chi \mathcal{D}\nu\int_{x(0)=x_i}^{x(1)=x_f} \mathcal{D}^4x\int \mathcal{D}^5\psi \nonumber\\
  &\times\int \frac{\mathcal{D}^4p}{(2\pi)^4}\exp\bigg\{-\frac{1}{4}\psi_\lambda(1)\psi_\lambda(0)-\frac{1}{4}\int^1_0 d\tau\psi_\lambda(\tau) \dot{\psi}_\lambda(\tau)+\int^1_0 d\tau \nu(\tau)\dot{\chi}(\tau)-i\int^1_0 d\tau\pi(\tau) \dot{\varepsilon}(\tau)\nonumber\\
    &-i\int^1_0 d\tau p_\mu(\tau)\dot{x}_\mu(\tau) -\int_0^1 d\tau H\bigg(p_\mu(\tau),x_\mu(\tau),\psi_\lambda(\tau),\varepsilon(\tau),\chi(\tau)\bigg)\bigg\}\bigg|_{\theta_\lambda=0}\,.
    \label{greenfunction_worldline}
\ea

This expression provides the exact worldline Hamiltonian representation of  the dressed fermion Green function in the background of an arbitrary Abelian gauge field $A_\mu(x)$. 
Shifting $p_\mu(\tau)+gA_\mu(x(\tau))= p_\mu'(\tau)$, and performing 
the path integral in the momentum worldline, we get the equivalent Lagrangian form,
\ba
&\bar{D}_F^A(x_n,x_0)=\frac{1}{N_5}\exp\bigg\{\bar{\gamma}_\lambda\frac{\partial}{\partial \theta_\lambda}\bigg\} \int_0^{\infty} d\varepsilon_0 \int d\chi_0 \int \mathcal{D}\varepsilon\frac{\mathcal{D}\pi}{2\pi} \int \mathcal{D}\chi \mathcal{D}\nu\int_{x(0)=x_i}^{x(1)=x_f} \mathcal{D}^4x\int \mathcal{D}^5\psi \nonumber\\
  &\times\exp\bigg\{-\frac{1}{4}\psi_\lambda(1)\psi_\lambda(0)+\int^1_0 d\tau \nu(\tau)\dot{\chi}(\tau)-i\int^1_0 d\tau\pi(\tau) \dot{\varepsilon}(\tau)-\mathcal{S}\bigg\}\bigg|_{\theta_\lambda=0}\,,\label{greenfunction_worldline_momentumintegrated}
\ea
where the reparametrization and super-gauge invariant  Euclidean action ${\cal S} =\int_0^1 d\tau \mathcal{L}(\tau)$ of the
relativistic spinning particle is given by the Lagrangian \cite{Berezin:1976eg,Barducci:1976qu} 
\ba
 &  \mathcal{L}(\tau)=m^2\epsilon(\tau)+\frac{\dot{x}^2_\mu(\tau)}{4\varepsilon(\tau)}+\frac{1}{4}\psi_\lambda(\tau) \dot{\psi}_\lambda(\tau)-\chi(\tau)\bigg\{m\psi_5(\tau)+\frac{i}{2\varepsilon(\tau)}\dot{x}_\mu(\tau)\psi_\mu(\tau)\bigg\}\nonumber\\
  &-ig\dot{x}_\mu(\tau) A_\mu\big(x(\tau)\big)+i\frac{g}{2}\epsilon(\tau)\psi_\mu(\tau)\psi_\nu(\tau) F_{\mu\nu}\big(x(\tau)\big)\,.
  \label{lagrangian_greenfunction_wordline_momentumintegrated}
\ea

To obtain the classical equations of motion of a single spinning charge in real time, we now need the Green function in Minkowski spacetime. Implementing in the expression above the Wick rotations in Eq.~\eqref{wick_rotation_coordinates} and Eq.~\eqref{wick_rotation_gammamatrices}, and replacing the propertime \textit{einbein} $\varepsilon(\tau)\to i\varepsilon(\tau)$ and its conjugate $\pi(\tau)\to-i\pi(\tau)$, we get 
\ba
&\bar{D}_F^A(x_n,x_0)=\frac{i}{N_5}\exp\bigg\{\bar{\gamma}_\lambda\frac{\partial}{\partial \theta_\lambda}\bigg\} \int_0^{\infty} d\varepsilon_0 \int d\chi_0 \int \mathcal{D}\varepsilon\frac{\mathcal{D}\pi}{2\pi} \int \mathcal{D}\chi \mathcal{D}\nu\int \mathcal{D}^4x\int \mathcal{D}^5\psi \nonumber\\
  &\times\exp\bigg\{-\frac{1}{4}\psi_\lambda(1)\psi^\lambda(0)+\int^1_0 d\tau \nu(\tau)\dot{\chi}(\tau)-i\int^1_0 d\tau\pi(\tau) \dot{\varepsilon}(\tau)-i\mathcal{S}_M\bigg\}\bigg|_{\theta_\lambda=0}\label{greenfunction_worldline_momentumintegrated_minkowski}\,,
\ea
the reparametrization and super-gauge invariant Minkowski action ${\cal S}_M =\int_0^1 d\tau \mathcal{L}_M(\tau)$, and the Langrangian
\ba
&\mathcal{L}_M(\tau) =m^2\varepsilon(\tau)+ \frac{\dot{x}^2_\mu(\tau)}{4\varepsilon(\tau)}-\frac{i}{4}\psi_\lambda(\tau) \dot{\psi}^\lambda(\tau)+\chi(\tau)\bigg\{m\psi_5(\tau)-\frac{1}{2\varepsilon(\tau)}\dot{x}_\mu(\tau)\psi^\mu(\tau)\bigg\}\nonumber\\
  &-g\dot{x}_\mu(\tau) A^\mu\big(x(\tau)\big)-i\frac{g}{2}\epsilon(\tau)\psi_{\mu}(\tau)\psi_{\nu}(\tau)F^{\mu\nu}\big(x(\tau)\big)\,.
  \label{lagrangian_greenfunction_wordline_momentumintegrated_minkowski}
\ea
Here Minkowski superscript labels are implicit everywhere on the r.h.s. 

The result for closed worldlines can be obtained as a special case of this result for the open worldline. Firstly, from 
inspection of  Eq.~\eqref{s_f_superpropertimes}, the fermion
determinant in Eq.~\eqref{trace_propertime} reduces to the special
case of $\chi=0$ in Eq.~\eqref{worldline_hamiltonian_openworldline}. Using Eq.~\eqref{greenfunction_worldline}, the r.h.s. of Eq.~\eqref{trace_propertime} 
can be rewritten as
\ba
 & \Tr\int_0^\infty \frac{d\varepsilon_0}{2\varepsilon_0}e^{-\hat{H}} = \frac{1}{N_4}\Tr\exp\bigg\{\gamma_\mu \frac{\partial}{\partial \theta_\mu}\bigg\}\int_0^\infty \frac{d\varepsilon_0}{2\varepsilon_0} \int \mathcal{D}\varepsilon\frac{\mathcal{D}\pi}{2\pi}\int \mathcal{D}^4x\int\frac{\mathcal{D}^4p}{(2\pi)^4}\int \mathcal{D}^4\psi\nonumber\\&
  \times\exp\bigg\{-\frac{1}{4}\psi_\mu(1)\psi_\mu(0)-\frac{1}{4}\int^1_0 d\tau\psi_\mu(\tau) \dot{\psi}_\mu(\tau)-i\int^1_0 d\tau\pi(\tau) \dot{\varepsilon}(\tau)\nonumber\\
    &-i\int^1_0 d\tau p_\mu(\tau)\dot{x}^\mu(\tau) -\int_0^1 d\tau H\bigg(p_\mu(\tau),x_\mu(\tau),\psi_\mu(\tau),\varepsilon(\tau),0\bigg)\bigg\}\bigg|_{\theta_\mu=0}\,,
\ea
with $\mu=0,1,2,3$ now. The trace over the spin degrees of freedom can be expressed as 
\ba
\frac{1}{N_4}\Tr_\gamma \exp\bigg\{\gamma_\mu \frac{\partial}{\partial \theta_\mu}\bigg\} = \frac{1}{N_4}\Tr_\gamma \bigg\{1+\gamma^\mu \frac{\partial}{\partial \theta^\mu}+\frac{1}{2}\gamma_\mu\frac{\partial}{\partial\theta_\mu} \gamma_\nu\frac{\partial}{\partial \theta_\nu}+\cdots\bigg\} = \frac{1}{N_4}\Tr 1 = 1\,,
\ea
since $\theta^2=0$ and $N_4=2^2$. Setting now $\theta_\mu=0$
in what remains produces in the boundary conditions
$\psi_\mu(1)+\psi_\mu(0)=2\theta_\mu=0$; hence the boundary term in the action $\psi_\mu(1)\psi_\mu(0)=-\psi_\mu(1)\psi_\mu(1)=0$
cancels. Finally taking the remaining trace over spatial positions, we get
\ba
\label{fermionloopdeterminant_worldline}
 &\Tr\int_0^\infty
 \frac{d\varepsilon_0}{2\varepsilon_0}e^{-\hat{H}}
 =\int_0^\infty \frac{d\varepsilon_0}{2\varepsilon_0}\int
 \mathcal{D}\varepsilon\frac{\mathcal{D}\pi}{2\pi}\int_{\rm P}
 \mathcal{D}^4x\int_{\rm AP}
 \mathcal{D}^4\psi\int\frac{\mathcal{D}^4p}{(2\pi)^4}\exp\bigg\{-i\int^1_0 d\tau\pi(\tau)
   \dot{\varepsilon}(\tau)
   \nonumber \\ &-\frac{1}{4}\int^1_0 dt\psi_\mu(\tau)
   \dot{\psi}_\mu(\tau)-i\int^1_0 d\tau p_\mu(\tau)\dot{x}^\mu(\tau) +i\int_0^1 d\tau
   H\bigg(p_\mu(\tau),x_\mu(\tau),\psi_\mu(\tau),\varepsilon(\tau),0\bigg)\bigg\}\,.
\ea

We emphasize that in the normalized open worldline expectation value open boundary conditions hold and the Grassmanian worldline has 5 components corresponding to the action of the 5 Dirac matrices, $\psi_\lambda \leftrightarrow \bar{\gamma}_\lambda$. In contrast, in the normalized closed worldline expectation value of the fermion loop determinant above, the Grassmann worldline has 4 components $\psi_\mu \leftrightarrow \gamma_\mu$.}
Integrating over the einbein variables leads to the well-known reparametrization gauge-fixed closed worldline form for the fermion loop determinant \cite{Strassler:1992zr} derived employing a wide variety of very different methods in the literature.
\section{Compact worldline result for $n^{\rm th}$-rank
polarization tensors}
\label{appendix_c}
We reproduce here a derivation of the master formula for the n$^{\rm th}$-rank polarization tensor,
\ba
 \Pi_{\mu_1\mu_2\cdots\mu_n}(k_1,k_2,\cdots,k_n)&=-\Big\langle i\tilde{J}_{\mu_1}(k_1)i\tilde{J}_{\mu_2}(k_{2})\cdots i\tilde{J}_{\mu_n}(k_{n}) \Big\rangle \medspace\,,
 \label{tensor_definitions}
\ea
by first considering the kinetic terms in the worldline fermion loop
in  Eq.~\eqref{w_l_definition}:
\ba
  &\Pi  = -\int^\infty_0 \frac{d\varepsilon_0}{2\varepsilon_0}e^{-\varepsilon_0m^2}\int_\mathrm{P}\mathcal{D}x \exp\bigg\{-\frac{1}{4\varepsilon_0}\int^1_0 d\tau \dot{x}_\mu^2(\tau)\bigg\}\int_\mathrm{AP} \mathcal{D}\psi \exp\bigg\{-\frac{1}{4}\int^1_0 d\tau \psi_\mu(\tau)\dot{\psi}_\mu(\tau)\bigg\}\,.
  \label{zero_rank_tensor}
\ea
The trace over the bosonic path integral, 
after transforming the path integral to momentum space, gives
\ba
\int d^4x(0) \int_{x(0)}^{x(0)} \mathcal{D}x \exp\bigg\{-\frac{1}{4\varepsilon_0}\int^1_0 d\tau \dot{x}_\mu^2(\tau)\bigg\} = (2\pi)^4\delta^{(4)}(0) \int\frac{d^4p}{(2\pi)^4}e^{-\varepsilon_0p^2}\,.
\ea
Likewise, the trace over the worldline fermionic 
degrees of freedom is simply, 
\ba
 \int^{-\psi(0)}_{\psi(0)} \mathcal{D}\psi \exp\bigg\{-\frac{1}{4}\int^1_0 d\tau\psi_\mu(\tau) \dot{\psi}_\mu(\tau)\bigg\}= \Tr\big(1_{4\times 4})=4\,.
\ea
Putting together the results of the two path integrals, we get
\ba
\Pi = -2(2\pi)^4\delta^{(4)}(0)\int^\infty_0 \frac{d\varepsilon_0}{\varepsilon_0}\int\frac{d^4p}{(2\pi)^4}e^{-\varepsilon_0(m^2+p^2)}\,.
\label{pi}  
\ea

We will next consider the tadpole tensor $\Pi_\mu(k_1)$  corresponding to the expectation value of a single worldline current. It, like all odd rank polarization tensors, will vanish. Shifting $y_\mu(\tau)=x_\mu(\tau)-x_\mu(0)$, and using Eq.~\eqref{normalized_expectation_value_closedworldlines}, it can be written as 
\ba
&\Pi_\mu(k_1) = -(2\pi)^4\delta^{(4)}(k_1)\int^\infty_0
\frac{d\varepsilon_0}{2\varepsilon_0}e^{-\varepsilon_0m^2}\int^{y(1)=0}_{y(0)=0}
\mathcal{D}y \exp\bigg\{-\frac{1}{4\varepsilon_0}\int^1_0 d\tau
\dot{y}_\mu^2(\tau)\bigg\}\label{tadpole_two}\\&\times\int_\mathrm{AP}
\mathcal{D}\psi\exp\bigg\{-\frac{1}{4}\int^1_0 d\tau
\psi_\mu(\tau)\dot{\psi}_\mu(\tau)\bigg\}\int^1_0 d\tau_1 e^{+ik_1\wc
  y(\tau_1)}  ig \bigg(\dot{y}_\mu(\tau_1)+i\varepsilon_0 
\psi_\mu(\tau_1)\psi_\nu(\tau_1)k_\nu^1\bigg)\nonumber\,,
\ea
with the trace over the bosonic zero-modes $x_\mu(0)$ yielding the 
delta function imposing momentum conservation.

We first rewrite
\ba
  &-\frac{1}{4\varepsilon_0}\int^1_0 d\tau \dot{y}_\mu^2(\tau)= \frac{1}{2\varepsilon_0}\int^1_0 d\tau\int^1_0 d\tau' y_\mu(\tau) \eta_{\mu\nu}G_B^{-1}(\tau,\tau') y_\nu(\tau')\,,
  \label{free_bosonic_action_rewritten}
\ea
where we integrated by parts and used $y(0)=y(1)$. Similarly, 
\ba
  -\frac{1}{4}\int^1_0 d\tau\psi_\mu(\tau) \dot{\psi}_\mu(\tau)=-\frac{1}{2}\int^1_0d\tau \int^1_0 d\tau' \psi_\mu(\tau)\eta_{\mu\nu}G_F^{-1}(\tau,\tau')\psi_\nu(\tau')\,,
  \label{free_fermionic_action_rewritten}
\ea
with  
\ba
  G_B^{-1}(\tau,\tau') =
  \frac{1}{2}\frac{d^2}{d\tau^2}\delta(\tau-\tau')\,,\medspace\medspace\medspace 
  G_F^{-1}(\tau,\tau') =
  \frac{1}{2}\frac{d}{d\tau}\delta(\tau-\tau')\,,
  \label{inverse_free_worldline_greenfunctions}
\ea
satisfying periodic and anti-periodic boundary conditions respectively. The corresponding free boson and fermion worldline Green functions are~\cite{Strassler:1992zr}
\ba
  G_B(\tau,\tau')=|\tau-\tau'|-(\tau-\tau')^2\,, \medspace\medspace\medspace  G_F(\tau,\tau')=\text{sign}(\tau-\tau')\,.
  \label{worldline_greenfunctions}
\ea
We will adopt here a strategy slightly different from our previous discussion in \cite{Tarasov:2019rfp}. This will allow us to resum 
the perturbation series in Eq.~\eqref{schematic_perturbative_expansion_vacuum} and simultaneously get rid of the path integrals 
in the QED partition function. The vertex at $\tau_1$ can be re-exponentiated as
\ba
  &ie^{+ik_1\wc y(\tau_1)}\Big(\dot{y}_{\mu}(\tau_1)+i\varepsilon_0\psi_{\mu}(\tau_1)\psi_\nu(\tau_1)k_\nu^1\Big)\nonumber\\
  &=\int d\bar{\theta}_1 d\theta_1\exp\bigg\{i\int^1_0 d\tau j_{\mu\rho}^{B,1}(\tau)y_\rho(\tau)-\int^1_0 d\tau j_{\mu\rho}^{F,1}(\tau)\psi_\rho(\tau)\bigg\}\,,
  \label{exponentiation_currents}
\ea
where $\bar{\theta}_1$ and $\theta_1$ are a pair of Grassmanian  auxiliary
variables, and we defined the boson and fermion sources,
\ba
j_{\mu\rho}^{B,1}(\tau)=\Big(\eta_{\mu\rho}{\theta}_1\bar\theta_1\frac{d}{d\tau_1}+k_\rho^1\Big)\delta(\tau-\tau_1)\,, \medspace\medspace\medspace\medspace j_{\mu\rho}^{F,1}(\tau)=\Big(\eta_{\mu\rho}\theta_1+\varepsilon_0k_\rho^1\bar{\theta}_1\Big)\delta(\tau-\tau_1)\,.
\label{auxiliary_currents_perturbative}
\ea
Plugging Eqs.~\eqref{free_bosonic_action_rewritten}, \eqref{free_fermionic_action_rewritten}, and
\eqref{exponentiation_currents} into Eq.~\eqref{tadpole_two} then gives,
\ba
&\Pi_{\mu_1}(k_1) = -(2\pi)^4\delta^{(4)}(k_1)\int^\infty_0
\frac{d\varepsilon_0}{2\varepsilon_0}e^{-\varepsilon_0m^2}\int d\bar{\theta}_1d\theta_1\int^1_0d\tau_1\\
&\times\int^{y(1)=0}_{y(0)=0}
\mathcal{D}y \exp\bigg\{\frac{1}{2\varepsilon_0}\int^1_0 d\tau \int^1_0 d\tau'
y_\rho(\tau)\eta_{\rho\sigma}G^{-1}_B(\tau,\tau')y_\sigma(\tau')+i\int^1_0 d\tau j_{\mu_1\rho}^{B,1}(\tau)y_\rho(\tau)\bigg\}\nonumber\\
&\times\int^{-\psi(0)}_{\psi(0)}
\mathcal{D}\psi\exp\bigg\{-\frac{1}{2}\int^1_0 d\tau\int^1_0 d\tau'
\psi_\rho(\tau)\eta_{\rho\sigma}G_F^{-1}(\tau,\tau'){\psi}_\sigma(\tau)-\int^1_0 d\tau j^{F,1}_{\mu_1\rho}(\tau)\psi_\rho(t)\bigg\} \nonumber\,.
\ea
The bosonic and fermionic path integrals are now quadratic and can be performed explicitly. We then get
\ba
&  \Pi_{\mu_1}(k_1) = -2g(2\pi)^4\delta^{(4)}(k_1)\int^\infty_0 \frac{d\varepsilon_0}{\varepsilon_0}\int\frac{d^4p}{(2\pi)^4}e^{-\varepsilon_0(m^2+p^2)}\int_0^1 d\tau_1 \int d\bar{\theta}_1d\theta_1\exp\bigg\{\frac{\varepsilon_0}{2}\int_0^1 d\tau\nonumber\\
  &\times  \int_0^1 d\tau'  j_{\mu_1\rho}^{B,1}(\tau)G_B(\tau,\tau')j_{\mu_1\rho}^{B,1}(\tau')-\frac{1}{2}\int_0^1 d\tau \int_0^1 d\tau'j_{\mu_1\rho}^{F,1}(\tau)G_F(\tau,\tau')j_{\mu_1\rho}^{f,1}(\tau')\bigg\}\,.
  \label{pi_mu}
\ea
(Note that there is not a summation in $\mu_1$ in the above expression, as it corresponds to the component of the polarization tensor.) Now noting from Eq.~\eqref{worldline_greenfunctions}, 
\ba
\frac{dG_B}{d\tau}=-\frac{dG_B}{d\tau'}\,, \medspace\medspace\medspace G_B(\tau_1,\tau_1)=0\,,
\ea
and $\bar{\theta}_1^2=\theta_1^2=0$ we get, for the bosonic phase,
\ba
  &\int^1_0 d\tau \int_0^1 d\tau'j_{\mu_1\rho}^{B,1}(\tau)G_B(\tau,\tau')j_{\mu_1\rho}^{B,1}(\tau')=\int^1_0d\tau \int^1_0 d\tau' \\
  &\times  \bigg\{k_{\mu_1}^1 \theta_1\bar{\theta}_1\frac{dG_B}{d\tau'}(\tau,\tau')+k_{\mu_1}^1 \theta_1\bar{\theta}_1\frac{dG_B}{d\tau}(\tau,\tau')+k_1^2G_B(\tau,\tau')\bigg\}\delta(\tau-\tau_1)\delta(\tau'-\tau_1)=0\nonumber\,.
\ea
Analogously, from Eq.~\eqref{worldline_greenfunctions} we have
$G_F(\tau,\tau)=0$. Hence
\ba
&\int_0^1 d\tau \int_0^1 d\tau'j_{\mu_1\rho}^{F,1}(\tau)G_F(\tau,\tau')j_{\mu_1\sigma}^{F,1}(\tau')=\bigg\{\eta_{\mu_1\rho}\theta_1+\varepsilon_0k_\rho\bar{\theta}_1\bigg\} G_F(\tau_1,\tau_1)\bigg\{\eta_{\mu_1\rho}\theta_1+\varepsilon_0k_\rho\bar{\theta}_1\bigg\}=0\,.
\ea
Thus as anticipated, the tadpole tensor vanishes:  $\Pi_{\mu_1}(k_1)=0$.

From Eq.~\eqref{tensor_definitions}, the vacuum polarization tensor is 
\ba
&\Pi_{\mu_1\mu_2}(k_1,k_2)= -\int \frac{d\varepsilon_0}{2\varepsilon_0}e^{-\varepsilon_0m^2}\int_\mathrm{P} \mathcal{D}^4x\exp\bigg\{-\frac{1}{4\varepsilon_0}\int^1_0 d\tau \dot{x}^2_\mu(\tau)\bigg\} \nonumber\\
&\times\int_\mathrm{AP} \mathcal{D}^4\psi \exp\bigg\{-\frac{1}{4}\int^1_0d\tau\psi_\mu(\tau)\dot{\psi}_\mu(\tau)\bigg\}
ig\int^1_0 d\tau_1 \big(\dot{x}_{\mu_1}(\tau_1)+\varepsilon_0\psi_{\mu_1}(\tau_1)\psi_{\nu_1}(\tau_1)k_{\nu_1}\big) e^{+ik_1\wc x(\tau_1)}\nonumber\\
&\times ig\int^1_0d\tau_2 \big(\dot{x}_{\mu_2}(\tau_2)+\varepsilon_0\psi_{\mu_2}(\tau_2)\psi_{\nu_2}(\tau_2)k_{\nu_2})e^{+ik_2\wc x(\tau_2)}
\ea
Again, performing the shift $y_\mu(\tau)=x_\mu(\tau)-x_\mu(0)$, rewriting the free worldline actions with
the help of Eqs.~\eqref{free_bosonic_action_rewritten} and \eqref{free_fermionic_action_rewritten}, and using Eq.~\eqref{exponentiation_currents} to exponentiate the vertices
at $\tau_1$ and $\tau_2$ we get,
\ba
&\Pi_{\mu_1\mu_2}(k_1,k_2)= -4g^2 (2\pi)^4\delta(k_1+k_2)\int \frac{d\varepsilon_0}{2\varepsilon_0}\int\frac{d^4p}{(2\pi)^4}e^{-\varepsilon_0(m^2+p^2)} \int^1_0d\tau_1\int^1_0 d\tau_2\int d\bar{\theta_1}d\theta_1 d\bar{\theta_2}d\theta_2 \nonumber\\
&\times\exp\bigg\{+\frac{\varepsilon_0}{2}\int^1_0d\tau \int^1_0d\tau' \Big(j_{\mu_1\rho}^{B,1}(\tau)+j_{\mu_2\rho}^{B,2}(\tau)\Big)G_{B}(\tau,\tau')\Big(j_{\mu_1\rho}^{B,1}(\tau')+j_{\mu_2\rho}^{B,2}(\tau')\Big)\bigg\}\nonumber\\
&\times\exp\bigg\{-\frac{1}{2}\int_0d\tau \int^1_0d\tau' \Big(j_{\mu_1\rho}^{F,1}(\tau)+j_{\mu_2\rho}^{F,2}(\tau)\Big)G_{F}(\tau,\tau')\Big(j_{\mu_1\rho}^{F,1}(\tau')+j_{\mu_2\rho}^{F,2}(\tau')\Big)\bigg\}\,.
\ea
Then performing the $\tau$ and $\tau'$ integrals gives, 
\ba
&i\Pi_{\mu_1\mu_2}(k_1,k_2)= -4g^2 (2\pi)^4\delta(k_1+k_2)\int \frac{d\varepsilon_0}{2\varepsilon_0}\int\frac{d^4p}{(2\pi)^4}e^{-\varepsilon_0(m^2+p^2)} \int^1_0d\tau_1\int^1_0 d\tau_2\int d\bar{\theta_1}d\theta_1 d\bar{\theta_2}d\theta_2 \nonumber\\
&\times\exp\bigg\{+\varepsilon_0 \eta_{\mu_1\mu_2}\frac{d^2G_B}{d\tau_1d\tau_2}\theta_1\bar{\theta}_1\theta_2\bar{\theta}_2+\varepsilon_0k_{2,\mu_1}\frac{dG_B}{d\tau_1}\theta\bar{\theta}_1+\varepsilon_0k_{1,\mu_2}\frac{dG_B}{d\tau_2}\theta_2\bar{\theta}_2+\varepsilon_0k_1\wc k_2 G_B(\tau_1,\tau_2)\nonumber\\
&-\Big(\eta_{\mu_1\mu_2}\theta_1\theta_2+\varepsilon_0k_{2,\mu_1}\theta_1\bar{\theta}_2+\varepsilon_0k_{1,\mu_2}\bar{\theta}_1\theta_2+\varepsilon_0^2k_1\wc k_2 \bar{\theta}_1\bar{\theta}_2\Big)G_F(\tau_1,\tau_2)\bigg\}\,.
\ea
For the rank-two polarization tensor, the  bosonic vertex contributions and the fermionic contributions do not mix, since the cross-products of the Grassmann variables vanish. Hence the spin and scalar terms in the second-order vacuum polarization tensor are separable. (Note that in loop graphs with $3$ or more photon insertions, 
this is no longer true.) Performing the Grassmanian integrals, we obtain,
\ba
\Pi_{\mu_1\mu_2}(k_1,k_2)=(2\pi)^4\delta^{(4)}(k_1+k_2)\Big(\Pi^{B}_{\mu_1\mu_2}(k_1)+\Pi^{F}_{\mu_1\mu_2}(k_1)\Big)\,,
\ea
where the bosonic and fermionic contributions, respectively, are 
\ba
&\Pi^{B}_{\mu_1\mu_2}(k)=-4g^2\int\frac{d\varepsilon_0}{2\varepsilon_0}\int\frac{d^4p}{(2\pi)^4}e^{-\varepsilon_0(p^2+m^2)}\\
&\times\int^1_0d\tau_1\int^1_0d\tau_2 \bigg\{\varepsilon_0\eta_{\mu_1\mu_2}\frac{d^2G_B}{d\tau_1d\tau_2}-\varepsilon_0^2k_{\mu_1}k_{\mu_2}\frac{dG_B}{d\tau_1}\frac{dG_B}{d\tau_2}\bigg\}e^{-\varepsilon_0k^2 G_B(\tau_1,\tau_2)}\nonumber\,,
\ea
and 
\ba
&\Pi^{F}_{\mu_1\mu_2}(k)=-4g^2\int\frac{d\varepsilon_0}{2\varepsilon_0}\int\frac{d^4p}{(2\pi)^4}e^{-\varepsilon_0(p^2+m^2)}\nonumber\\
&\times\int^1_0d\tau_1\int^1_0d\tau_2\Big(\eta_{\mu_1\mu_2}\varepsilon_0^2k^2G_F^2(\tau_1,\tau_2)-\varepsilon_0^2k_{\mu_1}k_{\mu_2}G_F^2(\tau_1,\tau_2)\Big)e^{-\varepsilon_0k^2G_B(\tau_1,\tau_2)}\,.
\ea
Lastly, performing the $\tau_1$ and $\tau_2$ integrals,
and rephrasing the results in terms of Schwinger parameters~\cite{Tarasov:2019rfp}, we can write the results as 
\ba
\Pi_{\mu_1\mu_2}^B(k)=2g^2\int\frac{d^4p}{(2\pi)^4}\bigg\{\frac{2\eta_{\mu_1\mu_2}}{p^2+m^2}-\frac{(2p_{\mu_1}+k_{\mu_1})(2p_{\mu_2}+k_{\mu_2})}{(p^2+m^2)((p+k)^2+m^2)}\bigg\}\,,
\ea
and
\ba
\Pi_{\mu_1\mu_2}^F(k)=2g^2\int\frac{d^4p}{(2\pi)^4}\frac{k_{\mu_1}k_{\mu_2}-\eta_{\mu_1\mu_2}k^2}{(p^2+m^2)((p+k)^2+m^2)}\,,
\ea
which combines to give for the polarization tensor the result, 
\ba
&\Pi_{\mu\nu}(q,k)=(2\pi)^4\delta^4(q+k)\Big(\Pi_{\mu\nu}^B(k)+\Pi_{\mu\nu}^F(k)\Big)\nonumber\\
&= (2\pi)^4\delta^4(q+k) 4g^2\int\frac{d^4p}{(2\pi)^4}\frac{\eta_{\mu\nu}(p^2+p\wc k+m^2)-p_{\mu}(p_{\nu}+k_{\nu})-p_{\nu}(p_{\mu}+k_{\mu})}{(p^2+m^2)((p+k)^2+m^2)}\,.
\label{pi_munu}
\ea
The structure of these tensors admits an easy generalization to higher orders
since in this formalism the currents for each new photon vertex are additive.
Thus to any $n$-th order in perturbation theory, the corresponding polarization tensor of rank $n$ is given by
\ba
 & \Pi_{\mu_1\ldots\mu_n}(k_1,\ldots,k_n)=-2\frac{g^n}{n!}(2\pi)^4\delta^{(4)}\bigg(\sum_{i=1}^n k_i\bigg) \int\frac{d^4p}{(2\pi)^4}\int^\infty_0 \frac{d\varepsilon_0}{\varepsilon_0}e^{-\varepsilon_0(p^2+m^2)}\nonumber\\
 &\times\prod_{i=1}^n\bigg\{\int^1_0 d\tau_i \int d\bar{\theta}_i d\theta_i\bigg\}
 \exp\bigg\{+\frac{\varepsilon_0}{2}\sum_{ij=1}^n \int^1_0 d\tau \int^1_0 d\tau'  j_{\mu_i\rho}^{B,i}(\tau)G_B(\tau,\tau') j_{\mu_j\rho}^{B,j}(\tau')\nonumber\\
 &-\frac{1}{2}\sum_{ij=1}^n\int^1_0 d\tau \int^1_0 d\tau' j_{\mu_i\rho}^{F,i}(\tau)G_F(\tau,\tau')j_{\mu_j\rho}^{F,j}(\tau')\bigg\}\,,
 \label{bern_kosower}
\ea
which (up to differences in choice of conventions) recovers the results in \cite{Bern:1991aq} (see also \cite{Strassler:1992zr}) for  the Abelian one-particle-irreducible fermion loop diagrams with $n$ photon vertices attached. 

In this construction of Abelian loop-diagrams, the integration in proper time of the terms $\sim k_1\wc k_2 G_B(\tau_1,\tau_2)$ in Eq.~\eqref{bern_kosower} containing no Grassmanian structures will lead to the standard denominator in a scalar loop integral in Feynman parameters. This piece carries no information on the spin and color degrees of freedom of the particles involved. The rest of the terms  (involving derivatives
of $G_B(\tau_1,\tau_2)$ and the fermion Green functions $G_F(\tau_1,\tau_2)$, accompanied always by Grassmanian variables) will contain the numerators of the loop diagrams with all the relevant information about the spin dependence of the particles in the loop. It is worth noting that in a conventional construction of these diagrams in Feynman perturbation theory, one has to go through lengthy algebraic manipulations (of matrix structures and the loop-momentum integrals) to extract spin-dependent pieces of the diagrams. 

The prior discussion can be extended to non-Abelian gauge theories. In this case, the gluon vertex involves the extra quadratic term in the gauge field in $F_{\mu\nu}$, which is introduced and evaluated in Eq.~\eqref{bern_kosower} through a \textit{pinching}  procedure \cite{Strassler:1992zr}, whereby different gluons are pinched to the same point $\tau_i=\tau_j$ in the fermion loop; the extra kinematic factor in the spin-dependent numerator in Eq.~\eqref{bern_kosower} can be extracted in a systematic way leading to the well-known Bern-Kosower rules \cite{Bern:1990cu,Bern:1991aq} to construct one-particle-irreducible fermion loop diagrams in QCD. 

\section{\label{appendix_d}Generalized Bargman-Michel-Telegdi  equations and QED worldline instantons}
The integration of the matter and gauge fields out of the action
formulates QED in terms of many-body first-quantized spinning charges described by even $x(t)$
and odd $\psi(t)$ 0+1-dimensional worldlines which interact through the Lorentz forces,
plus a set of \textit{einbein} functions containing the (super) gauge
symmetry constraints of the worldline action under propertime reparametrizations. Quantum dynamics is implemented via the sum of over all possible worldline paths and \textit{einbein}
configurations. 

The classical equations of motion of the spinning
particle
\cite{Berezin:1976eg,Brink:1976sz,Brink:1976uf,Barducci:1976qu,PhysRevD.17.3247}
are recovered as a special case of the full quantum scenario
\cite{Fradkin:1966zz,Fradkin:1991ci}. We briefly review here the
strategies to obtain so-called ``worldline instanton" configurations from variational
principles
\cite{Affleck:1981bma,Dunne:2005sx,BastianelliSchubertBook,BastianelliLectures,Mueller:2017arw,Mueller:2019gjj}
and the role of \textit{einbeins} in setting the energy-momentum and
helicity-momentum constraints. The steepest descent method applied to
the amplitudes we found yield a slight more complicated picture than
the case of a spinning particle moving in an arbitrary
background. Gauge fields are fully dynamical now; hence, the instanton
configuration of each particle back reacts on to the instanton
configuration of the others, entangled by the particular nature of the spin-spin couplings. 

To illustrate these points, we begin with the
amplitude for the propagation of an electron from $x_i$ to $x_f$ in the presence of $A_\mu(x)$. Following Eq.~\eqref{greenfunction_worldline}, the
worldline Lagrangian is
\ba
  \mathcal{L}=p^\mu \dot{x}_\mu+\pi\dot{\varepsilon}+i\nu\dot{\chi}-\frac{i}{4}\psi^n\dot{\psi}_n-H(p,x,\psi,\varepsilon,\nu)\,, \medspace\medspace\medspace\medspace\medspace\medspace 
  \tau\in[0,1] \,.
  \label{lagrangian_gaugeunfixed}
\ea
We look for the instanton configurations with boundary conditions
$x(1)=x_f$ and $x(0)=x_i$, and \textit{einbein} zero-modes
$\varepsilon(0)=\varepsilon_0$ and $\chi(0)=\chi_0$. The gauge-unfixed
Hamiltonian reads,
\ba
H=-\varepsilon\Big(m^2-(p_\mu+gA_\mu)^2-i\frac{g}{2}\psi_\mu\psi_\nu F^{\mu\nu}\Big)+i\chi\Big(m\psi_5-(p_\mu+gA_\mu)\psi^\mu\Big)\,.
\ea
A closed-loop electron is adequately described by the particular case
$\chi=0$ without loss of generality,
\ba
\mathcal{L}=-p^\mu \dot{x}_\mu-\pi\dot{\varepsilon}+\frac{i}{4}\psi^\mu\dot{\psi}_\mu+H(p,x,\psi,\varepsilon,0)\,, \medspace\medspace\medspace\medspace\medspace\medspace \tau\in[0,1]\,. \label{lagrangian_gaugeunfixed_chi0}
\ea
We consider first the instantons of the fermion determinant
containing the loop electrons with $\chi=0$. The \textit{einbein}
saddle point is trivial; from $\partial_\pi H=0$, we find
$\dot{\varepsilon}=0$. Thus $\varepsilon=\varepsilon_0$ is a constant of
motion. It is constant as well for all worldlines contributing to the
amplitude, since the gauge-fixing term, trivially integrated in $\pi$,
produces $\dot{\varepsilon}=0$ for all configurations. From
\ba
\frac{\partial H}{\partial \varepsilon} =(p_\mu+gA_\mu)^2-m_R^2=0, \medspace\medspace\medspace\medspace\medspace\medspace  m_R^2=m^2-i\frac{g}{2}\psi_\mu\psi_\nu F^{\mu\nu}\,,
\label{energy_momentum_relation_chi0}
\ea
we get the energy-momentum constraint. Similarly,
\ba
\frac{\partial H}{\partial p_\mu} = \dot{x}^\mu = 2\varepsilon (p^\mu+gA^\mu) \medspace\medspace\medspace \to \medspace\medspace\medspace\medspace  p^\mu= \frac{\dot{x}^\mu}{2\varepsilon}-gA^\mu\,,
\label{momentum_velocity_chi0}
\ea
allows one to interpret Eq.~\eqref{energy_momentum_relation_chi0} as fixing the \textit{einbein}:
\ba
\frac{\dot{x}_\mu^2}{4\varepsilon^2}=m_R^2 \medspace\medspace\medspace \to \medspace\medspace\medspace\medspace \varepsilon= \frac{\sqrt{\dot{x}^2}}{2m_R}=\frac{1}{2m_R}\frac{d\tau'}{d\tau}\sqrt{\frac{dx_\mu}{d\tau'}\frac{dx_\mu}{d\tau'}}\,. \label{varepsilon_chi0}
\ea
In the last step above, we changed to arbitrary parametrizations
$\tau'(\tau)$. If $\tau'$ is chosen as the propertime $s$ then $\dot{x}^2(s)=1$; substituting Eq.~\eqref{varepsilon_chi0} into
Eq.~\eqref{lagrangian_gaugeunfixed_chi0}, we recover the gauge-fixed action
in \cite{Barducci:1976qu},
\ba
S=\int^{s_f}_{s_i} ds \Bigg(p_\mu\frac{dx^\mu}{ds}-\frac{i}{4}\psi_\mu\frac{d\psi^\mu}{ds}-H_{cl} \Bigg), \medspace\medspace\medspace\medspace\medspace\medspace \medspace\medspace\medspace  H_{cl}=-\frac{1}{2m_R}\Big(m_R^2-(p_\mu+gA_\mu)^2\Big)\,.
\ea
The gauge-fixed Lagrangian as a function of $\dot{x}_\mu$ can be
obtained now by substituting Eq.~\eqref{momentum_velocity_chi0} in
Eq.~\eqref{lagrangian_gaugeunfixed_chi0}, and substituting the expression for the fixed 
$\varepsilon$ in Eq.~\eqref{varepsilon_chi0}. Since $H_{cl}$ vanishes when this is imposed, we get 
\ba
\mathcal{L}_{cl}=p_\mu\dot{x}^\mu -\frac{i}{4}\psi_\mu\dot{\psi}^\mu = m_R\sqrt{\dot{x}^2}-g\dot{x}_\mu A^\mu-\frac{i}{4}\psi_\mu \dot{\psi}^\mu \,,\medspace\medspace\medspace\medspace\medspace\medspace \tau\in[0,1]\,,
\label{lagrangian_gaugefixed_chi0}
\ea
which is the classical Lagrangian for the spinning charge in a closed loop \cite{Barducci:1976qu}. Helicity-momentum constraints are not present since the fermion determinant has been squared without loss of generality. The gauge-fixed equations of motion are
\ba
  \frac{d}{d\tau}\Bigg(m_R\frac{dx^\rho}{d\tau'}\bigg/\sqrt{\frac{dx_\mu}{d\tau'}\frac{dx^\mu}{d\tau'}}\Bigg)=g\frac{dx_\mu}{d\tau'}F^{\rho\mu}-\frac{1}{2m_R}\sqrt{\frac{dx_\mu}{d\tau'}\frac{dx^\mu}{d\tau'}}\frac{g}{2}\sigma_{\mu\nu}\frac{\partial F^{\mu\nu}}{\partial x_\rho}\,,
\ea
where we defined $\sigma_{\mu\nu}=i[\psi_\mu,\psi_\nu]/2$ and used
again an arbitrary $\tau'(\tau)$. For the propertime, this simplifies
to
\ba
\frac{d}{ds}\Bigg(m_R\frac{dx^\rho}{ds}\Bigg)=g\frac{dx_\mu}{ds}F^{\rho\mu}-\frac{g}{4m_R}\sigma_{\mu\nu}\frac{\partial F^{\mu\nu}}{\partial x_\rho}\,,
\label{classical_equations_externalfield_x}
\ea
and the spin dynamics of the worldlines is described by 
\ba
\frac{d\psi^\rho}{ds}=-\frac{g}{m_R}\psi_\nu F^{\rho\nu}\label{classical_equations_externalfield_psi}\,.
\ea
For homogeneous backgrounds,
Eqs.~\eqref{classical_equations_externalfield_x} and
\eqref{classical_equations_externalfield_psi} reduce to the Bargman-Michel-Telegdi (BMT)
equations \cite{PhysRevLett.2.435} in covariant form, with wide
applications, an example being their fundamental importance in describing the properties of highly relativistic beams of leptons and hadrons in accelerator physics 
\cite{Derbenev:1973ia,CHAO198129,MONTAGUE19841,doi:10.1146/annurev.nucl.50.1.525}. For arbitrary backgrounds, Eqs.~\eqref{classical_equations_externalfield_x} and
\eqref{classical_equations_externalfield_psi} contain corrections to
the BMT equations. The generalization to colored fields are the Wong equations \cite{Wong:1970fu,Barducci:1976wc}, and in gravity, they are the Papapetrou-Mathisson-Dixon equations~\cite{Papapetrou:1951pa,Mathisson:1937zz,Dixon:1970zza}. 

This worldline instanton
configuration can be used to evaluate the action. Since $H$ vanishes
for these configurations, the integration in $\varepsilon_0$ introduces a
trivial (infinite) normalization to the fermion determinant times a
pure phase factor, with the latter given by the coupling of the scalar charged current
to the background $\dot{x}_\mu A^\mu$, as expected. 

We consider now
the instanton configurations of real electrons ($\chi\neq 0$),
\ba
\frac{\partial H}{\partial p_\mu}=2\varepsilon(p^\mu+gA^\mu)-i\chi \psi^\mu=\dot{x}^\mu\,, \medspace\medspace\medspace\medspace\medspace\medspace \frac{\partial H}{\partial\varepsilon } = (p_\mu+gA_\mu)^2-m_R^2=0\,. \label{momentum_velocity_chi}
\ea
Hence in this case,
\ba
\varepsilon=\frac{1}{2m_R}\sqrt{(\dot{x}_\mu+i\chi\psi_\mu)^2}\,.
\label{varepsilon_chi}
\ea
Inserting back Eqs.~\eqref{varepsilon_chi} and \eqref{momentum_velocity_chi}
into Eq.~\eqref{lagrangian_gaugeunfixed_chi0}, using 
$\partial_{\nu}H=\dot{\pi}=0$ and noting that the energy-momentum
relation cancels the even $\varepsilon$ contribution of the Hamiltonian, we obtain
\ba
\mathcal{L}=m_R\sqrt{\big(\dot{x}_\mu-i\chi\psi_\mu\big)^2}-g\dot{x}_\mu A^\mu-\frac{i}{4}\psi_n\dot{\psi}^n -i\chi m\psi^5\,,
\ea
which is the Lagrangian of a classical spinning charge with helicity
constraints \cite{Barducci:1976qu}. Further expanding the odd
contributions in the Lagrangian in $\chi$, and noting $\chi^2=0$, we get
\ba
\mathcal{L}_{cl}=m_R \sqrt{\dot{x}^2}-g\dot{x}_\mu A^\mu - \frac{i}{4}\psi_n\dot{\psi}^n-i\chi\bigg(m\psi^5-\frac{m_R\dot{x}_\mu\psi^\mu}{\sqrt{\dot{x}^2}}\bigg)\,.
\ea
The saddle-point configuration for the odd \textit{einbein}
produces the constraint
\ba
i\frac{\partial L}{\partial \chi} = m\psi^5-\frac{m_R\dot{x}_\mu\psi^\mu}{\sqrt{\dot{x}^2}}= 0 \medspace\medspace \to \medspace\medspace  m\psi^5-  m_R \frac{dx_\mu}{ds}\psi^\mu=0\,,
\ea
with $s(\tau)$ propertime, so that the instanton configurations are
those of the Lagrangian in Eq.~\eqref{lagrangian_gaugefixed_chi0}
supplemented now with the helicity constraint. The helicity-momentum  constraint, is as expected, 
 modified in presence of background fields with magnetic
components for which $m_R\neq m$.

We address now the most general case
in which the background field is sourced by  dynamical
worldlines. The worldline action in the $\ell$-th loop order
contribution to the functionally averaged $n$-point function is, 
\ba
  S &=\sum_{\alpha=1}^n \int^1_0 d\tau_\alpha \bigg\{ \pi_\alpha \dot{\varepsilon}_\alpha+\varepsilon_\alpha m^2+\frac{\dot{x}_\alpha^2}{4\varepsilon_\alpha}+i\nu_\alpha\dot{\chi}_\alpha+i\chi_\alpha\bigg(\frac{\dot{x}^\alpha_\mu \psi^\mu_\alpha}{2\varepsilon_\alpha}-m\psi^5_\alpha\bigg)-\frac{i}{4}\psi_\alpha^n\dot{\psi}^\alpha_n\bigg\}\\
  &+\sum_{\alpha=n+1}^{n+\ell}\int^1_0 d\tau_\alpha \bigg\{\pi_\alpha\dot{\varepsilon}_\alpha+\varepsilon_\alpha m^2+\frac{\dot{x}_\alpha^2}{4\varepsilon_\alpha}-\frac{i}{4}\psi_\mu^\alpha\dot{\psi}^\mu_\alpha\bigg\}\nonumber\\
  &+\frac{1}{2}\sum_{\alpha,\beta=1}^{n+\ell}\int^1_0 d\tau_\alpha \bigg\{g\dot{x}_\mu^\alpha+ig\varepsilon_\alpha\psi^\alpha_\rho\psi^\alpha_\mu\frac{\partial}{\partial x_\rho^\mu}\bigg\}\int^1_0 d\tau_\beta \bigg\{g\dot{x}_\nu^\beta+ig\varepsilon_\beta\psi_\sigma^\beta\psi_\nu^\beta\frac{\partial}{\partial x_\sigma^\beta}\bigg\} D_B^{\mu\nu}(x_\alpha-x_\beta)\nonumber
\ea
The first term is the free action of the $n$ external particles, the
second the free action of the $\ell$ loop particles and the last term
all their Lorentz interactions. The Lagrangian is non-local hence the
saddle points will be solutions of integral equations. Expanding the action
\ba
S = S\big|_{\text{cl}}+ \delta S\big|_{\text{cl}} + \delta^2
S\big|_{\text{cl}}+\cdots
\ea
we get to first order
\ba
  \delta S &= \sum_{\alpha=1}^n \int_0^1 d\tau \Bigg\{\bigg[m_R^2-\frac{1}{4\varepsilon_\alpha^2}(\dot{x}_\alpha^\mu+i\chi_\alpha\psi_\alpha^\mu)^2\bigg]\delta \varepsilon_\alpha+i\bigg[\frac{\dot{x}_\mu^\alpha\psi^\mu_\alpha}{2\varepsilon_\alpha}-m\psi^5_\alpha\bigg]\delta \chi_\alpha\nonumber\\
    &+i\bigg[\chi_\alpha m-\frac{\psi_\alpha^5}{2}\bigg]\delta\psi_5^\alpha-i\bigg[\frac{\dot{\psi}^\rho_\alpha}{2}+\chi_\alpha\frac{\dot{x}_\alpha^\rho}{2\varepsilon_\alpha}-g\psi_\nu F^{\rho\nu}_{all}\bigg]\delta \psi^\alpha_\rho\nonumber\\
    &-\bigg[\frac{d}{d\tau}\bigg(\frac{1}{2\varepsilon_\alpha}(\dot{x}_\alpha^\rho+i\chi_\alpha\psi^\rho_\alpha)\bigg)-g\dot{x}_\mu^\alpha F^{\rho\mu}_{all}-i\frac{g}{2}\varepsilon_\alpha\psi_\mu^\alpha\psi_\nu^\alpha\frac{\partial F_{all}^{\mu\nu}}{\partial x_\rho^\alpha}\bigg]\delta x_\rho^\alpha\Bigg\}\nonumber\\
    &+\sum_{\alpha=n+1}^{n+\ell} \int_0^1 d\tau \Bigg\{\bigg[m_R^2-\frac{\dot{x}_\alpha^2}{4\varepsilon_\alpha^2}\bigg]\delta \varepsilon_\alpha-i\bigg[\frac{\dot{\psi}^\rho_\alpha}{2}-g\psi_\nu F^{\rho\nu}_{all}\bigg]\delta \psi^\alpha_\rho\nonumber\\
    &-\bigg[\frac{d}{d\tau}\bigg(\frac{\dot{x}_\alpha^\rho}{2\varepsilon_\alpha}\bigg)-g\dot{x}_\mu^\alpha F^{\rho\mu}_{all}-i\frac{g}{2}\varepsilon_\alpha\psi_\mu^\alpha\psi_\nu^\alpha\frac{\partial F_{all}^{\mu\nu}}{\partial x_\rho^\alpha}\bigg]\delta x_\rho^\alpha\Bigg\}\label{semiclassical_expansion_action}
\ea
where the field strength tensor is given recursively in terms of the
scalar and spinorial charged currents and the \textit{einbeins}
$F^{\mu\nu}_{all}(x):=\partial^\mu A^\nu_{all}(x)-\partial^\nu
A^\mu_{all}(x)$ with
\ba
A^\mu_{all}(x)= g \sum_{\beta=1}^{n+\ell}\int^1_0 d\tau D_B^{\mu\nu}\big(x-x_\beta\big)\dot{x}_\beta+ig\sum_{\beta=1}^{n+\ell}\int^1_0 d\tau \frac{\partial}{\partial x_\rho^\beta } D_B^{\mu\nu}\big(x-x_\beta(t)\big)\varepsilon_\beta \psi^\beta_{\rho} \psi^\beta_\nu
\ea
and the renormalized mass has been defined
\ba
m_R^2=m^2+ig\sum_{\beta=1}^{n+\ell}\int^1_0 d\tau \frac{\partial}{\partial x_\rho^\beta } D_B^{\mu\nu}\big(x-x_\beta\big)\varepsilon_\beta \psi^\beta_{\rho} \psi^\beta_\nu
\ea
The condition $\delta S=0$ yields the classical equations of motion
satisfied by $n+\ell$ spinning charges in interaction (the ones
present in the $\ell$-th order contribution to the functionally
averaged $n$-point function) plus the energy-momentum and
helicity-momentum constraints. However, unlike the case of an external
field, in which the interactions are given for all particles at once,
the motion described here addresses a much more involved
scenario. Since the worldlines themselves - including the classical
self-interaction term $\beta=\alpha$ - do participate in creating the
field at all retarded and advanced times, the classical equations for
the $\alpha$-th particle worldlines $x_\alpha(\tau)$ and $\psi_\alpha(\tau)$
are given by integral equations. 
 
\section{\label{appendix_e}Computing the $\mathrm{R}^{nm}$ and $\Phi^{nm}$ dressings.}
We will compute here the integrals in the photon momentum $\v{k}$ found in Section \ref{section_5_2}, that are required for the evaluation of the real and imaginary parts of the IR dressings, $\mathrm{R}^{nm}$ and $\Phi^{nm}$. 
In $\mathrm{R}^{nm}$ in Eq.~\eqref{Rnm} and Eq.~\eqref{r_nm_asymptotic}, the required 
integrals are of the form
\ba 
I(\beta_p,\beta_q) =\int_\lambda^\Lambda \frac{d^3\v{k}}{(2\pi)^3} \frac{1}{2\omega_k}\frac{\beta_p\wc \beta_q}{(k\wc\beta_p)(k\wc\beta_q)} = \int_\lambda^\Lambda\frac{\omega_k^2d\omega_k}{2\pi}\int\frac{d\Omega_k}{(2\pi)^2}\frac{1}{2\omega_k}\frac{\beta_p\wc \beta_q}{(k\wc\beta_p)(k\wc\beta_q)} \label{I_bp_bq}\,,
\ea 
where $d\Omega_k=\sin\theta_kd\theta_kd\phi_k$, $k=\omega_k(1,\hat{\bm{k}})$, $\beta=(1,\bm{\beta})$ and $k\wc \beta = \omega_k(1-\hat{\bm{k}}\wc\bm{\beta})$. 
 Introducing a Schwinger parameter, the angular integral can be performed as follows:
\ba
 &\int d\Omega_k \frac{1}{(k\wc \beta_q) (k\wc \beta_p)}=\int d\Omega_k \int_0^1  \frac{dy}{\big(y\medspace k\wc \beta_q+(1-y)k\wc \beta_p\big)^2}\nonumber\\
 &= \frac{2\pi}{\omega_k^2}\int^1_0 dy\int^{+1}_{-1}\frac{d\cos\theta_k}{\big(1-\cos\theta_k|\bm{\beta}_qy+(1-y)\bm{\beta}_p|\big)^2} = \frac{4\pi}{\omega_k^2}\int^1_0 \frac{dy}{1-\big(\bm{\beta}_qy+\bm{\beta}_p(1-y)\big)^2}\,.
\ea 
Noticing $1-\big(\bm{\beta}_qy+\bm{\beta}_p(1-y)\big)^2=\big(\beta_qy+\beta_p(1-y)\big)^2$ and collecting all the terms, Eq.~\eqref{I_bp_bq} gives
\ba 
&I(\beta_p,\beta_q)= \frac{1}{4\pi^2} \int_{\lambda}^\Lambda \frac{d\omega_k}{\omega_k} \int^{1}_0 \frac{dy}{\big(\beta_qy+\beta_p(1-y)\big)^2}\label{I_bp_bq_result}\\
&= \frac{1}{4\pi^2} \frac{\beta_q\wc \beta_p}{\sqrt{(\beta_q\wc\beta_p)^2-\beta_q^2\wc \beta_p^2}}\text{Atanh}\Bigg(\frac{\sqrt{(\beta_q\wc\beta_p)^2-\beta_q^2 \beta_p^2}}{\beta_q\wc \beta_p}\Bigg)\log \frac{\Lambda}{\lambda}=\frac{1}{4\pi^2}\gamma^{pq}\cosh\gamma^{pq}\log \frac{\Lambda}{\lambda}\nonumber\,,
\ea
where the relative cusp angle between the 4-vectors $\beta_p$ and $\beta_q$ is defined in Eq.~\eqref{relative_angle} of Section \ref{section_5_2}.

On the other hand, the integrals required to compute $\Phi^{nm}$ in Eqs.~\eqref{phi_nm} and \eqref{phi_nm_asymptotic} are of the form
\ba 
J(\beta_p,\beta_q)=\lim_{\epsilon\to 0} \bigg\{J_+^\epsilon(\beta_p,\beta_q)-J_-^\epsilon(\beta_p,\beta_q)\bigg\}\,,
\ea 
where
\ba
J_{\pm}^\epsilon(\beta_q,\beta_p)=\int^\Lambda_{\lambda}\frac{d^3\v{k}}{(2\pi)^3}\frac{1}{2\omega_k} \frac{1}{k\wc\beta_p}\frac{\beta_p\wc\beta_q}{k\wc (\beta_p-\beta_q)\pm i\epsilon}\,.
\ea 
Introducing a Schwinger parameter to perform the angular integration, in the same way as for Eq.~\eqref{I_bp_bq}, this gives
\ba 
J_{\pm}^\epsilon(\beta_q,\beta_p) = \int^1_0 dy \int^\Lambda_{\lambda}\frac{d^3\v{k}}{(2\pi)^3}\frac{1}{2\omega_k} \frac{1}{k\wc\beta_p}\frac{\beta_p\wc\beta_q}{(1-y)k\wc \beta_p+yk\wc (\beta_p-\beta_q)\pm iy\epsilon} = \frac{1}{8\pi^2}\int^\Lambda_{\lambda}d\omega_k\nonumber\\\times\int^1_0 dy \frac{\beta_p\wc\beta_q}{|\v{\beta}_p-y\v{\beta}_q|}\bigg\{\frac{1}{(1-y)\omega_k-\omega_k|\v{\beta}_p-y\v{\beta}_q|\pm iy\epsilon}-\frac{1}{(1-y)\omega_k+\omega_k|\v{\beta}_p-y\v{\beta}_q|\pm iy\epsilon}\bigg\}\,.
\ea 
Using that
\ba 
2\pi\delta(q)= \lim_{\epsilon\to 0} \int^\infty_0 dte^{-iqt-\epsilon t}+\int_{-\infty}^0 dt e^{-iqt+\epsilon t} = \lim_{\epsilon\to 0} \frac{2\epsilon}{q^2+\epsilon^2} \,,
\ea 
then
\ba 
&J(\beta_p,\beta_q) = \frac{i}{8\pi^2}(\beta_p\wc\beta_q)\int^\Lambda_\lambda d\omega_k\int^1_0 \frac{ dy}{y} \frac{1}{|\v{\beta}_p-y\v{\beta}_q|} \nonumber\\
&\times \bigg\{2\pi\delta\bigg(\frac{\omega_k}{x}\Big(1-y+|\v{\beta}_p-y\v{\beta}_q|\Big)\bigg)-2\pi\delta\bigg(\frac{\omega_k}{x}\Big(1-y-|\v{\beta}_p-y\v{\beta}_q|\Big)\bigg)\bigg\}\,.
\ea 
The first of these two Dirac deltas above does not contribute, since the argument is always positive. The second can be rewritten as
\ba 
2\pi\delta\bigg(\frac{\omega_k}{x}\Big(1-y-|\v{\beta}_p-y\v{\beta}_q|\Big)\bigg) =\Bigg| \frac{y_0(1-y_0)}{\omega_k \sqrt{(\beta_p\wc\beta_q)^2-\beta_p^2\beta_q^2}}\Bigg|2\pi\delta(y-y_0)\,,
\ea 
where $y_0$ is the solution of $|\v{\beta}_p-y_0\v{\beta}_q|=(1-y_0)$. Then finally
\ba 
J(\beta_p,\beta_q)=-\frac{i}{4\pi}\frac{\beta_p\wc\beta_q}{\sqrt{(\beta_p\wc\beta_q)^2-\beta_p^2\beta_q^2}}\int^\Lambda_\lambda \frac{d\omega_k}{\omega_k} = -\frac{i}{4\pi}\coth\gamma^{pq}\log\frac{\Lambda}{\lambda}\,,\label{J_bp_bq_result}
\ea 
where the relative cusp angle is the one in Eq.~\eqref{relative_angle} of Section \ref{section_5_2} again. 
%

%

\bibliography{main}
\bibliographystyle{utphys.bst}

\end{document}